\documentclass[10pt, twocolumn]{article}   	
\usepackage{geometry}                		
\pdfoutput=1                 		
\usepackage{graphicx}
\usepackage{subfigure}				
\usepackage{amssymb}
\usepackage{url}
\title{Classification of the Extremely Low Frequency Magnetic Field Radiation Measurement from the Laptop Computers}
\author{Darko Brodi\'c$^1$, Alessia Amelio$^2$}
\date{\small $^1$University of Belgrade, Technical Faculty in Bor, Vojske Jugoslavije 12, 19210 Bor, Serbia, dbrodic@tf.bor.ac.rs\\$^2$Institute for High Performance Computing and Networking, National Research Council of Italy, CNR-ICAR, Via P. Bucci 41C, 87036, Rende (CS), Italy,\\amelio@icar.cnr.it}							

\begin{document}
\twocolumn[
\maketitle

\begin{@twocolumnfalse}
\begin{center}
{\bf Abstract}
\end{center}
{\bf \small The paper considers the level of the extremely low-frequency magnetic field, which is produced by the laptop computers. The magnetic field, which is characterized by the extremely low frequencies up to 300 Hz is measured due to its hazardous effects to the laptop user's health. The experiment consists of testing 13 different laptop computers in normal operation conditions. The measuring of the magnetic field is performed in the adjacent neighborhood of the laptop computers. The measured data are presented and then classified. The classification is performed by the K-Medians method in order to determine the critical positions of the laptop. At the end, the measured magnetic field values are compared with the critical values suggested by different safety standards. It is shown that some of the laptop computers emit a very strong magnetic field. Hence, they must be used with extreme caution.\\

}
\end{@twocolumnfalse}
\vspace{1cm}
]

\section{Introduction}
The static and extremely low frequency electromagnetic energy occurs naturally or in association with the generation and transmission of electrical power and with the use of power in some appliances \cite{[1]}. Static and low-frequency electric and magnetic fields are produced by both natural and man-made sources. The natural fields are static or very slow in variation. However, the main source of their peaks represents man-made sources connected with the electrical power. Most man-made sources radiate at extremely low frequencies (ELF), which are defined as frequencies up to 300 Hz \cite{[2]}. Due to ELF fields interaction with biological systems, the potential hazards to laptop users health are in the focus.

Laptop is a portable personal computer that is used in a different location. It can be powered by alternative current or battery. The benefit of having a battery is that it can provide power to the laptop if there is no alternative current power supply. Also, it has an additional external component called alternative current (AC). It enables that laptop to be powered by AC supply and to charge a battery.

In the last decade, the use of laptops has rapidly expanded. Because the laptop is a portable computer, it is typically in close contact with the user's body. Consequently, the laptop is in contact with the areas of skin, blood, lymph, bones, etc. Common use of the laptop in such circumstances might cause negative effects to the user's health. Hence, raised concern about the adequate use of the laptop is mandatory \cite{[3]}. It is mainly based on the effect of the extremely low magnetic field radiation at a frequency below 300 Hz. The risk of the extremely low frequency magnetic exposure of laptop users has been investigated in ref. \cite{[4]}. It states that  the power supply induces strong intracorporal electric current densities in the fetus and in the adult subject. Consequently, they are from 182 to 263\% and from 71 to 483\% higher than ICNIRP 98 basic restriction recommended in order to prevent adverse health effects. Also, some research was conducted on EMF exposure of biological systems like birds, which confirmed  generally changes in their behavior, reproductive success, growth and development, physiology and endocrinology, and oxidative stress \cite{[5]}. Furthermore, EMFs have negative effects by   increasing the risks of the illnesses such as leukemia \cite{[6]}, brain cancer \cite{[7]}, amyotrophic lateral sclerosis \cite{[8]}, clinical depression \cite{[9]}, and Alzheimer�s disease \cite{[10]}.

The critical level of the radiation above which the environmental conditions can be unsafe for laptop users is defined as the reference limit level. In \cite{[11]}, it is defined as 0.3 $\mu$T. It can be noted that the international commission for the non-ionized radiation ICNIRP put the EMF reference limit value differently for the people and for the employee. Accordingly, the reference limit level is 5/$f$ for the people and 25/$f$ for the employee \cite{[12]}, where $f$ represents the frequency of the EMF radiation.

In this paper, we address the problem of the magnetic field radiation received from the laptop computers characterized by the extremely low frequencies up to 300 Hz. The measurement of the EMR is carried out on 13 different laptop computers. The laptops are tested in so-called �normal� operation condition (typical office work). Then, the magnetic field measurement results are presented. Apart from previously published papers \cite{[4]}, \cite{[13]} which do not classify data, we propose an unsupervised classification of measured data by K-Medians method in order to determine the critical positions of the laptop. This process segments measured data into classes leading to the establishment of different levels of dangerous and non-dangerous zones in the laptop neighborhood. Furthermore, it defines the areas at the laptop which are more or less dangerous to be in direct contact with the users. Then, the risk assessment of the low frequency magnetic induction from laptops to their users according to the proposed reference limit level given in literature \cite{[6]} is discussed. These results are invaluable for the laptop users to prevent their exposure to dangerous level of magnetic field radiation.

The paper is organized as follows. Section 2 describes the magnetic field measurement and classification procedure. Section 3 defines the experiment. Section 4 presents the measurement results. Then, it performs data classification by K-Medians method and discusses the obtained classification results. Section 5 makes conclusions and points out the future research work direction.

\section{Subject \& Methods}

\subsection{Magnetic field}
The method consists of measuring the extremely low-frequency magnetic field produced by the laptop computers. The laptop represents an all-in-one design with the functionality of a desktop computer. It includes the following components: computer, monitor, keyboard, touchpad, camera, microphone, speaker and battery. 

In order to perform properly, the laptop components are supplied by current $I$. As a consequence of the current $I$ flowing through electronic or electrical components, the magnetic-field is induced. According to the Biot-Savart law, the magnetic field $B$ is generated by a steady current $I$. It is given as:

\begin{equation}
B=\frac{\mu_0 I}{4\pi} \int_{wire} \frac{dl \cdot \hat{r}}{r^2}
\end{equation}

where the integral sums over the wire length, the vector $dl$ is the vector line element with direction as the current $I$, $\mu_0$ is the magnetic constant, $r$ is the distance between the location of $dl$ and the location where the magnetic field is calculated, and $\hat{r}$ is a unit vector in the direction of $r$. 

\begin{figure}[!ht]
\begin{center}
\includegraphics[width=7cm, height=7cm, keepaspectratio]{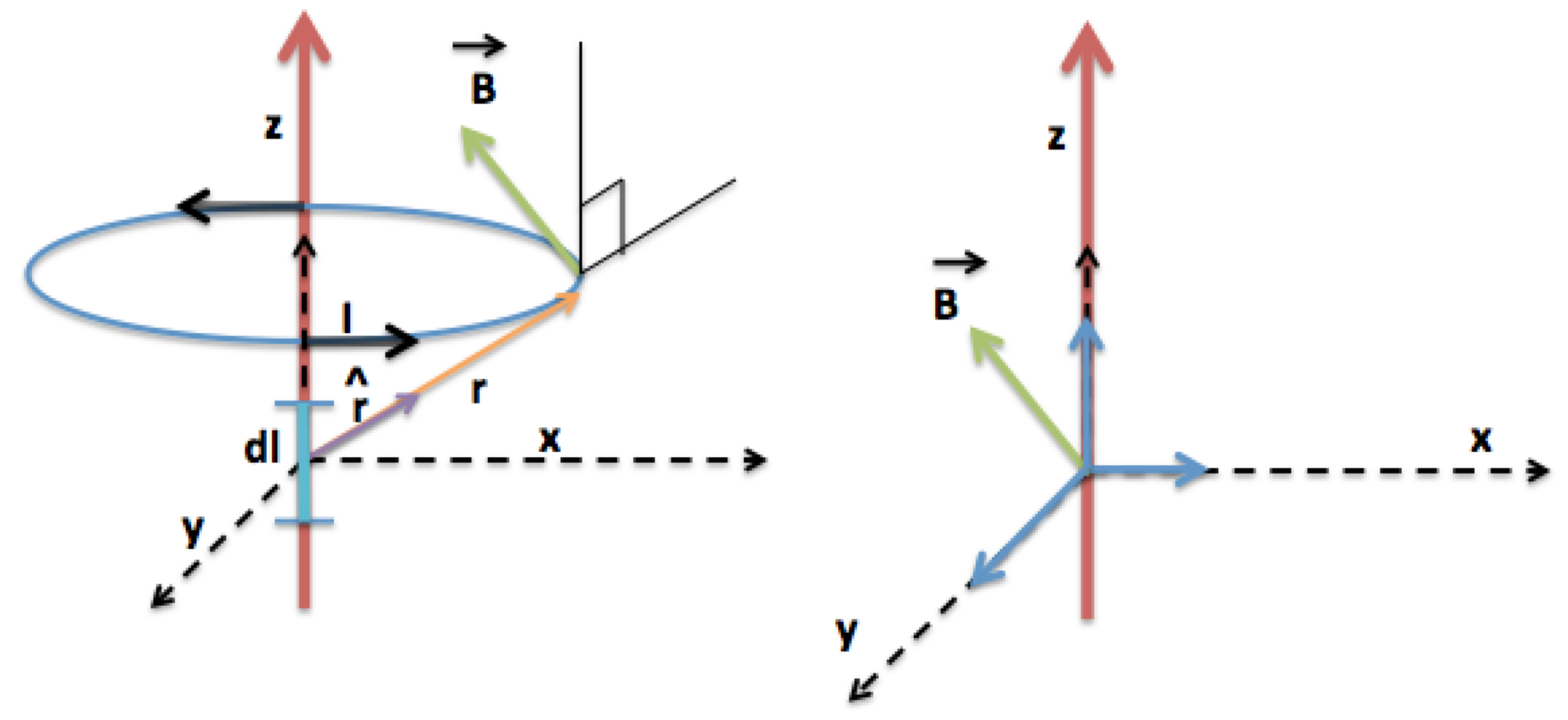}
\caption{The Biot-Savart law (a) and the three components (in blue) of the magnetic induction $B$ along the axes $x$, $y$, $z$ (b).}
\label{Figure1}
\end{center}
\end{figure}

While working with the laptop computers, the users are exposed to the magnetic field. In the circumstances of uniform magnetic, the time dependence on the field is the same in all points of the exposed subjects \cite{[4]}. The measurement of the magnetic induction $B$ representing a vector can be split into 3 independent scalar fragments parallel to each one, i.e. $B_x$, $B_y$ and $B_z$ in the direction of the axes $x$, $y$ and $z$. It is given as \cite{[4]}:
\begin{equation}
B(r,t) = B(t) = B_x(t) \hat{x} + B_y(t) \hat{y} + B_z(t) \hat{z}
\end{equation}

where $t$ is the time, and $\hat{r}$ is a unit vector in the direction of $r$ sometimes called position vector. Fig.\ref{Figure1} depicts the Biot-Savart law (a) and the three components of the magnetic induction along the axes $x$, $y$, $z$ (b).

The measuring devices usually register the scalar components of the magnetic induction: $B_x$, $B_y$ and $B_z$. Taking into account these values, the root mean square (RMS) of magnetic induction $B$ is calculated as:

\begin{equation}
B = \sqrt{B_x^2 + B_y^2 + B_z^2}
\end{equation}

\subsection{Measuring devices}

The measurement of magnetic field is performed by using 3D EMF tester at the position where the influence of the environment magnetic field is negligible. It implies a magnetic field lower or equal to 0.01 $\mu$T. Magnetic field measurement is performed by EMF measuring device Lutron EMF-828 with separate probe, including sensing head. The calibration of the measurement device is performed according to ISO 9001 by the producer of the equipment. Lutron EMF-828 measures the scalar components of the magnetic induction $B_x$, $B_y$ and $B_z$ from 0.01 $\mu$T to 2 mT in the extremely low-frequency range, i.e. between 30 and 300 Hz. It has three measurement extents: 20 $\mu$T, 200 $\mu$T and 2000 $\mu$T. The precision of the measurement is of the order of 0.01 $\mu$T for the measurement extent of 20 $\mu$T, 0.1 $\mu$T for the measurement extent of 200 $\mu$T and 1 $\mu$T for the measurement extent of 2000 $\mu$T. 

The magnetic field is additionally tested with EMF measuring device Aaronia Spectran NF-5010, whose frequency range is from 1 Hz to 1 MHz to confirm the results obtained from Lutron EMF-828. This device is fully compliant to TCO1 (from 500 Hz to 2 kHz) and TCO2 (from 2kHz to 400 kHz). It also allows measurement of extremely low frequency EMF values from 1 Hz to 500 Hz, and higher frequency values from 400 kHz to 1 MHz.

\subsection{Feature description}

Identification of laptop's points with similar magnetic field value is of great interest in this context. In fact, discovering zones where high-risk points are located, with very high magnetic field value, is a really important information for users who are daily in touch with their laptops. For each laptop, we consider the typical points determined by the experiment, which are on the top and on the bottom of its body. The RMS of the magnetic field $B$ measured for each of those points is considered as a feature. So, we have different features from the top and the bottom points of the laptop. Each feature is given as a one-dimension real value.

\subsection{Classification}
We adopted automatic classification by center-based clustering for detection of classes of laptop's points exhibiting a similar magnetic field value. Two algorithms belonging to this category are K-Medians \cite{[14]} and K-Means \cite{[15]}. In one-dimension, similar approaches turned out to be a good methodology in different domains for measured values quantization into a certain number of categories \cite{[16]}, \cite{[17]}, \cite{[18]}, \cite{[19]}. We chose to adopt the K-Medians algorithm because it is more robust to outliers than K-Means and produces more compact clusters \cite{[20]}. Our classification is unsupervised. We don't have any information on the classes of magnetic field value. We find them by clustering the laptop's points with associated magnetic field value. Then, for each cluster, we evaluate if the magnetic field value of its points corresponds to a reasonable "hazard level." It is performed by considering that the proposed limit for safe use of the electronic or computer is 0.3 $\mu$T \cite{[11]}. Finally, we discuss the correspondence between the found clusters of points and the zones of the laptop where these points are located for discovery of dangerous and no dangerous zones.

\subsection{Algorithm description}
K-Medians is an effective and fast algorithm whose aim is to group the points represented by the magnetic field value in a certain number $k$ of non-uniform clusters. The number $k$ of clusters is an input parameter, and it is fixed a priori. 

The first step consists in choosing $k$ centroids, one for each cluster. It is critical, because different choices of the centroids can determine different final results in clusters. The second step is to associate each laptop's point with its nearest centroid. It produces a first raw partitioning of the points in clusters.  The third step is to generate $k$ new centroids as the "median points" of the previously detected clusters. After that, a new re-assignment of the points with the new nearest centroids is performed. The second and third steps are repeated multiple times as centroids change their location, until no more centroid modifications occur. 

K-Medians minimizes an objective function as:
\begin{equation}
J = \sum_{j=1}^k \sum_{i=1}^n || x_i^{(j)} - c_j ||
\end{equation}
where $|| x_i^{(j)} - c_j ||$ is the Manhattan distance measure between the laptop's point magnetic field value $x_i^{(j)}$ in cluster $j$ and the centroid $c_j$ in cluster $j$. 

\begin{figure}[!ht]
\begin{center}
\subfigure[Iteration 1]{
\includegraphics[width=6cm, height=6cm, keepaspectratio]{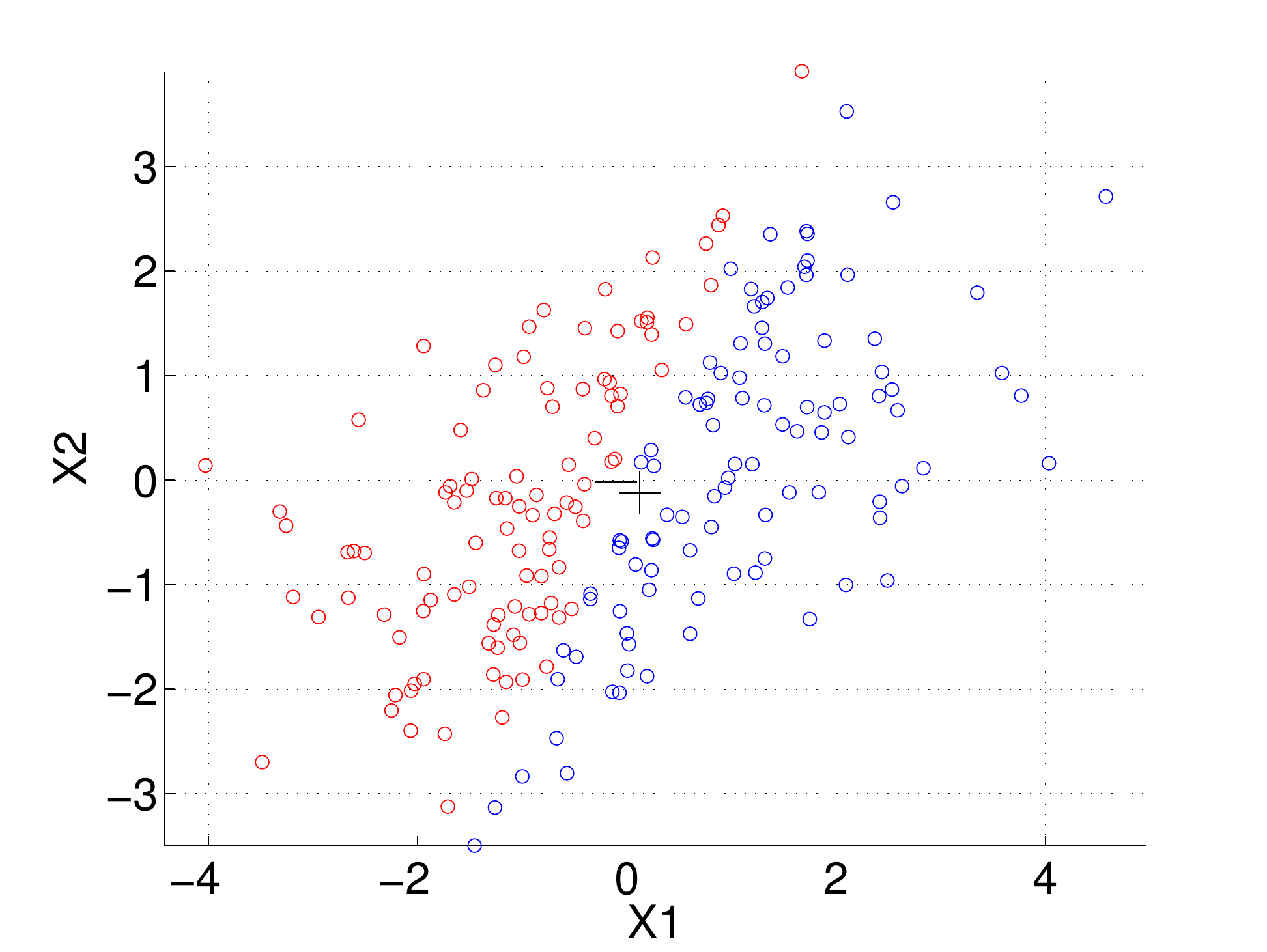}
}
\subfigure[Iteration 2]{
\includegraphics[width=6cm, height=6cm, keepaspectratio]{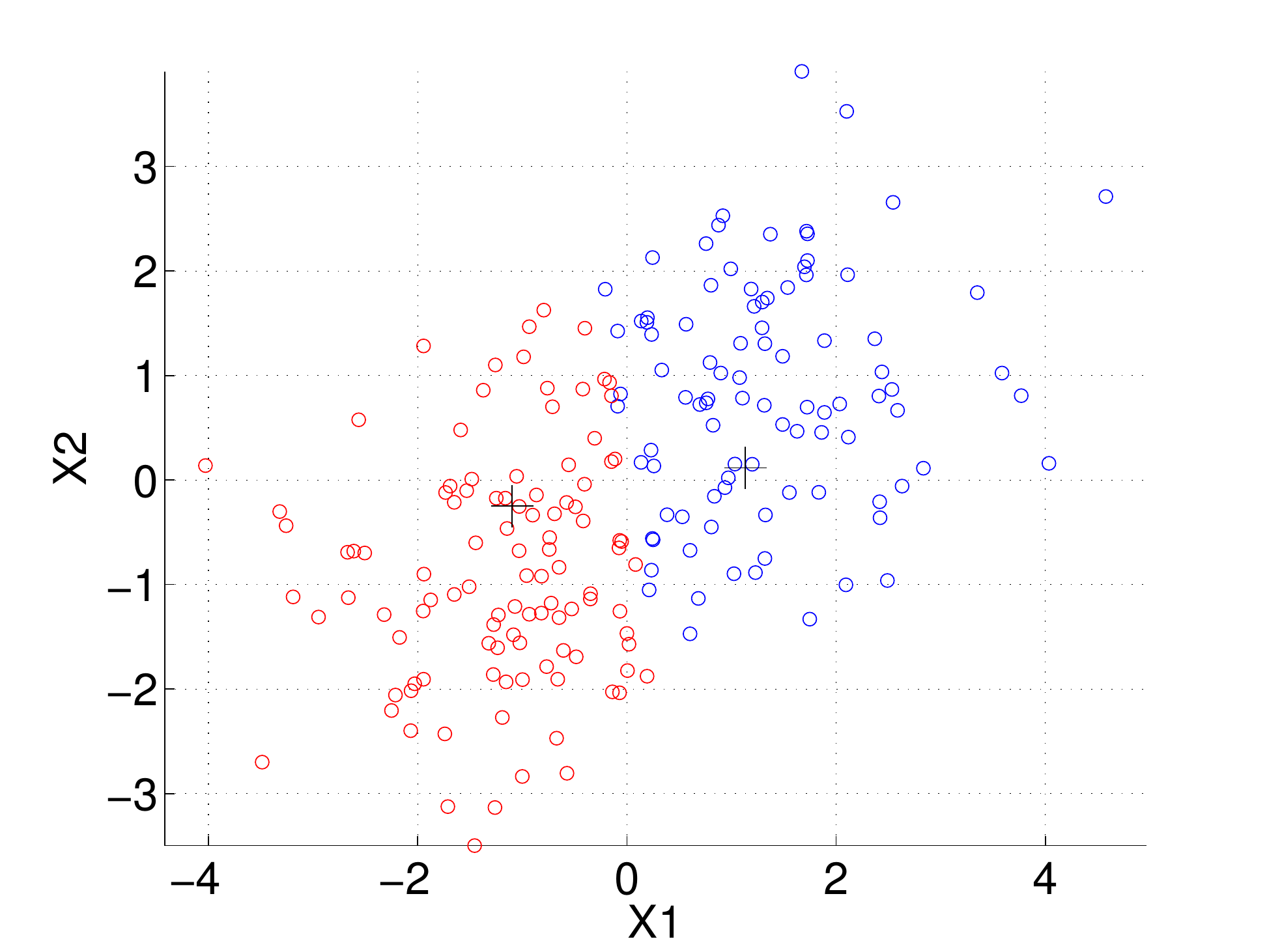}
}
\subfigure[Iteration 3]{
\includegraphics[width=6cm, height=6cm, keepaspectratio]{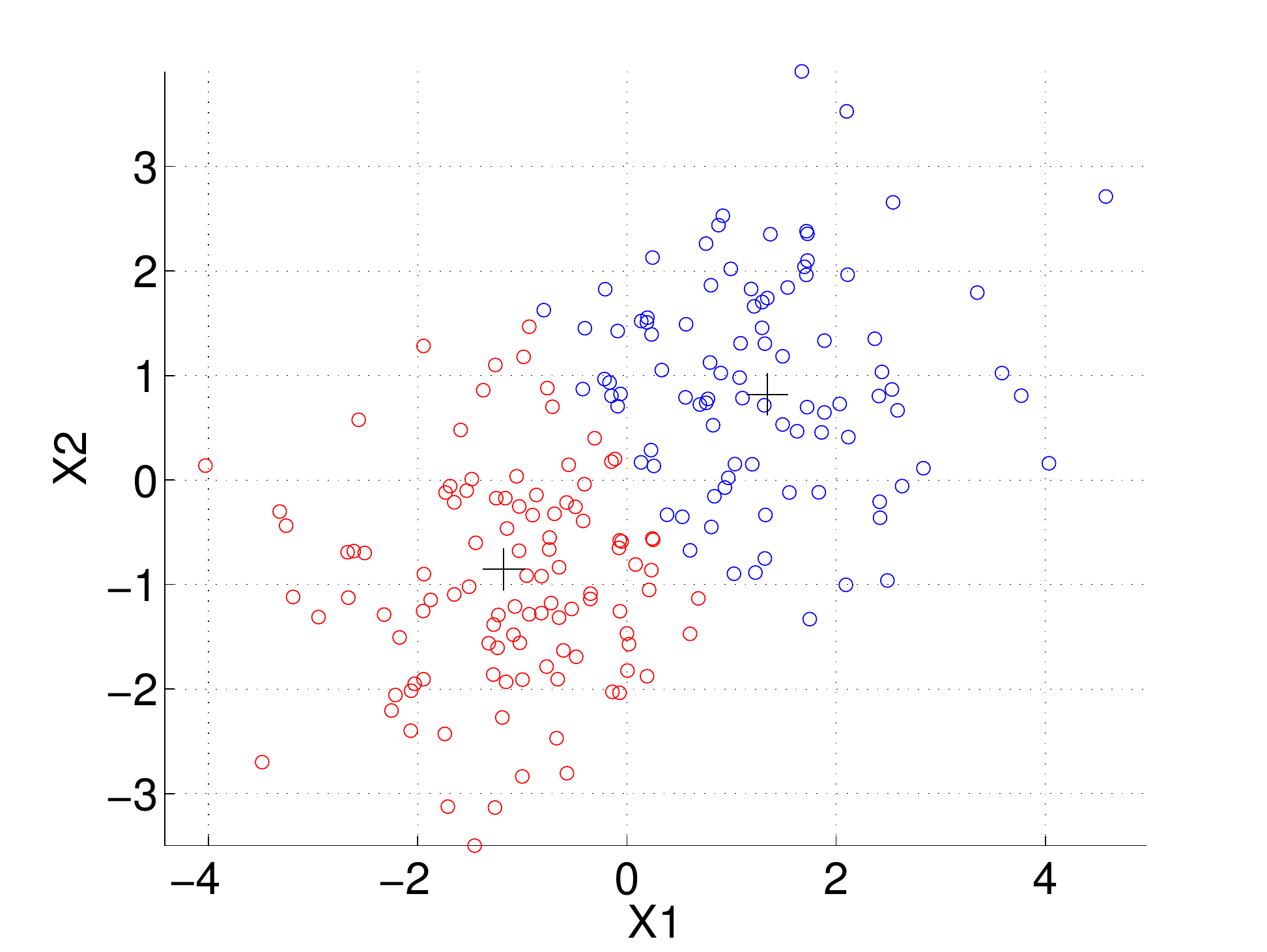}
}
\caption{Three iterations of the K-Medians algorithm on a dataset of 200 data points. Each data point has two dimensions X1 and X2. The big crosses in black represent the centroids of the two found clusters (red and blue).}
\label{Figure2}
\end{center}
\end{figure}

\begin{figure}[!ht]
\begin{center}
\subfigure[Iteration 1]{
\includegraphics[width=6cm, height=6cm, keepaspectratio]{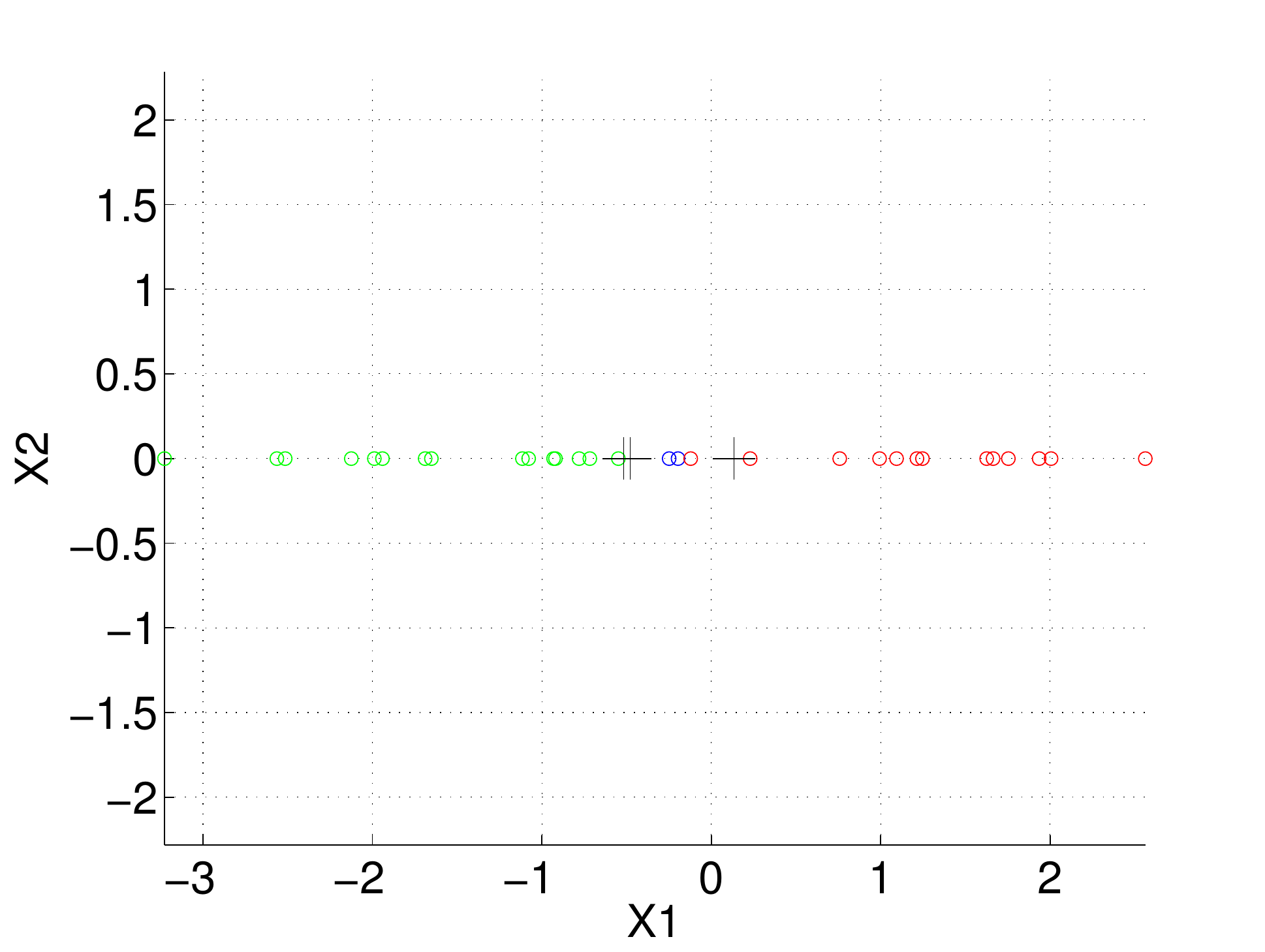}
}
\subfigure[Iteration 2]{
\includegraphics[width=6cm, height=6cm, keepaspectratio]{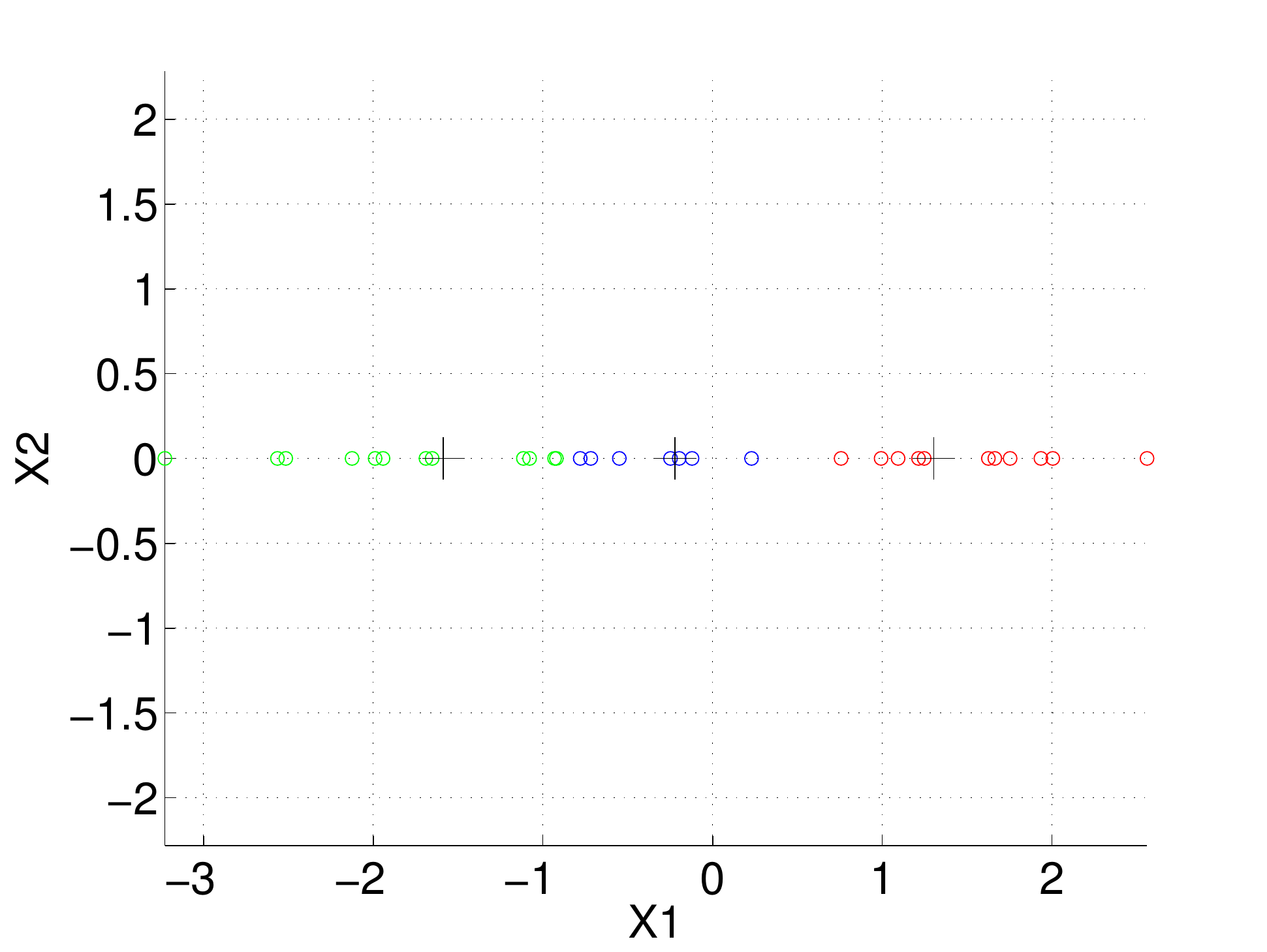}
}
\subfigure[Iteration 3]{
\includegraphics[width=6cm, height=6cm, keepaspectratio]{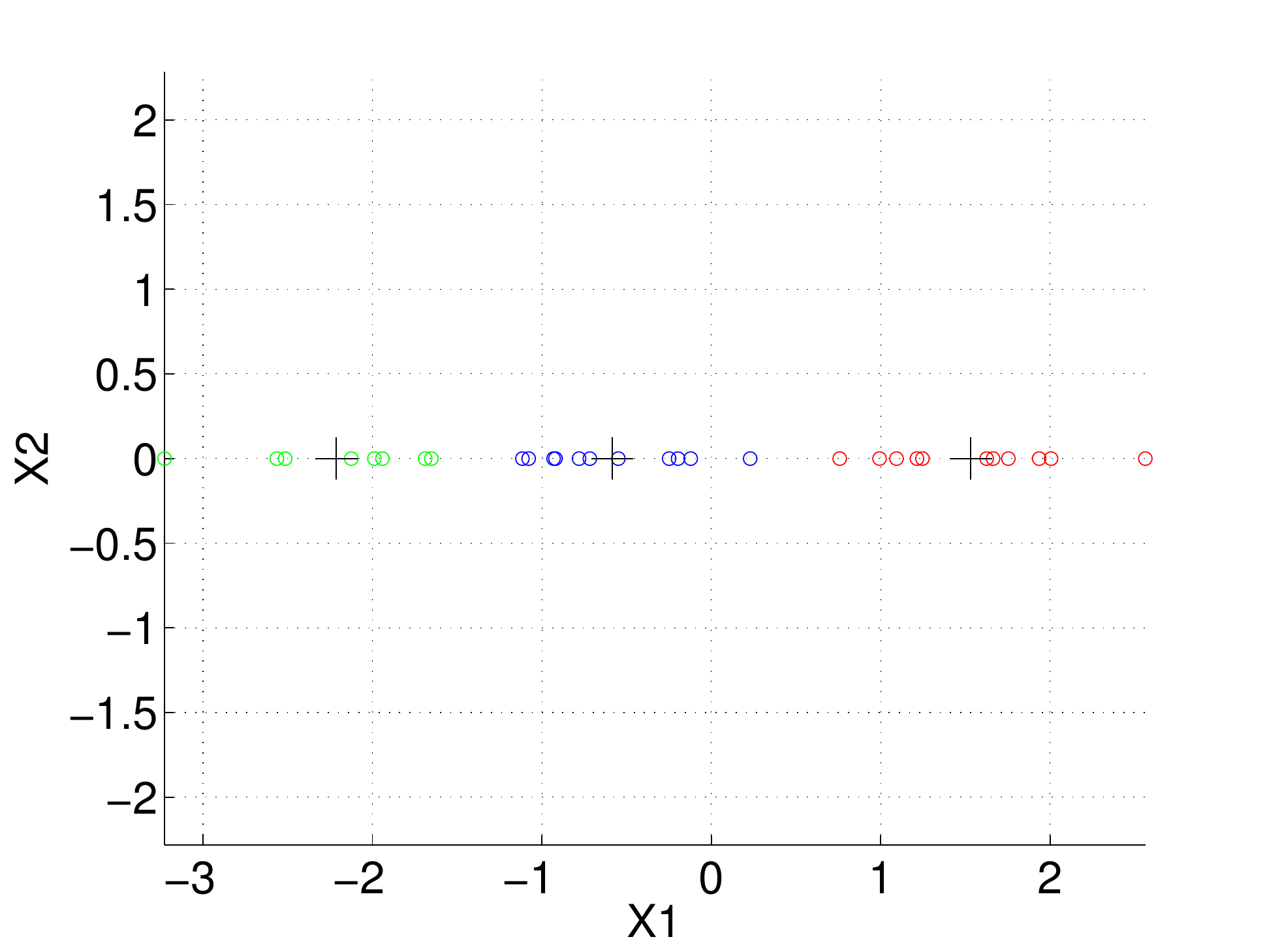}
}
\caption{Three K-Medians iterations on a dataset of 30 data points, where each data point has one dimension X1. The big crosses in black represent the centroids of the found clusters. Each cluster is in a different color.}
\label{Figure3}
\end{center}
\end{figure}

The main steps of the algorithm are the following:
\begin{enumerate}
\item Choose $k$ initial centroids.
\item Assign each point to the partition having the nearest centroid, in terms of Manhattan distance.
\item After assignment of all points to the groups, compute the new $k$ centroids.
\item Repeat Steps 2 and 3 until no centroid changes occur. 
It determines a partitioning of the points into groups of similar magnetic field from which the $J$ function to be minimized can be computed. 
\end{enumerate}
Fig.\ref{Figure2} shows an example of three iterations of the K-Medians algorithm ($k$ = 2), on 200 two-dimensional samples generated by a uniform random distribution, while Fig.\ref{Figure3} depicts an example of three iterations of the K-Medians algorithm ($k$ = 3), on 30 one-dimensional samples generated by a uniform random distribution. 

It is interesting to observe the variation in position of the centroid data samples (big crosses in black) and the clusters, where each color represents a different cluster, along the iterations.

\section{Experiment}

The proposed method is evaluated on custom-oriented datasets of laptop's points. They are composed of the magnetic field measures from 13 laptops of different typology and screen size: 17'', 15.6'', 14'', 13.3'' and 11.6''. Measures have been collected when laptops were supplied by alternating current (AC) and battery. 

Measurement points are divided into two groups only: top body measurement points (tbmp1-tbmp9), and bottom body measurement points (bbmp1-bbmp9). Previous EMF measurement showed that the EMF radiation near the screen is negligible \cite{[21]}. Fig.\ref{Figure4} illustrates the laptop's measurement points.

The laptops were tested under so-called normal operating condition. It means that laptop operates with typical programs like Word, Excel, Internet browsing, etc. 

Four independent experiments have been performed. The first experiment considers the 9 magnetic field values of the top body points for each of the 13 laptops supplied by AC with extracted battery. The second experiment evaluates the 9 magnetic field values of the top body points for each of the 13 laptops supplied by the battery. The third experiment collects the 9 magnetic field values from the bottom body points for each of the 13 laptops supplied by AC with extracted battery. The last experiment considers the 9 magnetic field values from the bottom body points for each of the 13 laptops supplied by the battery. Each of the four experiments determines a dataset of 117 feature values. 

Experiments have been executed by using MatlabR2013b, on a desktop computer quad-core with 8 Gbyte of RAM and operating system Windows 7.

\section{Results and Discussion}
Magnetic field radiation of typical laptop in the frequency range between 50 Hz and 400 kHz is obtained by measuring device AARONIA NF-5010. Fig.\ref{Figure5} shows magnetic field radiation in this frequency range.

\begin{figure}[!ht]
\begin{center}
\includegraphics[width=7cm, height=7cm, keepaspectratio]{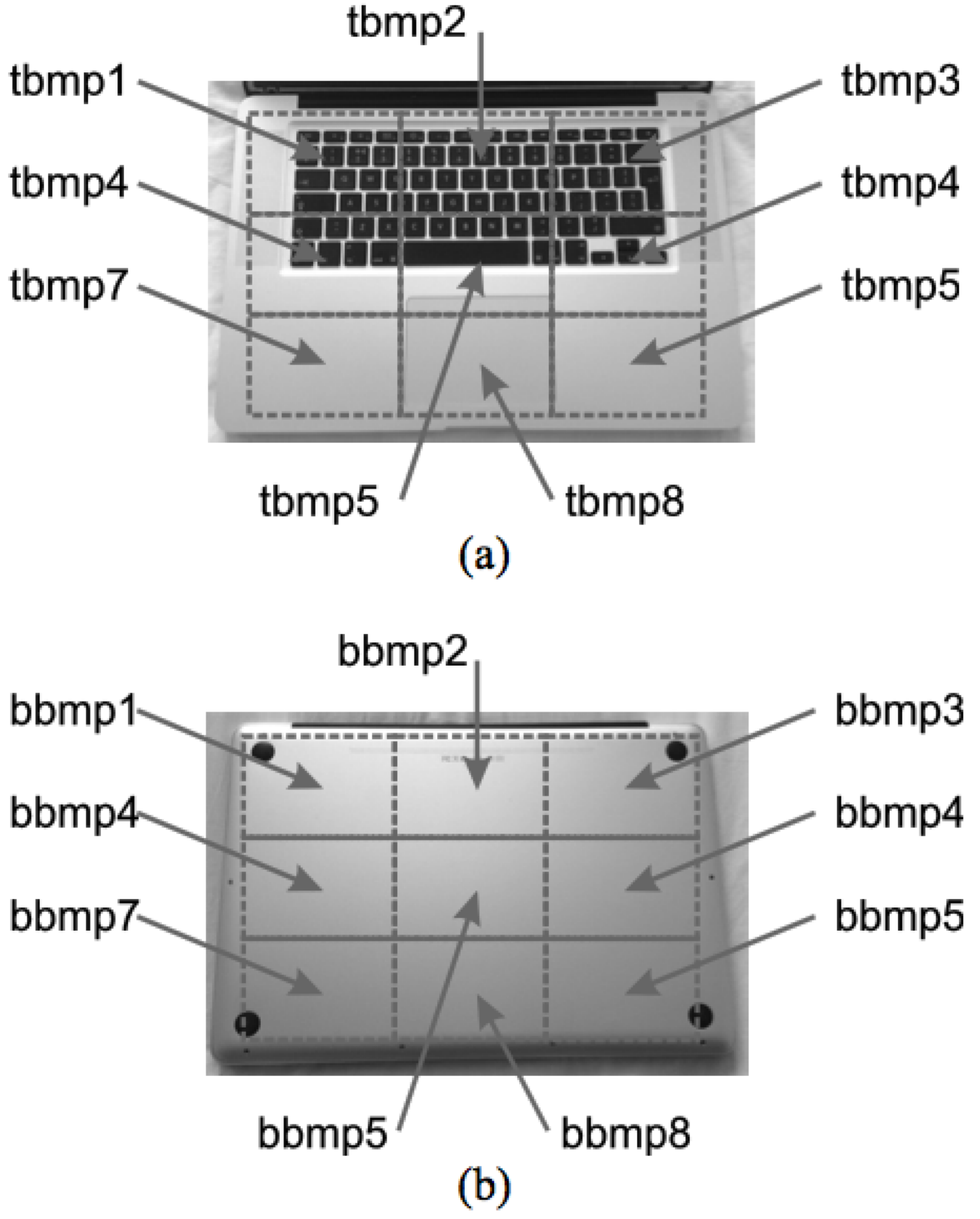}
\caption{Laptop's measurement points of the magnetic field radiation (a) top part, (b) bottom part.}
\label{Figure4}
\end{center}
\end{figure}

\begin{figure}[!ht]
\begin{center}
\includegraphics[width=7.4cm, height=7.4cm, keepaspectratio]{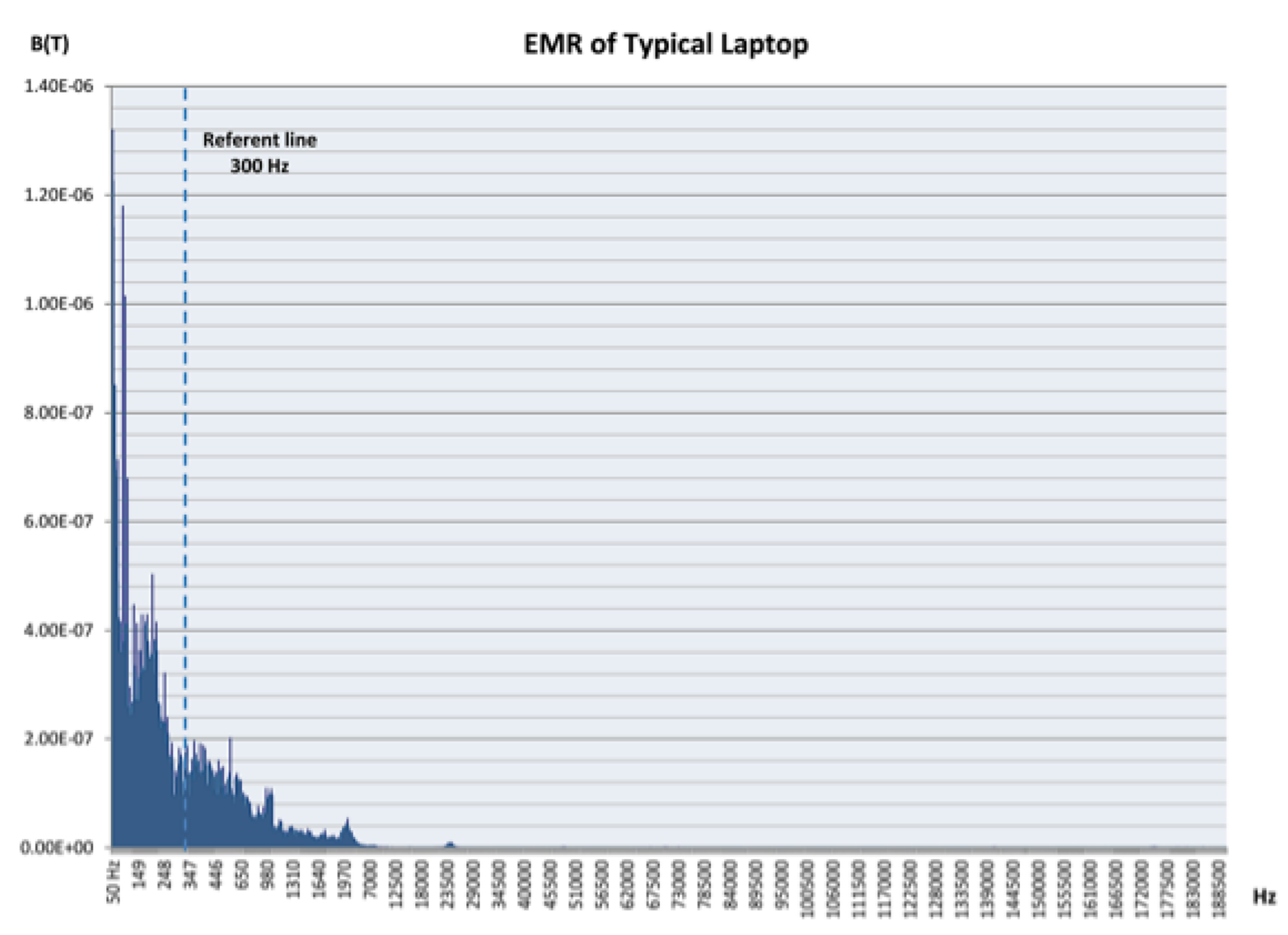}
\caption{Magnetic field radiation of typical laptop in the frequency range between 50 Hz and 400 kHz obtained by measuring device AARONIA NF-5010 (blue line represents the border frequency of the measuring device Lutron EMF 828)}
\label{Figure5}
\end{center}
\end{figure}
Fig.\ref{Figure5} shows the three frequency measurement extents: (i) extremely low frequency range (from 50 Hz to 500 Hz), (ii) TCO1 range (from 500 Hz to 2 kHz), and (iii) TCO2 range (from 2 kHz to 400 kHz). If we observe the results, it is clear that the magnetic field radiation peaks are primarily in the area below 300 Hz. Furthermore, the TCO safe limits for the higher frequencies, i.e. TCO2 frequency range, are much stricter, i.e. $\le$ 25 nT \cite{[22]}. However, the measured results show that laptop's magnetic field radiation in this frequency range is much lower than TCO safe limit. Hence, the measurement below 300 Hz is mandatory, especially due to hazardous effects to the laptop users. Still, the magnetic field radiation is present in the high frequency domain up to 4 GHz. It is a consequence of the WLAN. However, it is out of the scope of this paper, because we explore and measure only extremely low frequency magnetic field radiation.

For each dataset, the number $k$ of clusters of the K-Medians algorithm has been fixed to five. We chose this value for testing the ability of the algorithm to receive classes of magnetic field values such as \emph{high}, \emph{low high}, \emph{middle}, \emph{safe} and \emph{very safe}. We identify the points with magnetic field values higher than 0.3 $\mu$T as dangerous points and the points with magnetic field values lower than 0.3 $\mu$T as no dangerous points.

\begin{figure*}[t]
\begin{center}
\subfigure[]{
\includegraphics[width=3cm, height=3cm, keepaspectratio]{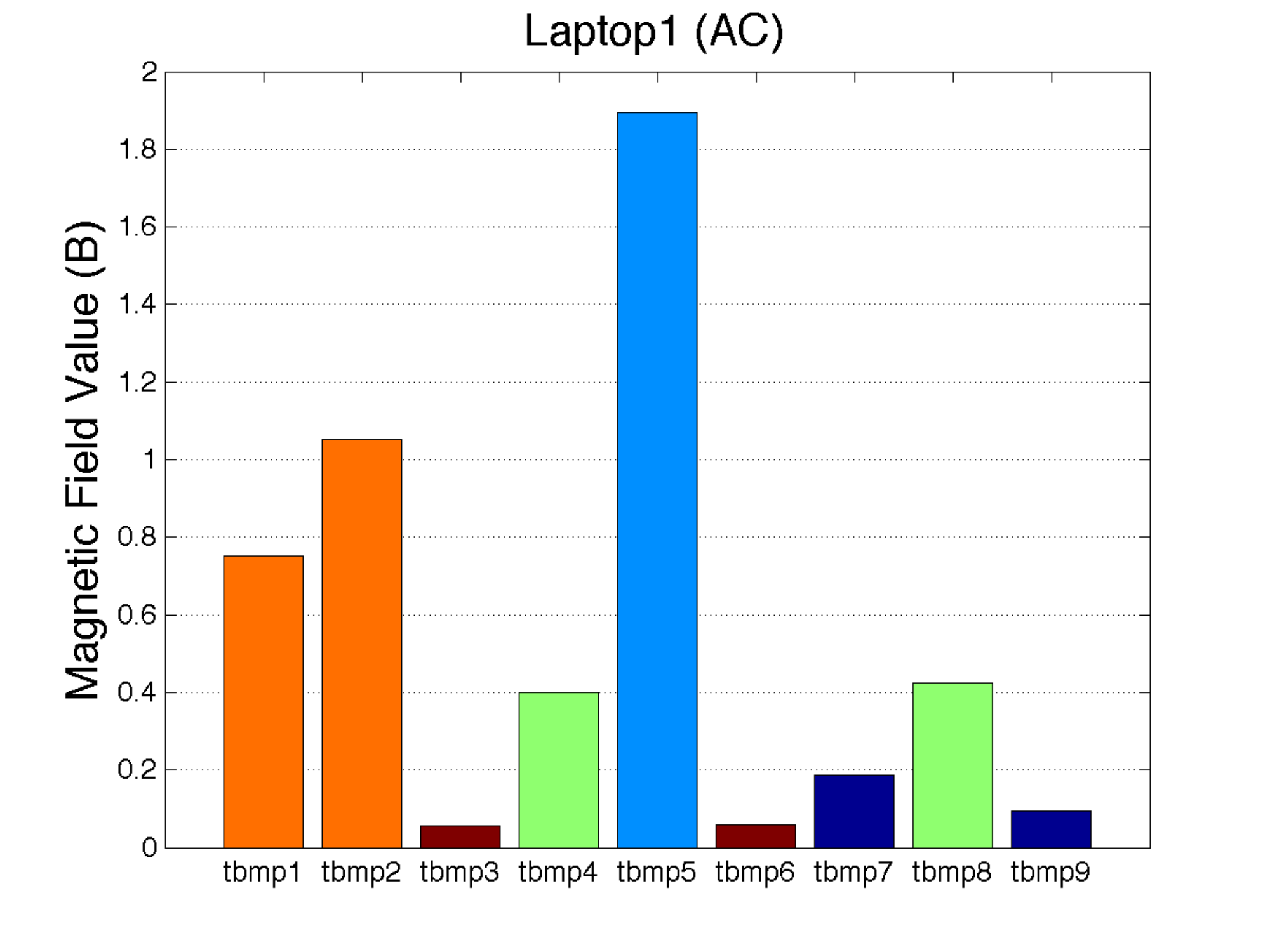}
}
\subfigure[]{
\includegraphics[width=3cm, height=3cm, keepaspectratio]{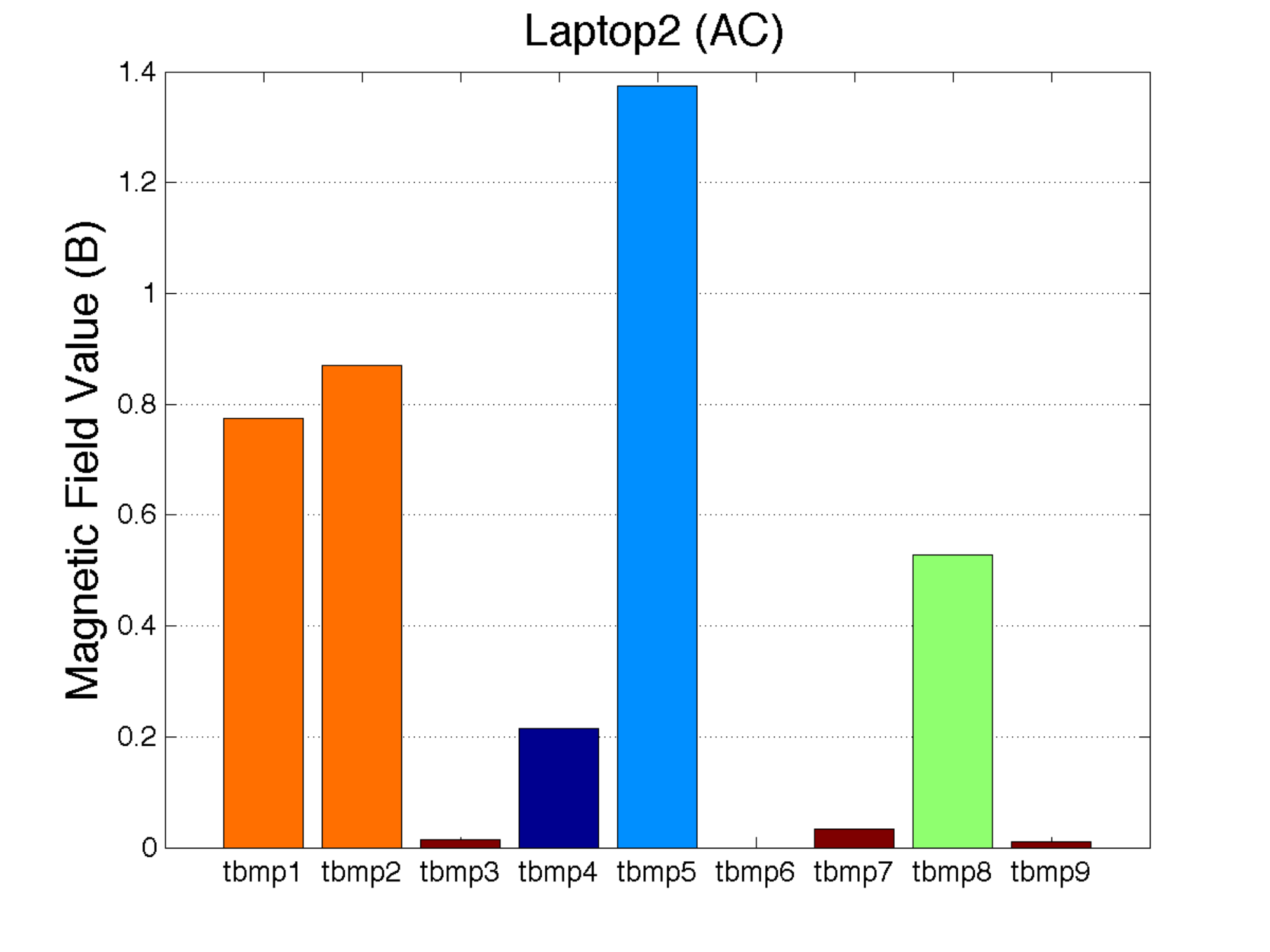}
}
\subfigure[]{
\includegraphics[width=3cm, height=3cm, keepaspectratio]{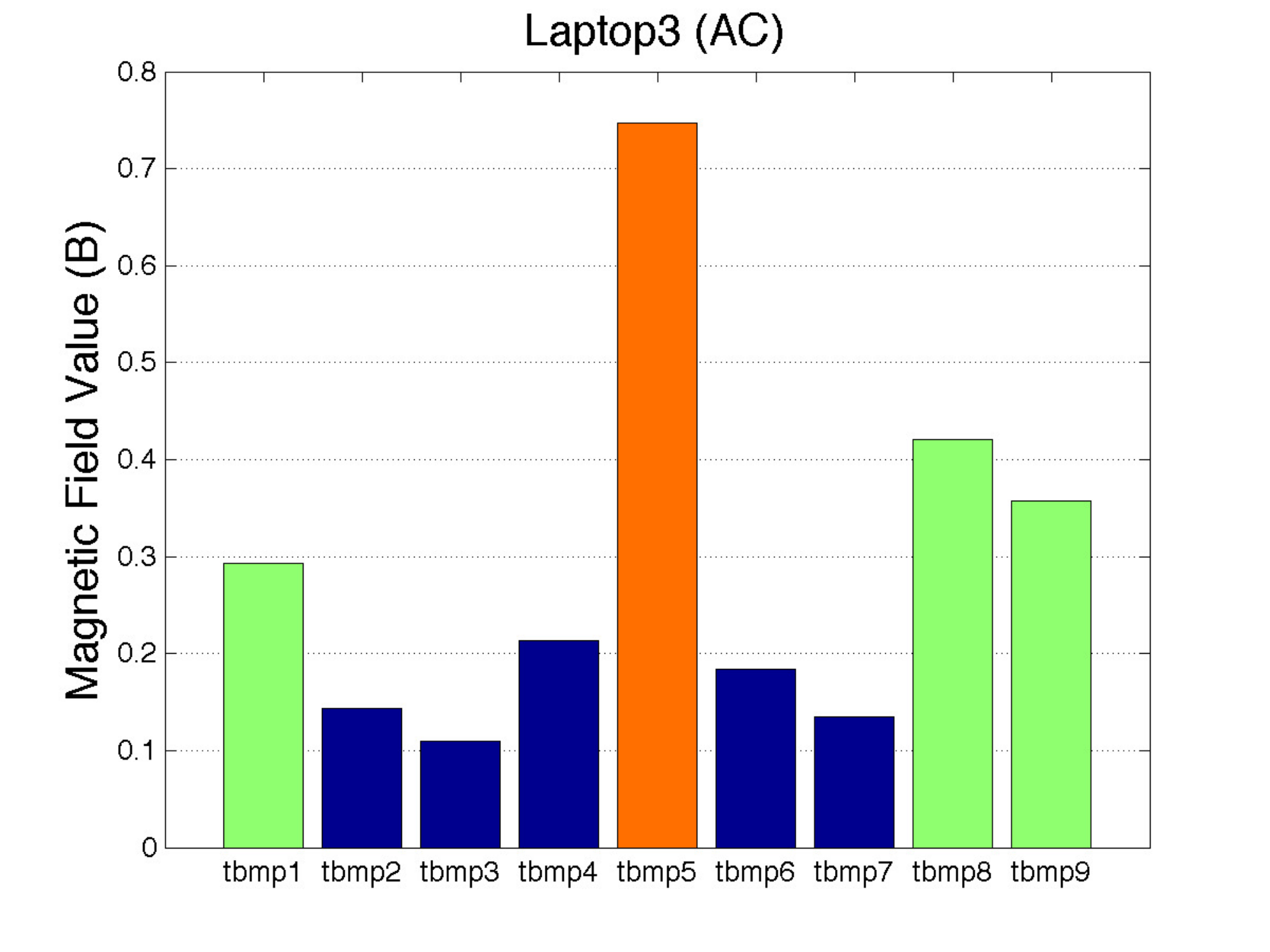}
}
\subfigure[]{
\includegraphics[width=3cm, height=3cm, keepaspectratio]{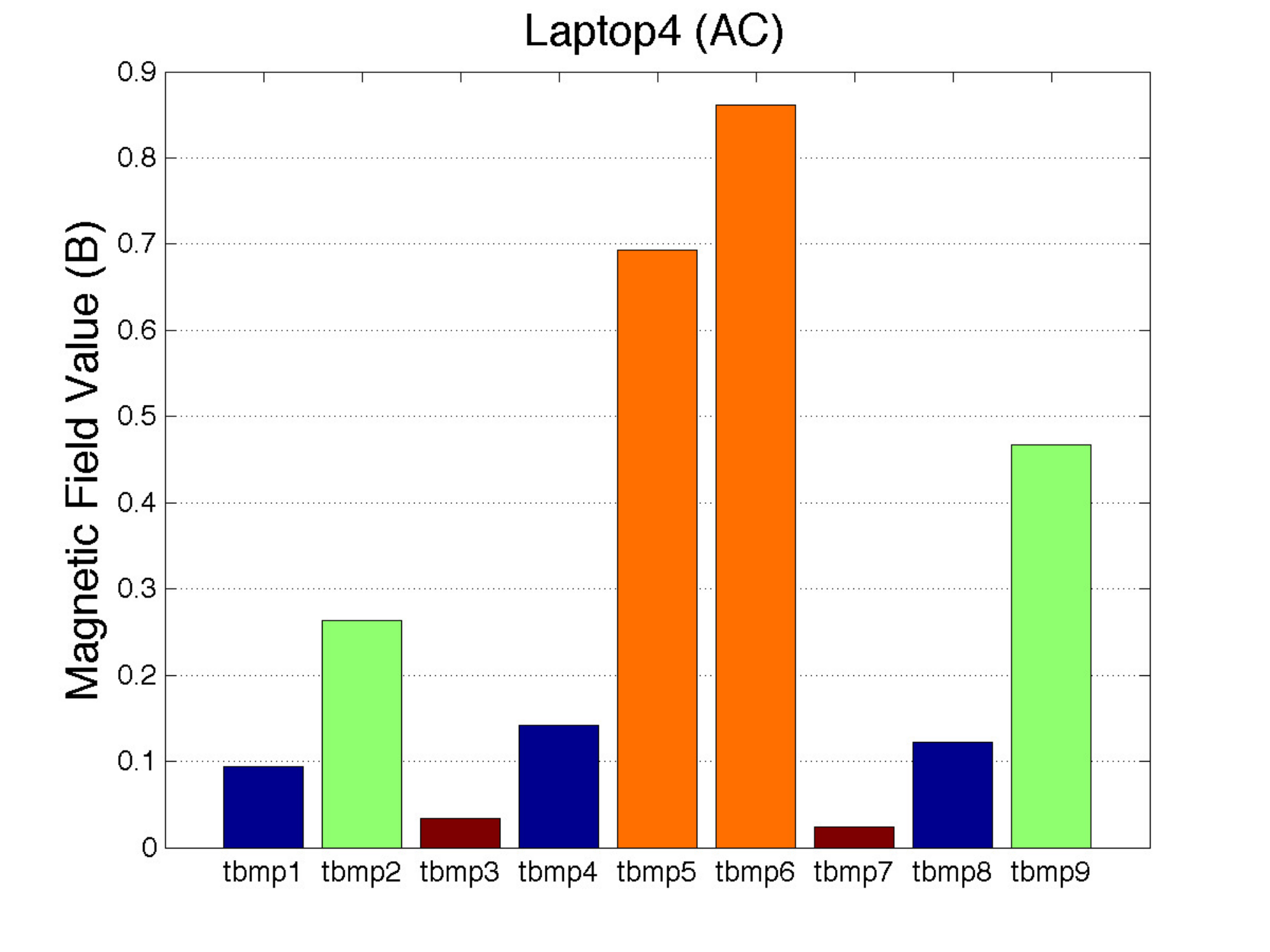}
}
\subfigure[]{
\includegraphics[width=3cm, height=3cm, keepaspectratio]{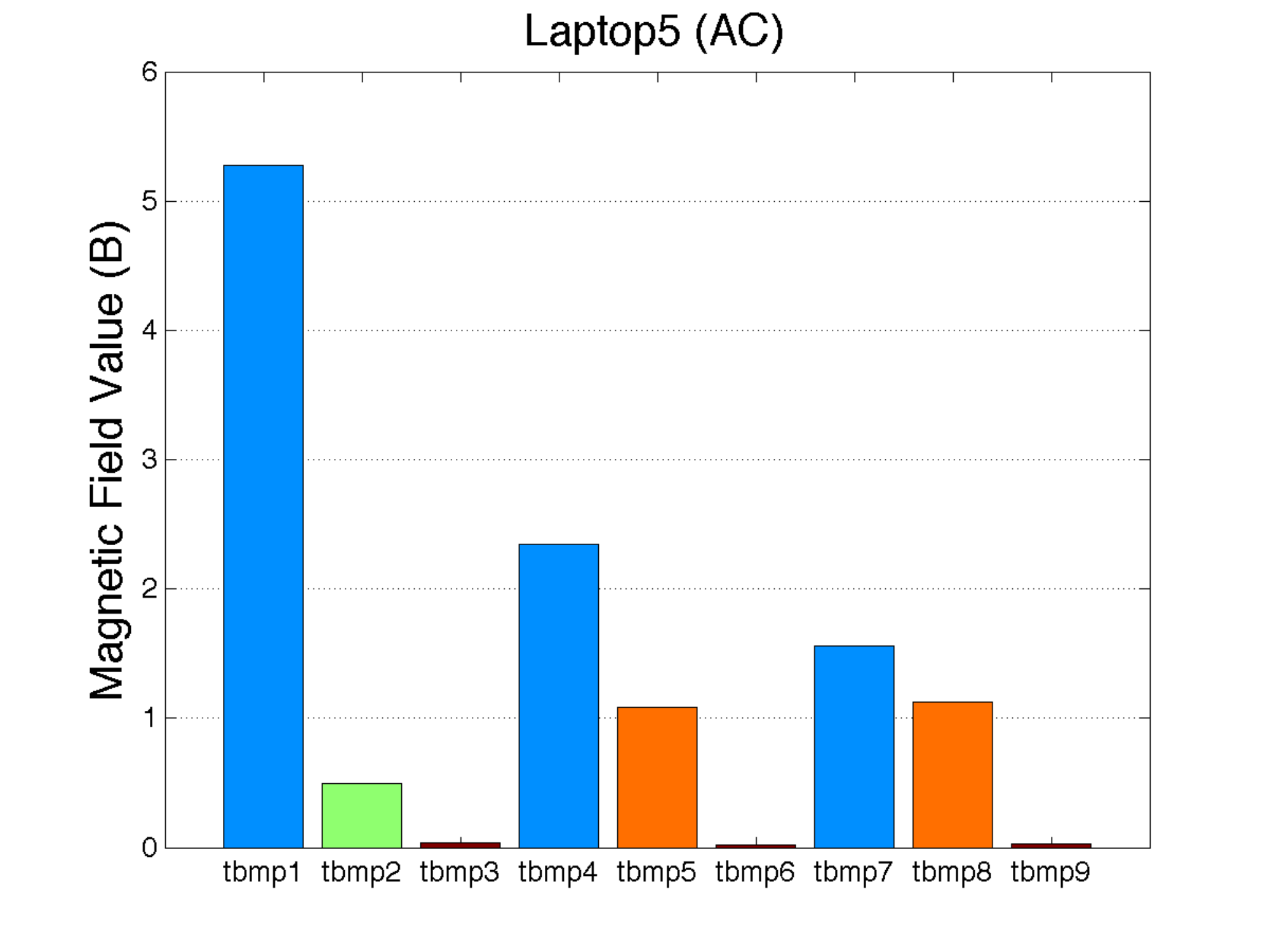}
}
\subfigure[]{
\includegraphics[width=3cm, height=3cm, keepaspectratio]{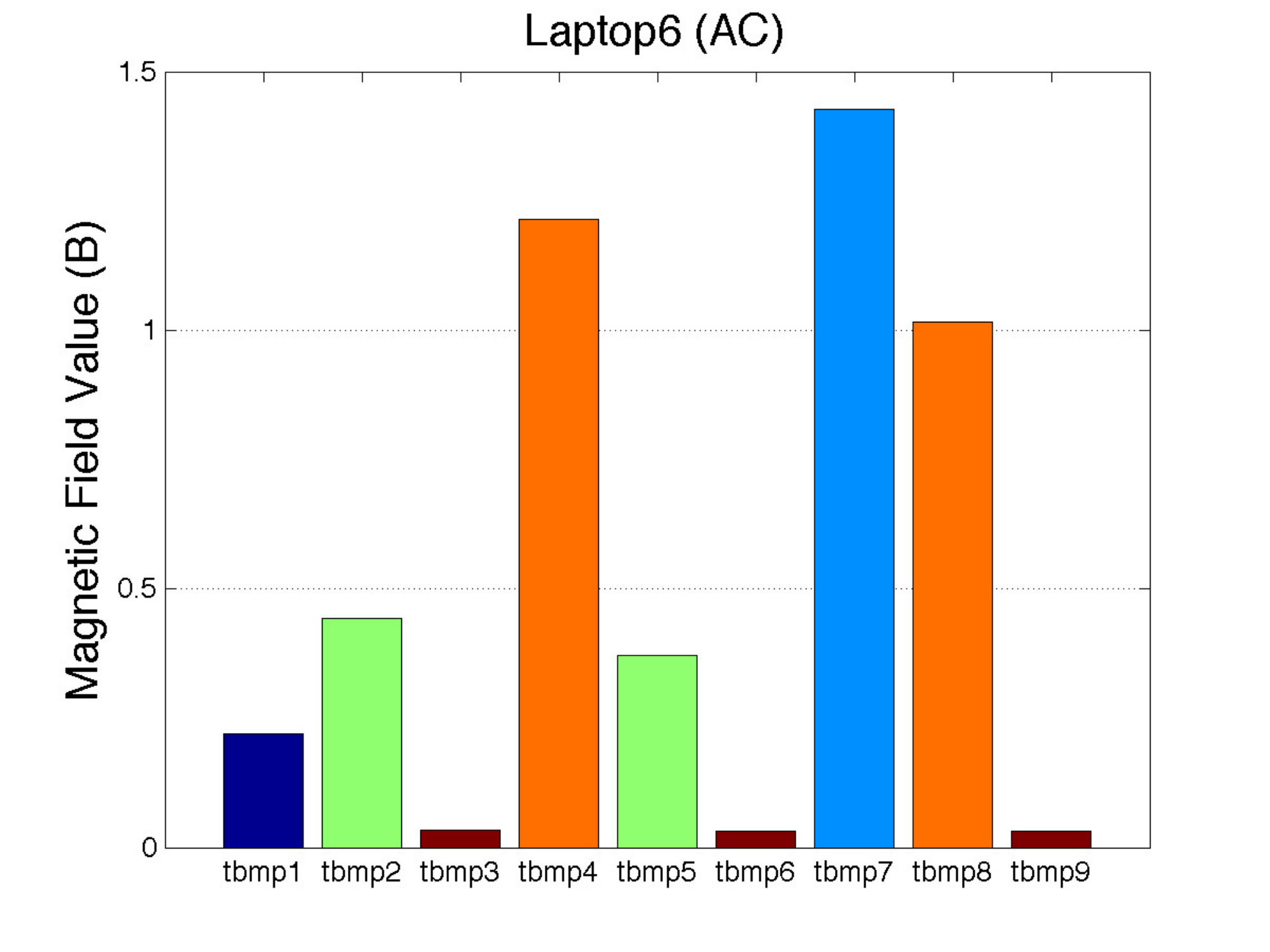}
}
\subfigure[]{
\includegraphics[width=3cm, height=3cm, keepaspectratio]{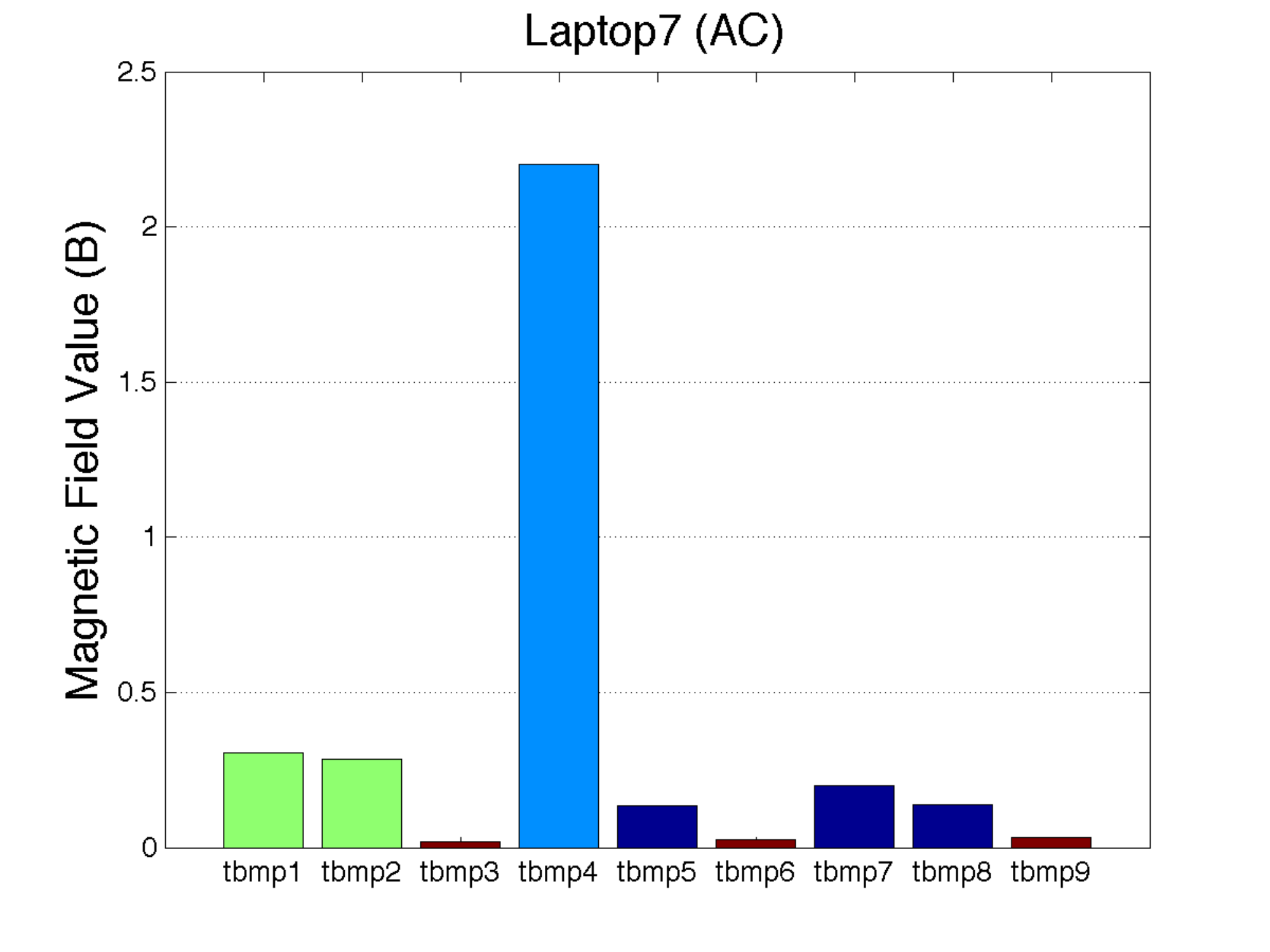}
}
\subfigure[]{
\includegraphics[width=3cm, height=3cm, keepaspectratio]{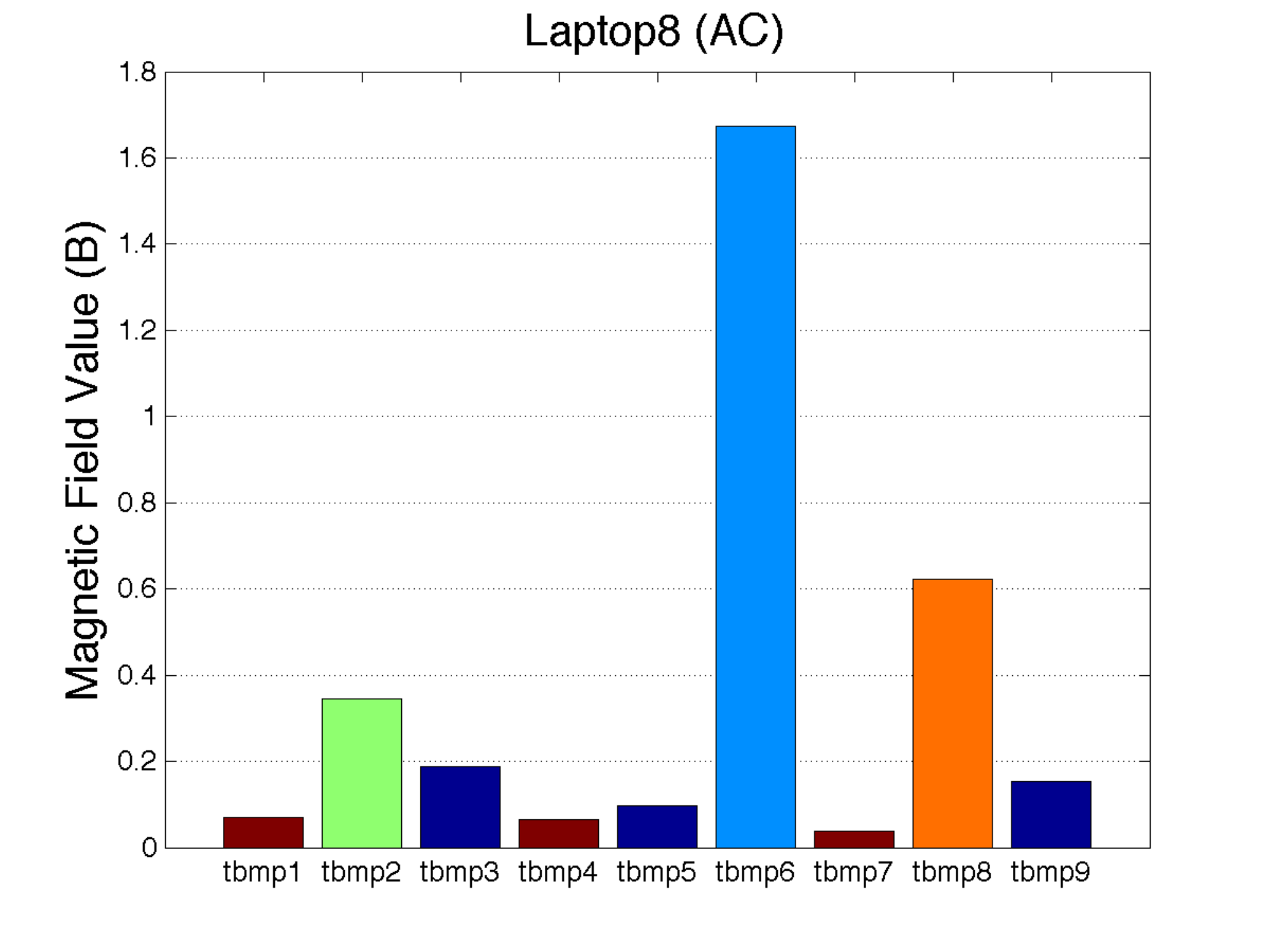}
}
\subfigure[]{
\includegraphics[width=3cm, height=3cm, keepaspectratio]{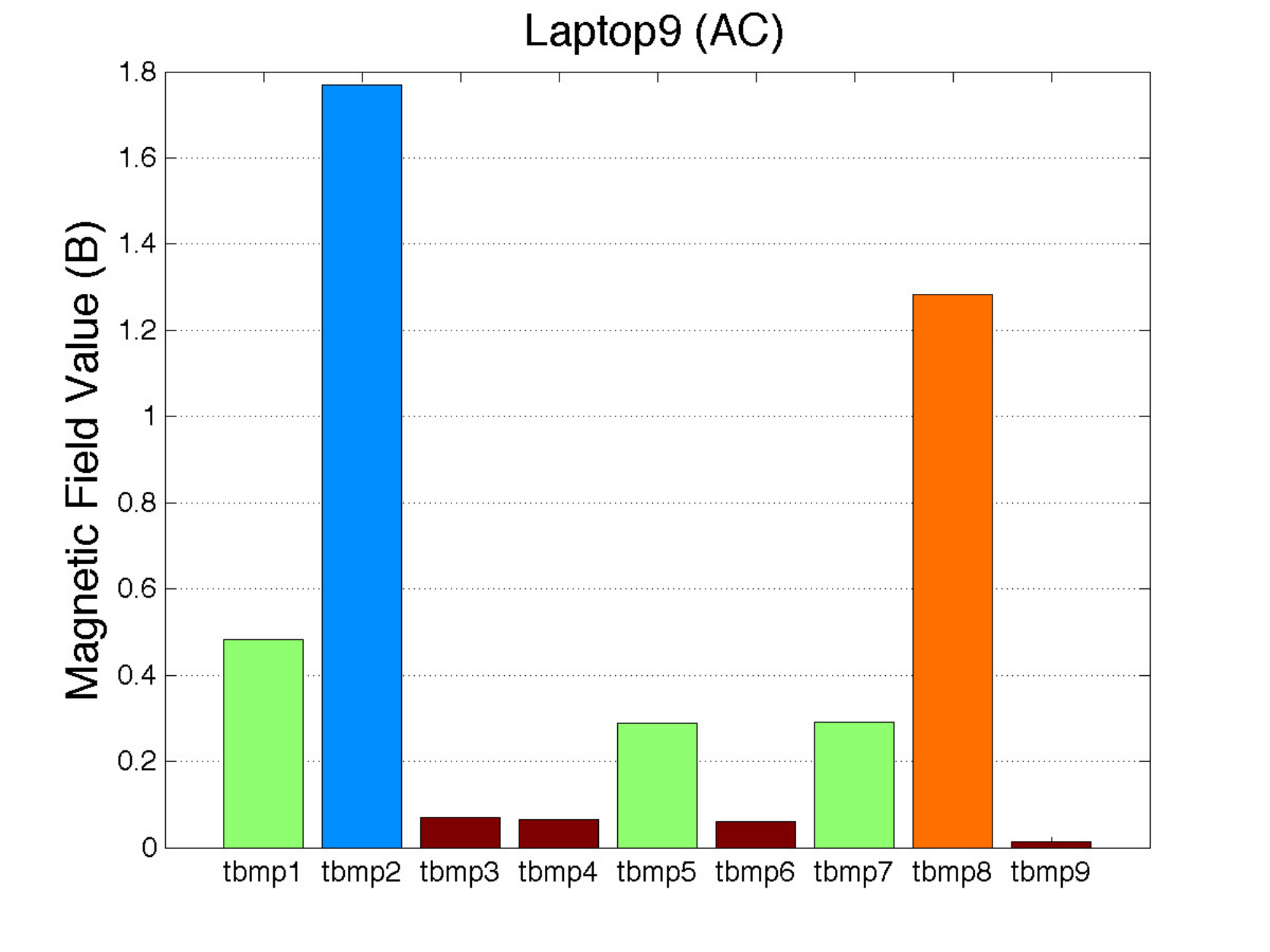}
}
\subfigure[]{
\includegraphics[width=3cm, height=3cm, keepaspectratio]{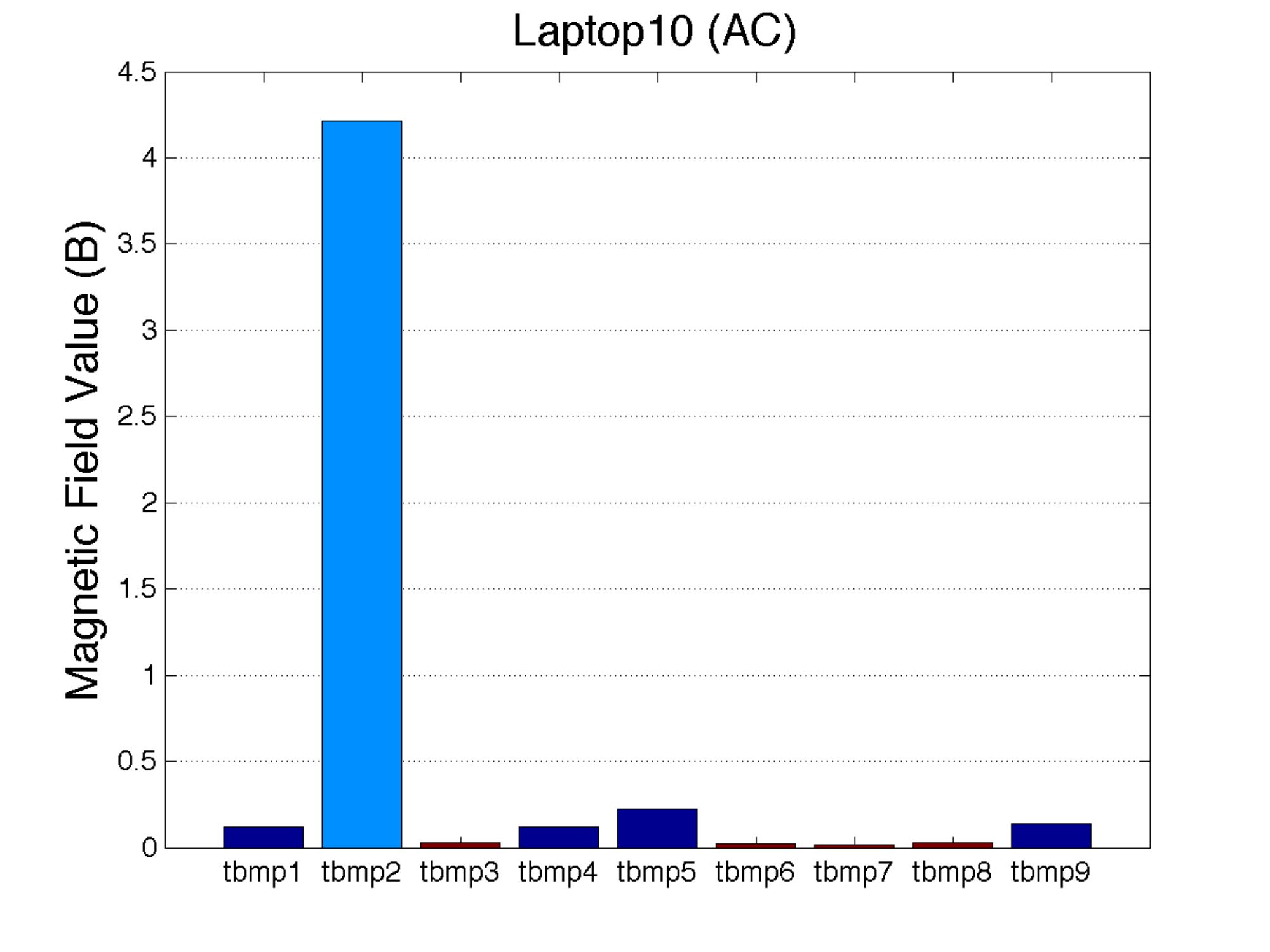}
}
\subfigure[]{
\includegraphics[width=3cm, height=3cm, keepaspectratio]{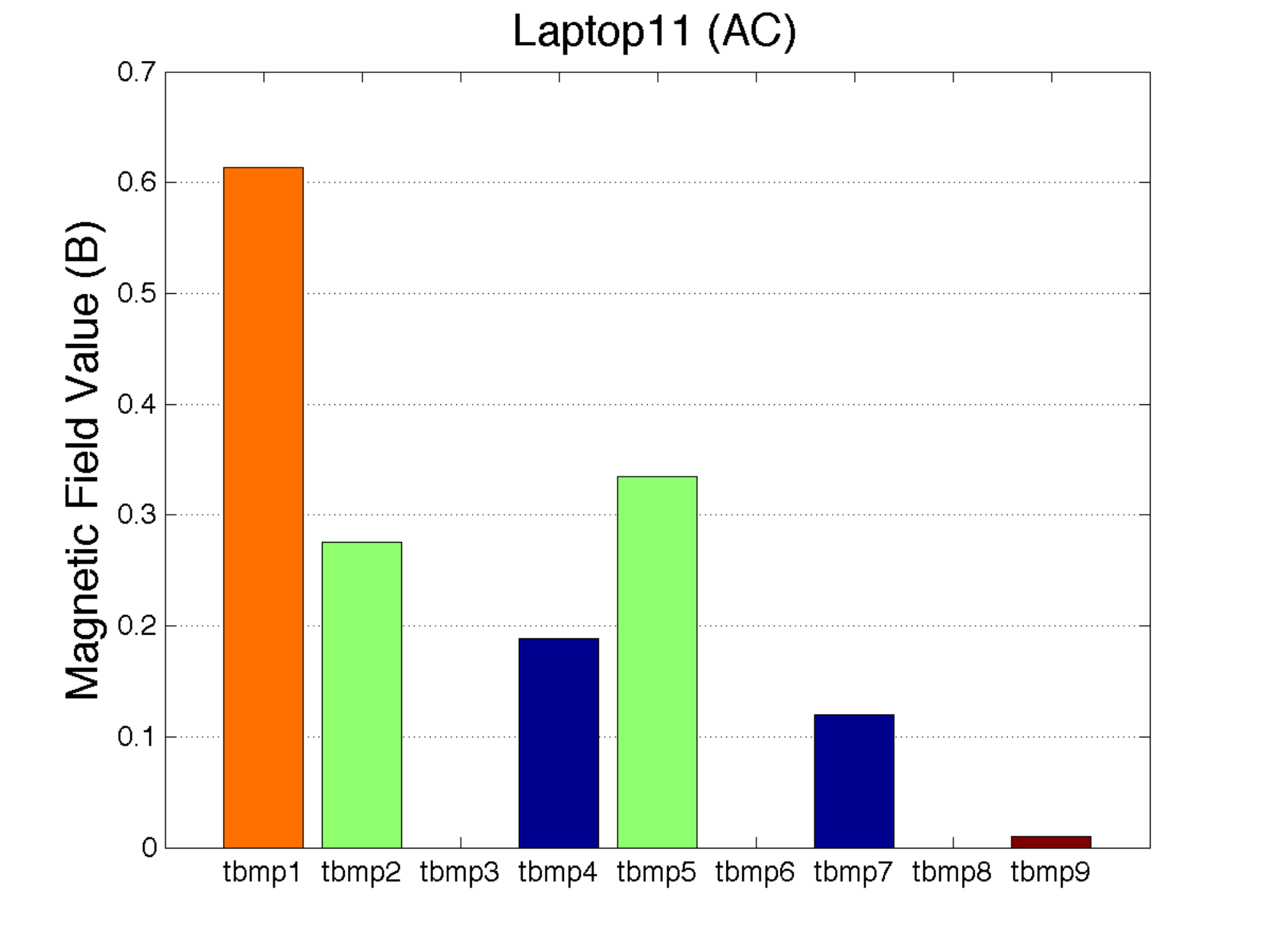}
}
\subfigure[]{
\includegraphics[width=3cm, height=3cm, keepaspectratio]{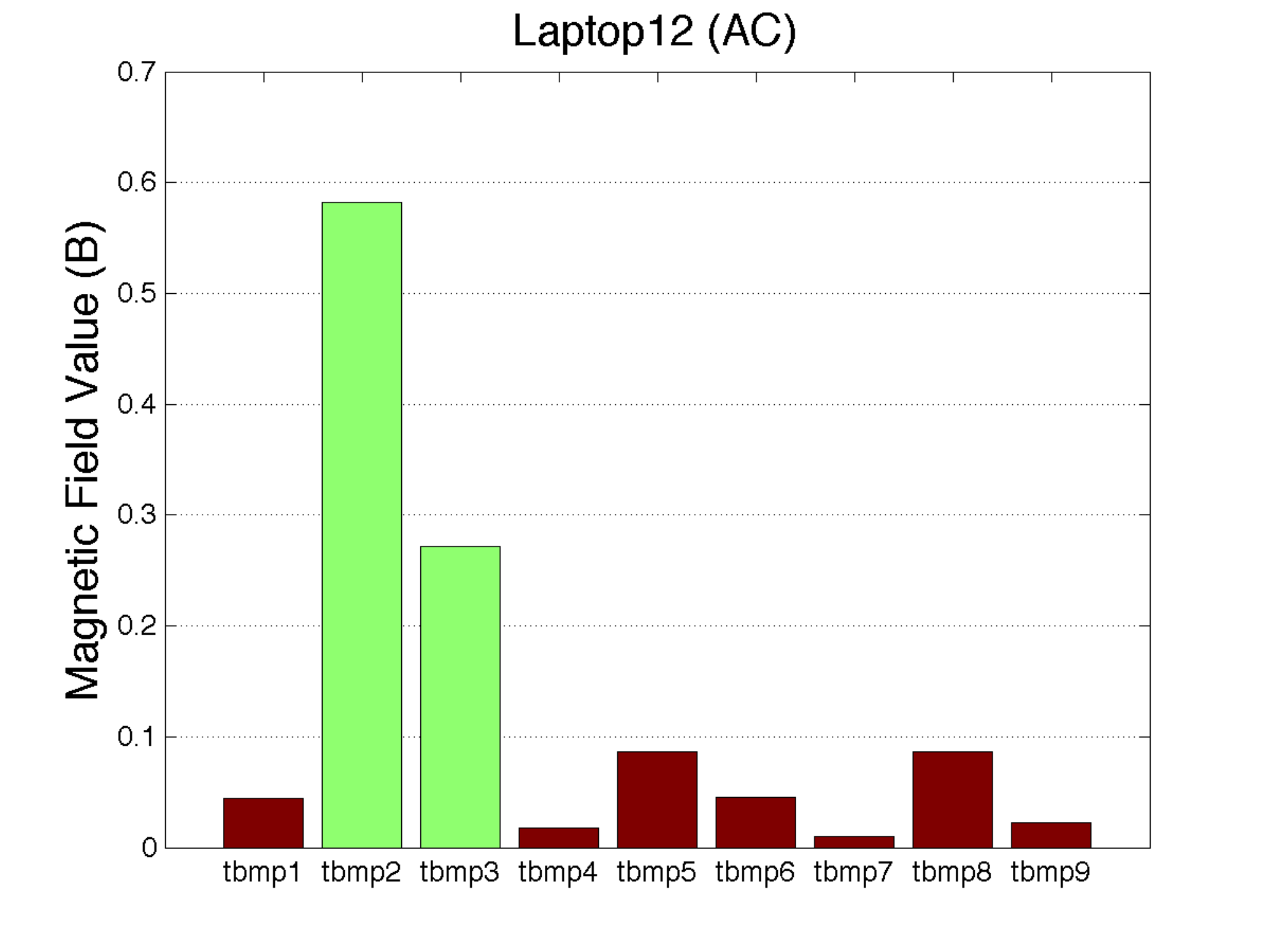}
}
\subfigure[]{
\includegraphics[width=3cm, height=3cm, keepaspectratio]{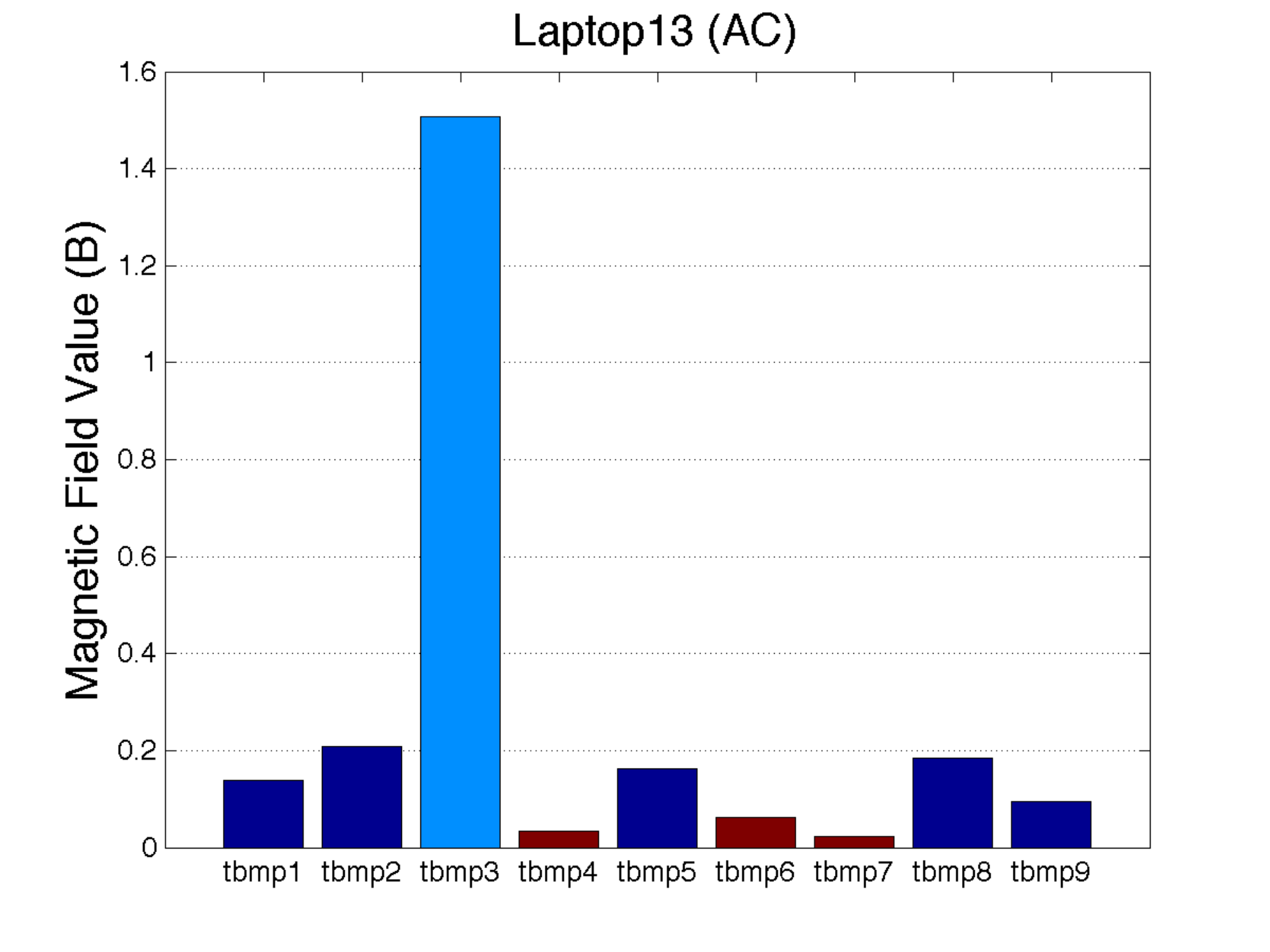}
}
\caption{Results of K-Medians on the dataset of top body points for the 13 laptops supplied by AC (a-m). Each color along the figures represents a cluster of top body points. The names of the top body points included in the cluster are reported in correspondence to the $x$ axis. $Y$ axis gives the magnetic field values measured at the top body points belonging to that cluster.}
\label{Figure6}
\end{center}
\end{figure*}

\begin{figure*}[t]
\begin{center}
\subfigure[]{
\includegraphics[width=3cm, height=3cm, keepaspectratio]{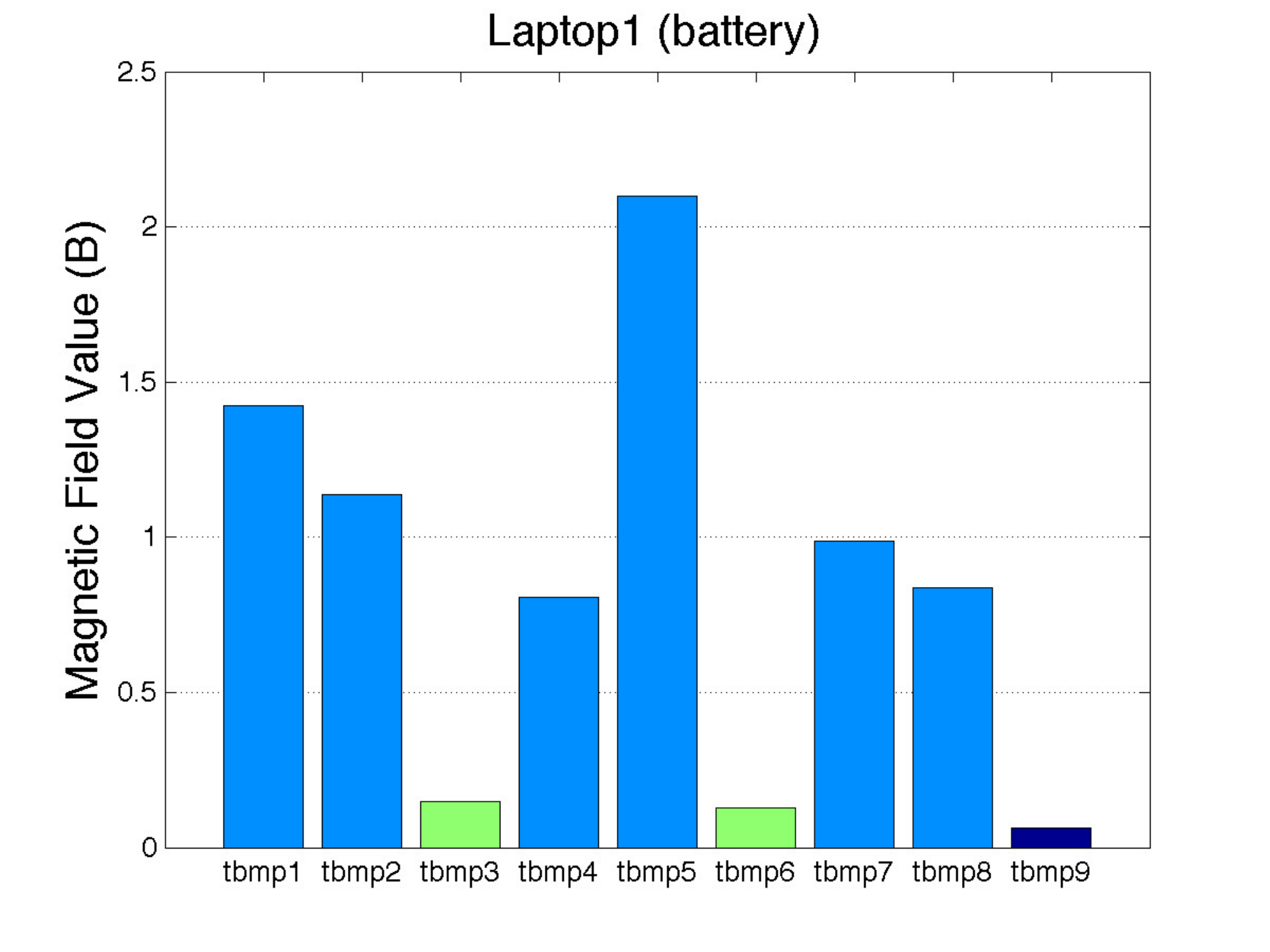}
}
\subfigure[]{
\includegraphics[width=3cm, height=3cm, keepaspectratio]{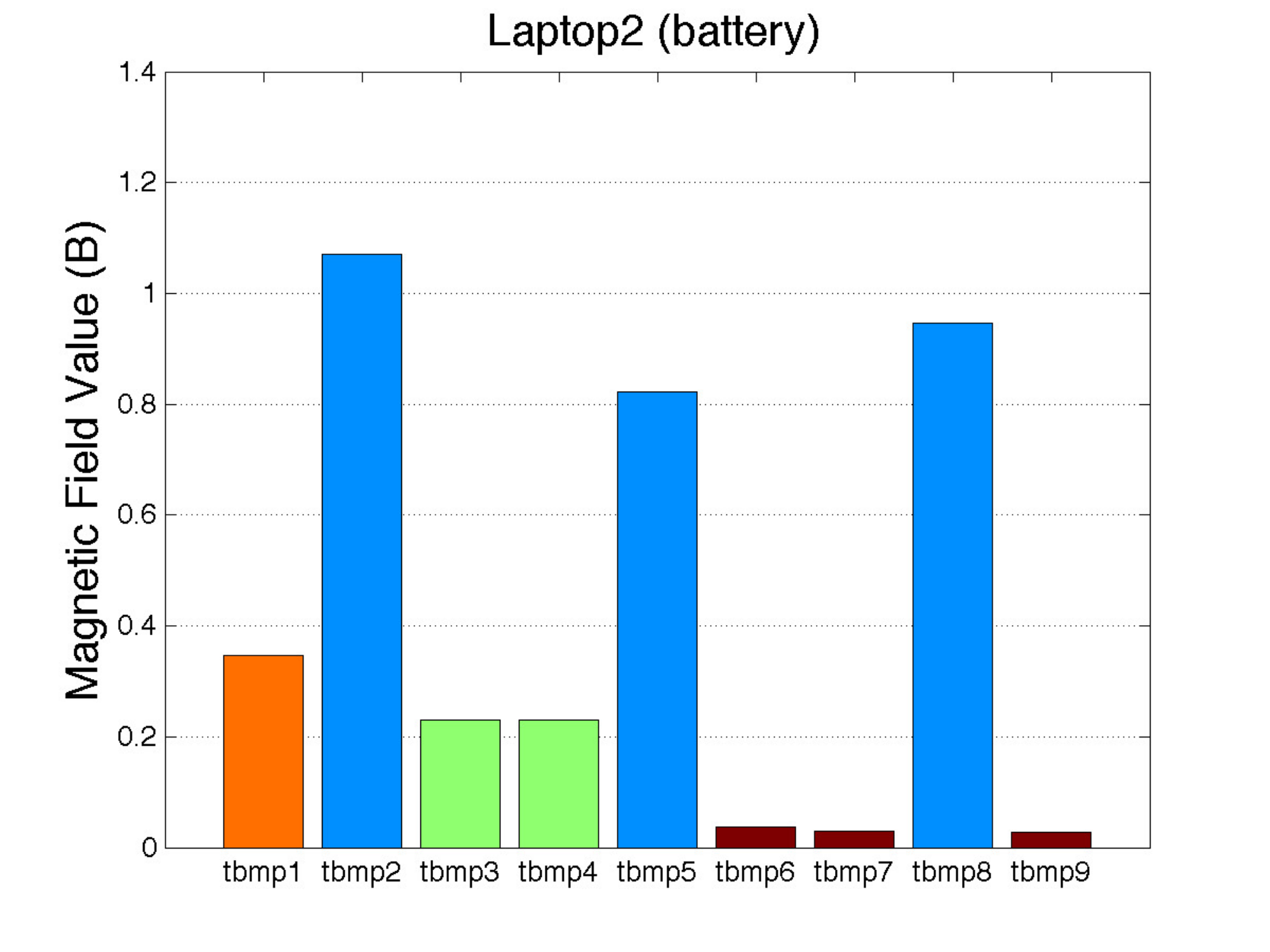}
}
\subfigure[]{
\includegraphics[width=3cm, height=3cm, keepaspectratio]{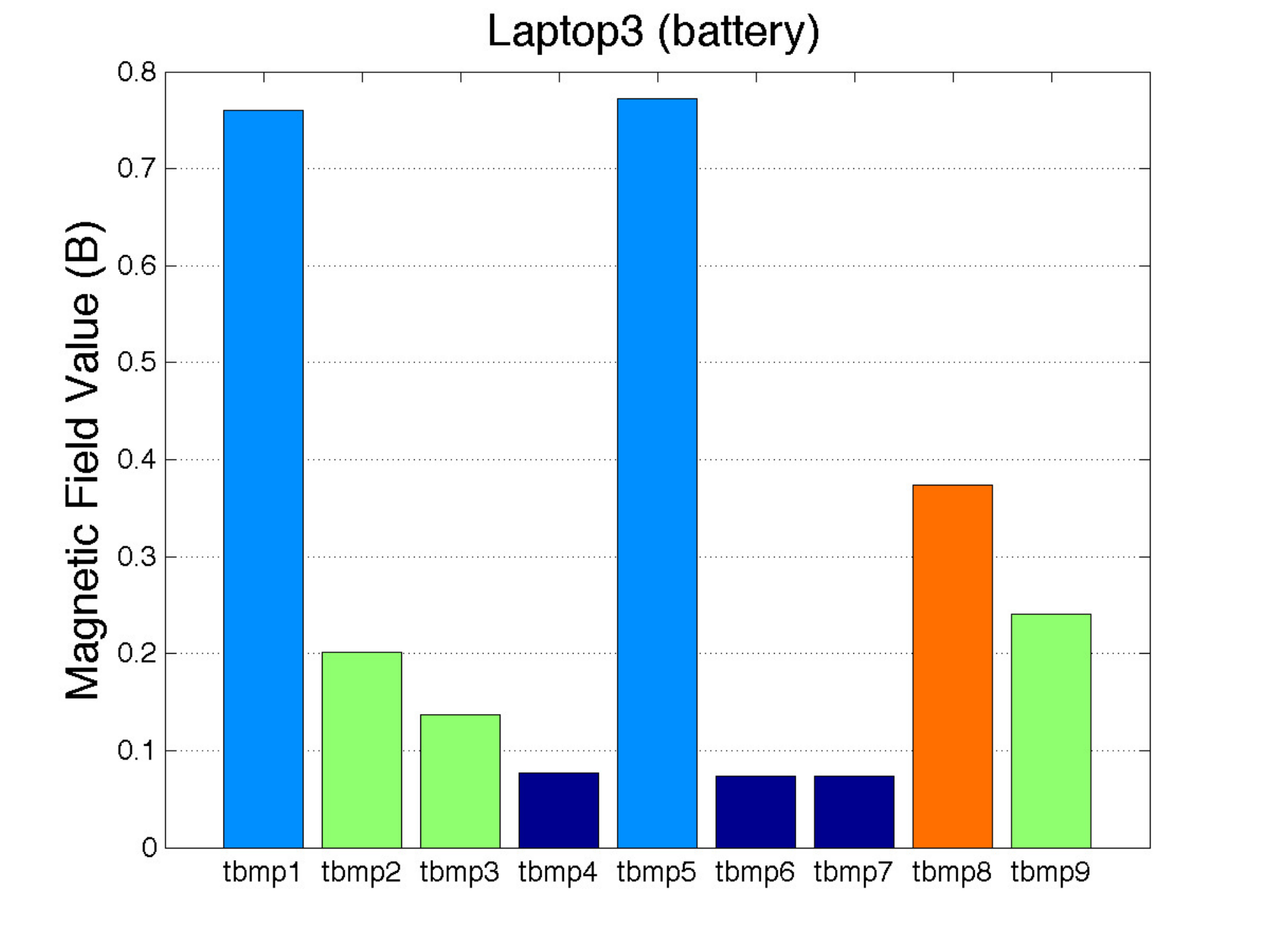}
}
\subfigure[]{
\includegraphics[width=3cm, height=3cm, keepaspectratio]{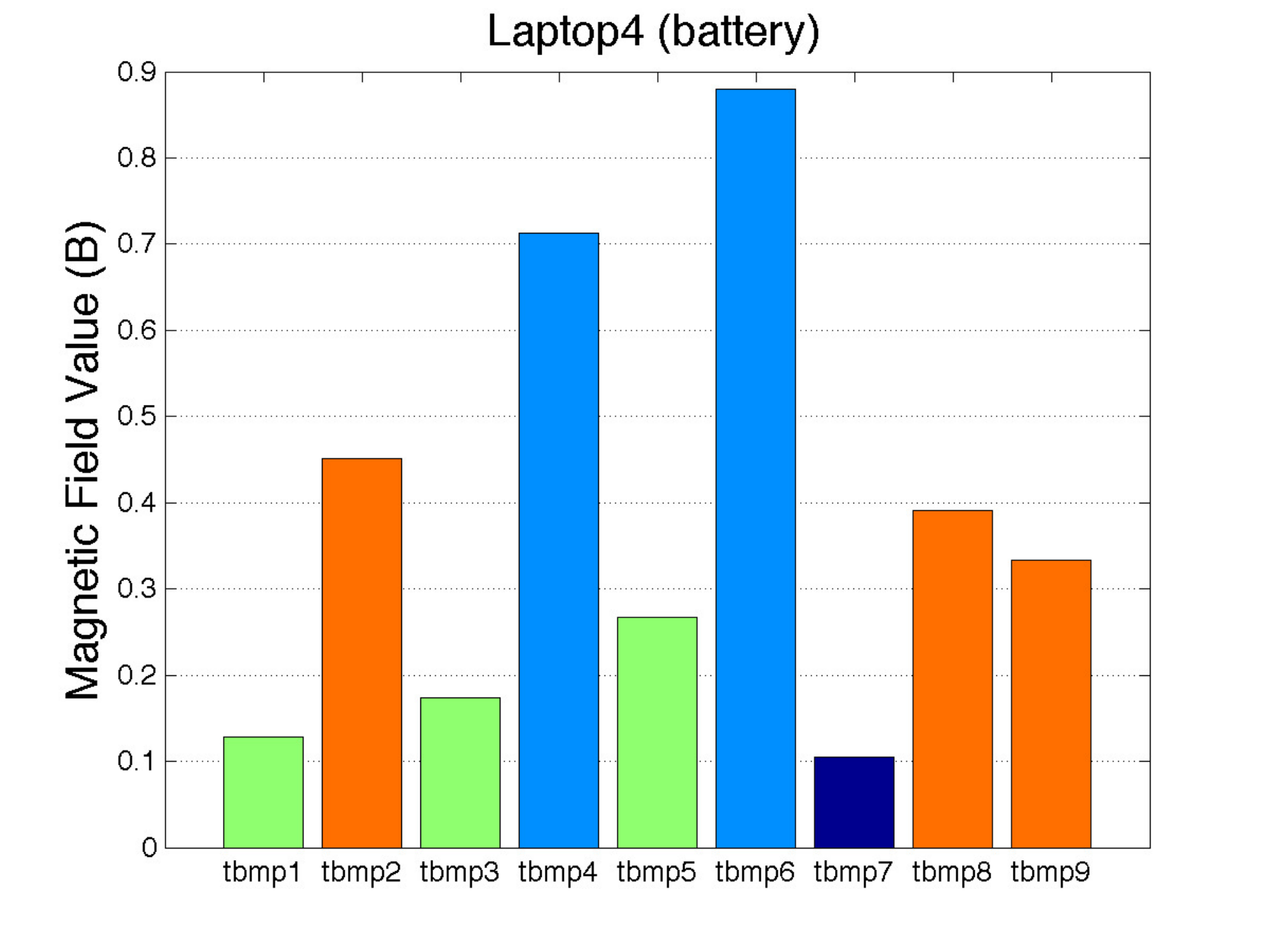}
}
\subfigure[]{
\includegraphics[width=3cm, height=3cm, keepaspectratio]{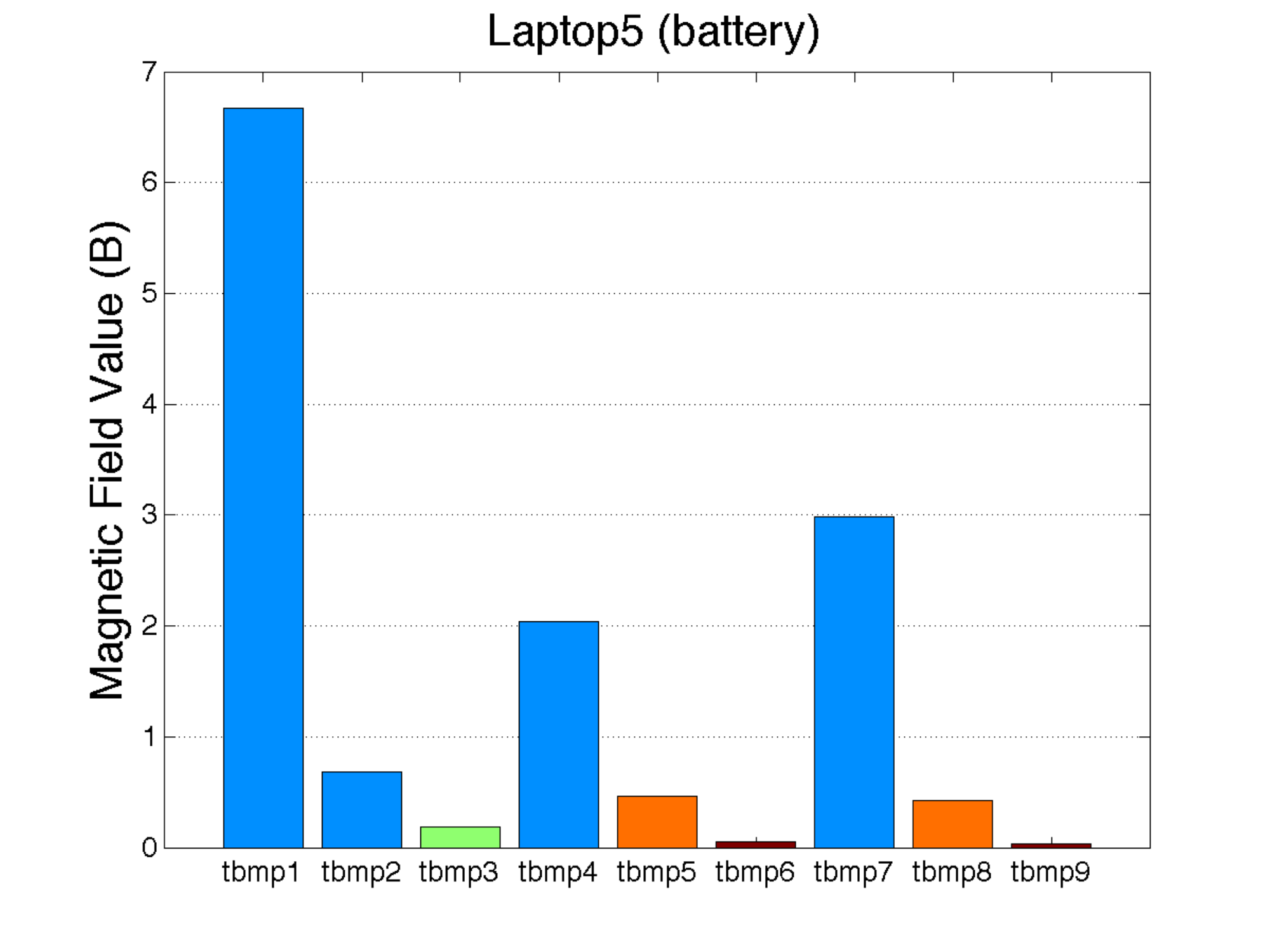}
}
\subfigure[]{
\includegraphics[width=3cm, height=3cm, keepaspectratio]{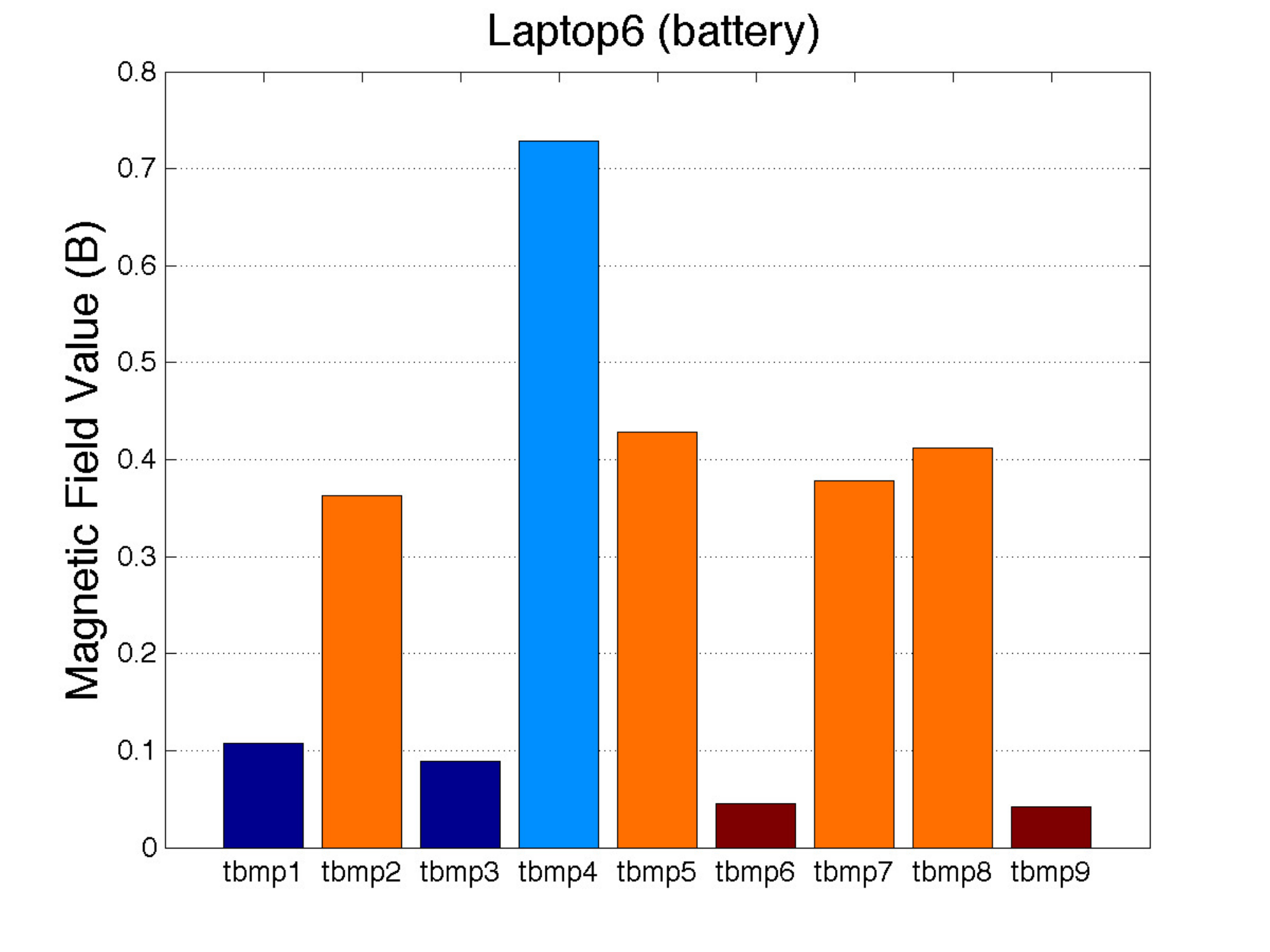}
}
\subfigure[]{
\includegraphics[width=3cm, height=3cm, keepaspectratio]{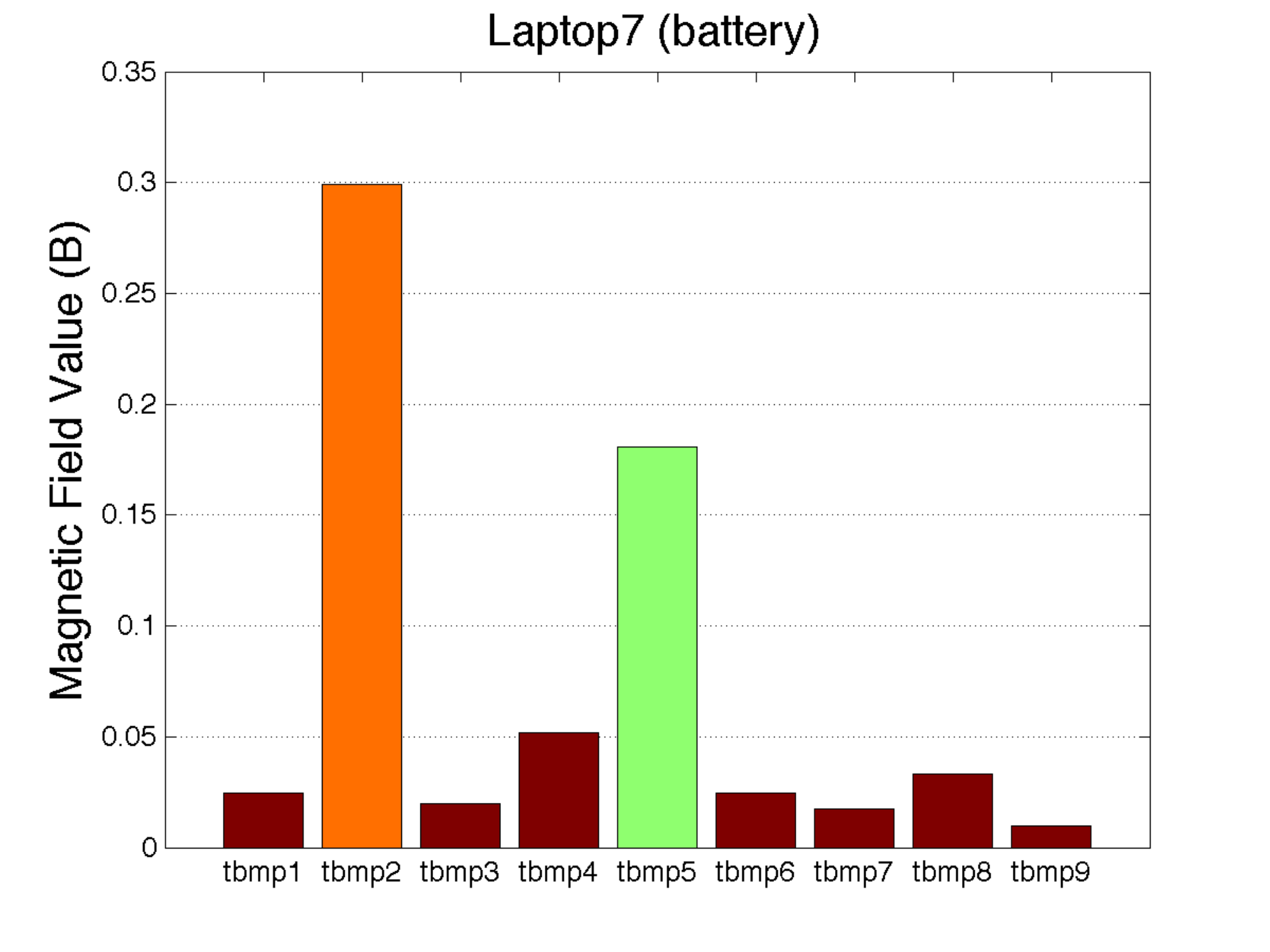}
}
\subfigure[]{
\includegraphics[width=3cm, height=3cm, keepaspectratio]{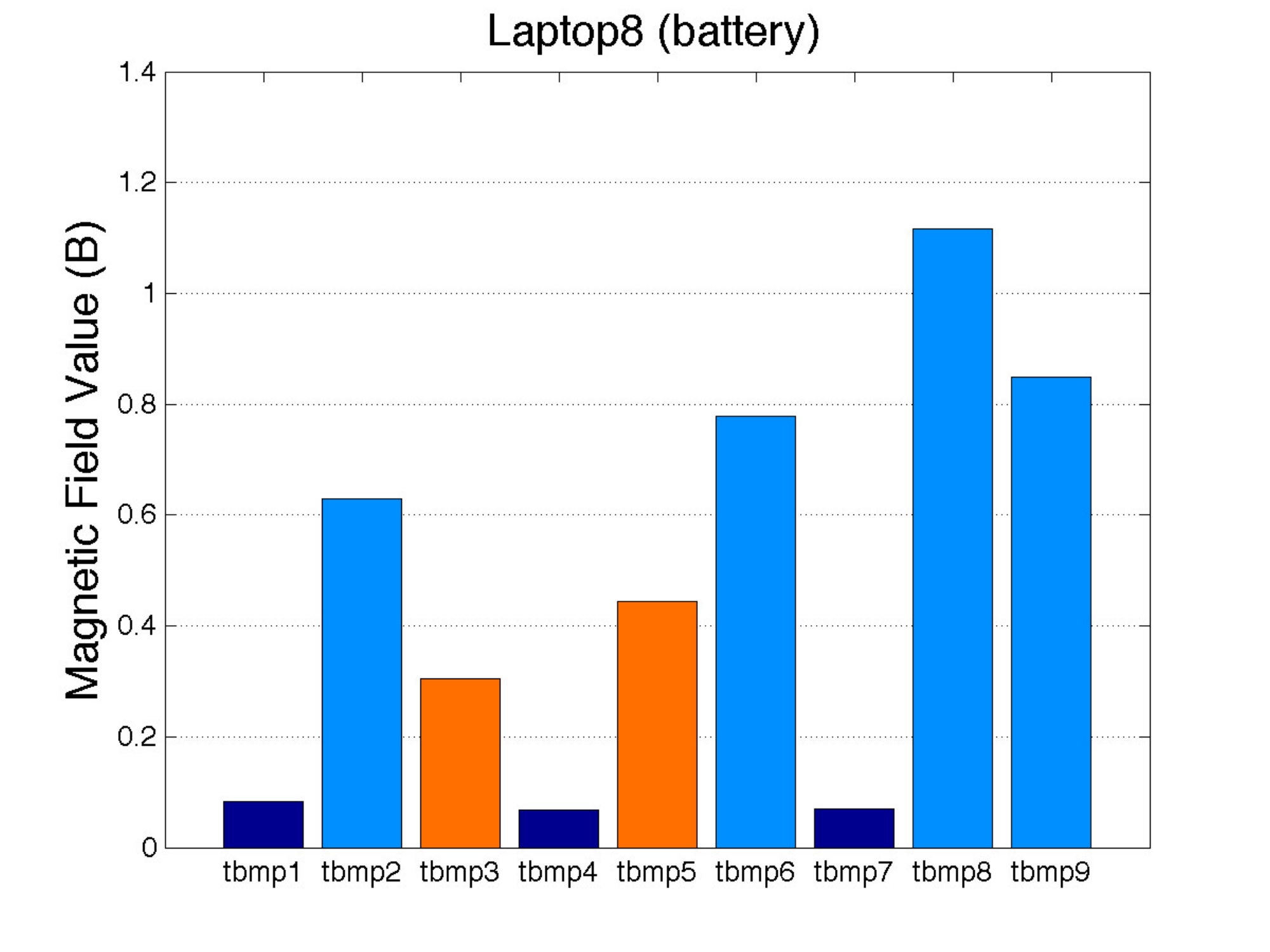}
}
\subfigure[]{
\includegraphics[width=3cm, height=3cm, keepaspectratio]{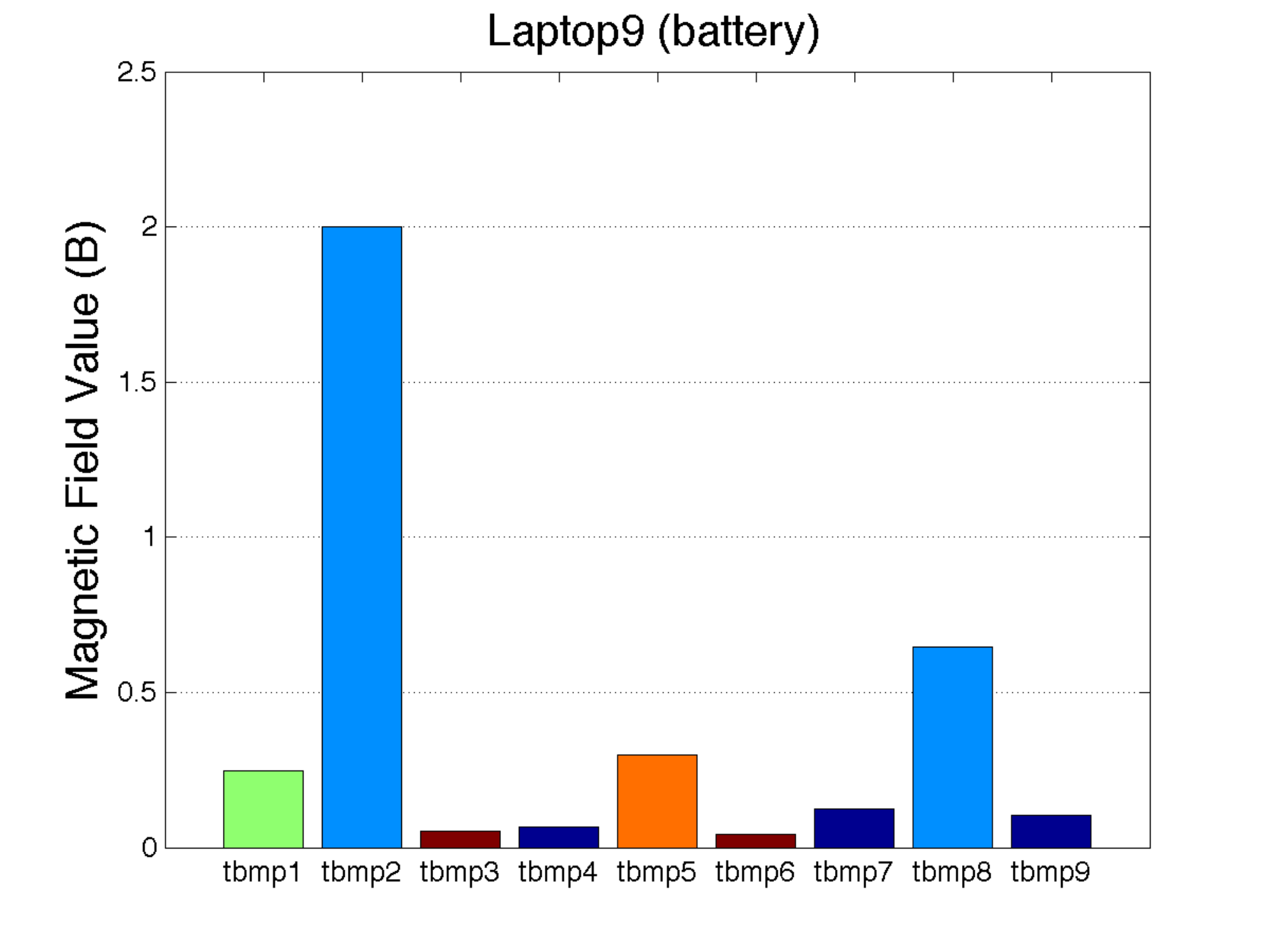}
}
\subfigure[]{
\includegraphics[width=3cm, height=3cm, keepaspectratio]{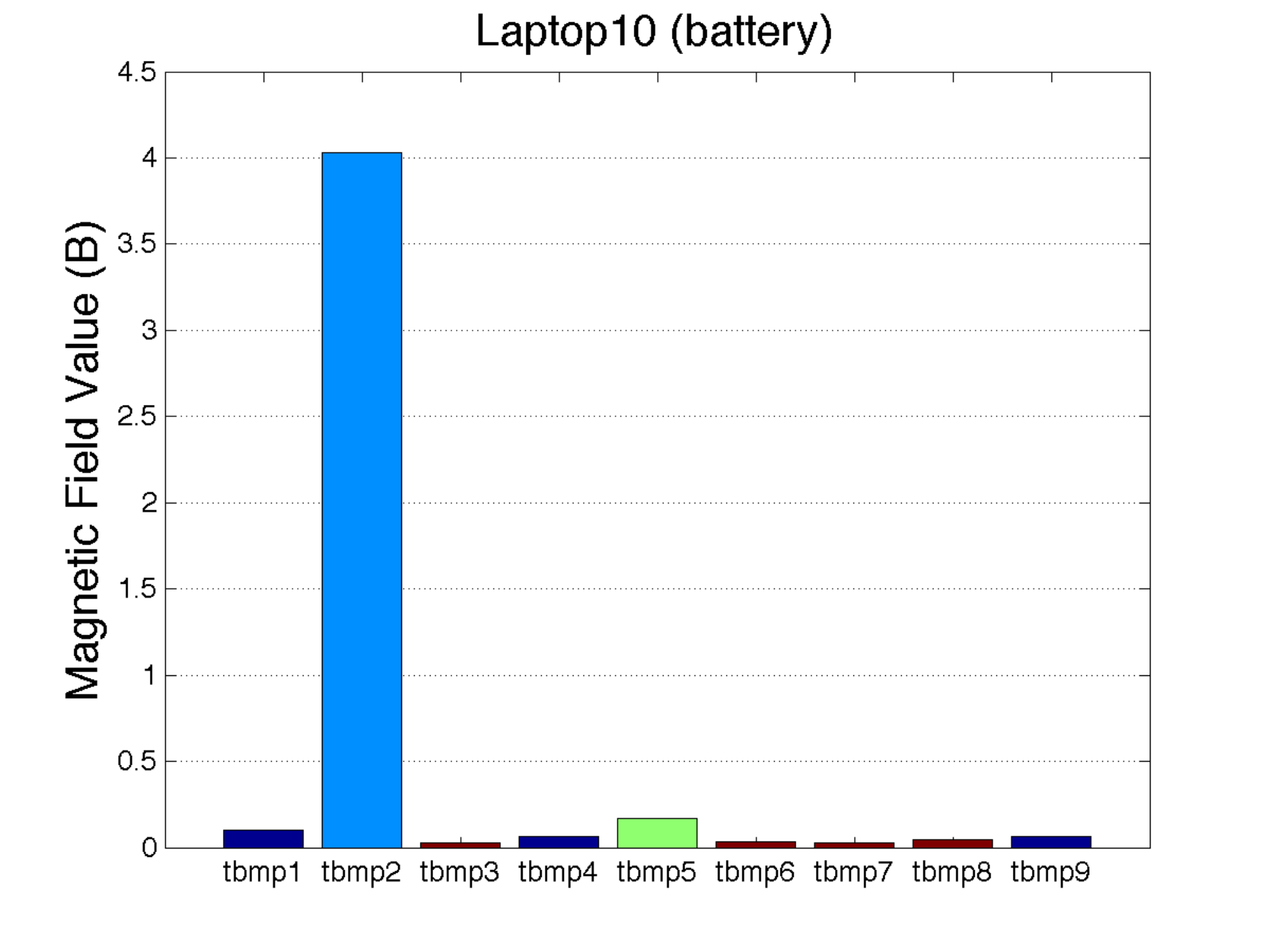}
}
\subfigure[]{
\includegraphics[width=3cm, height=3cm, keepaspectratio]{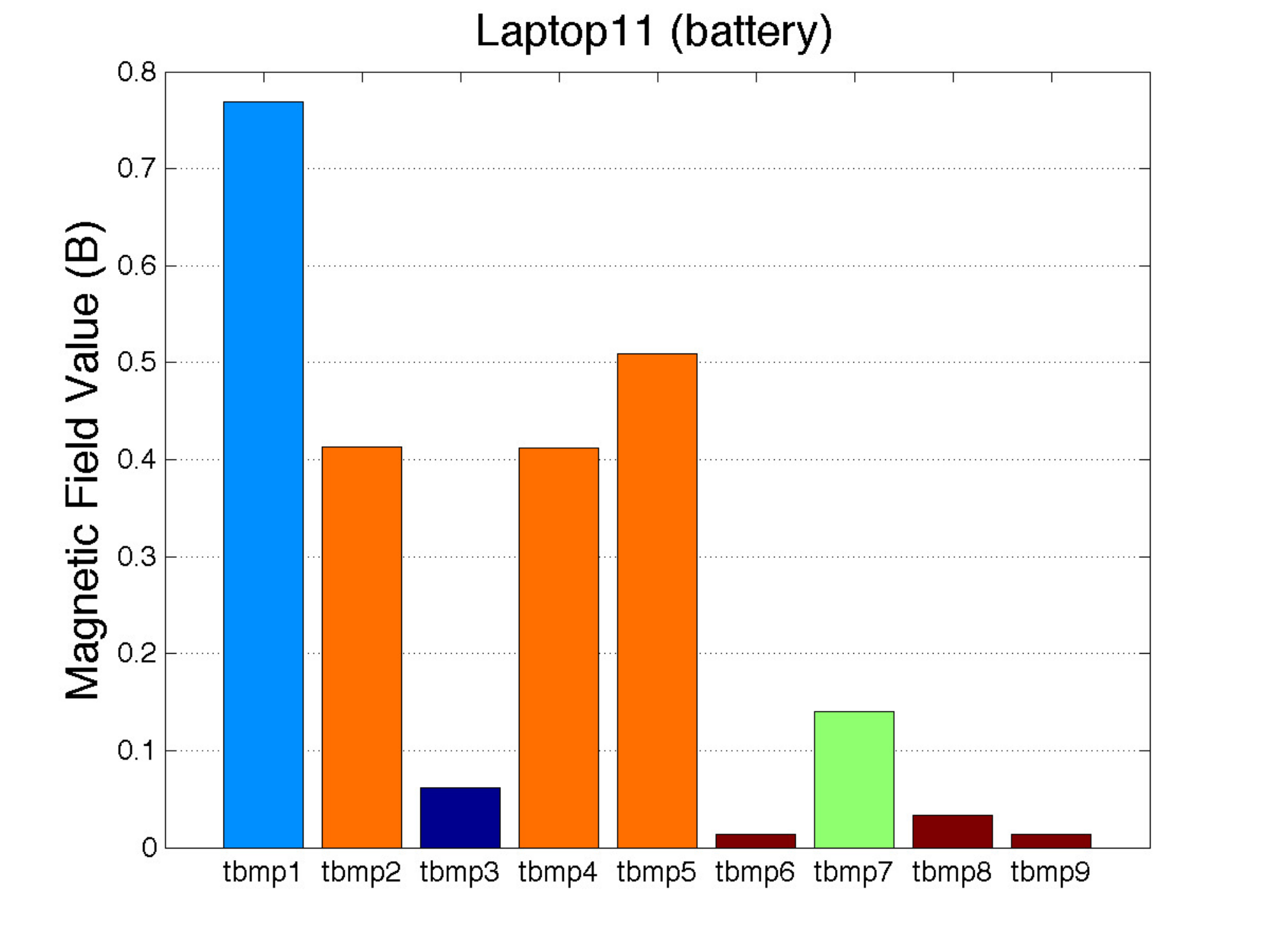}
}
\subfigure[]{
\includegraphics[width=3cm, height=3cm, keepaspectratio]{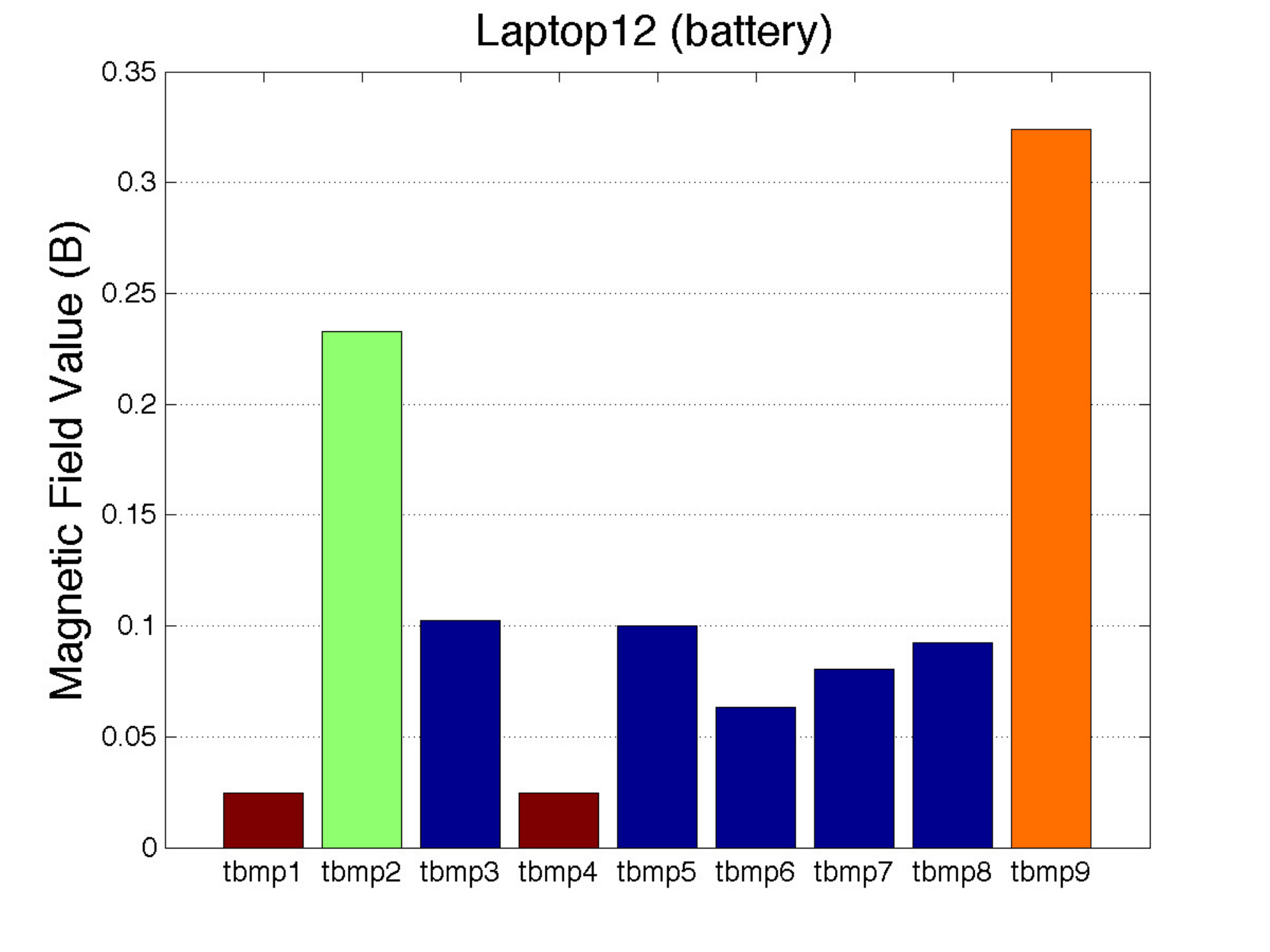}
}
\subfigure[]{
\includegraphics[width=3cm, height=3cm, keepaspectratio]{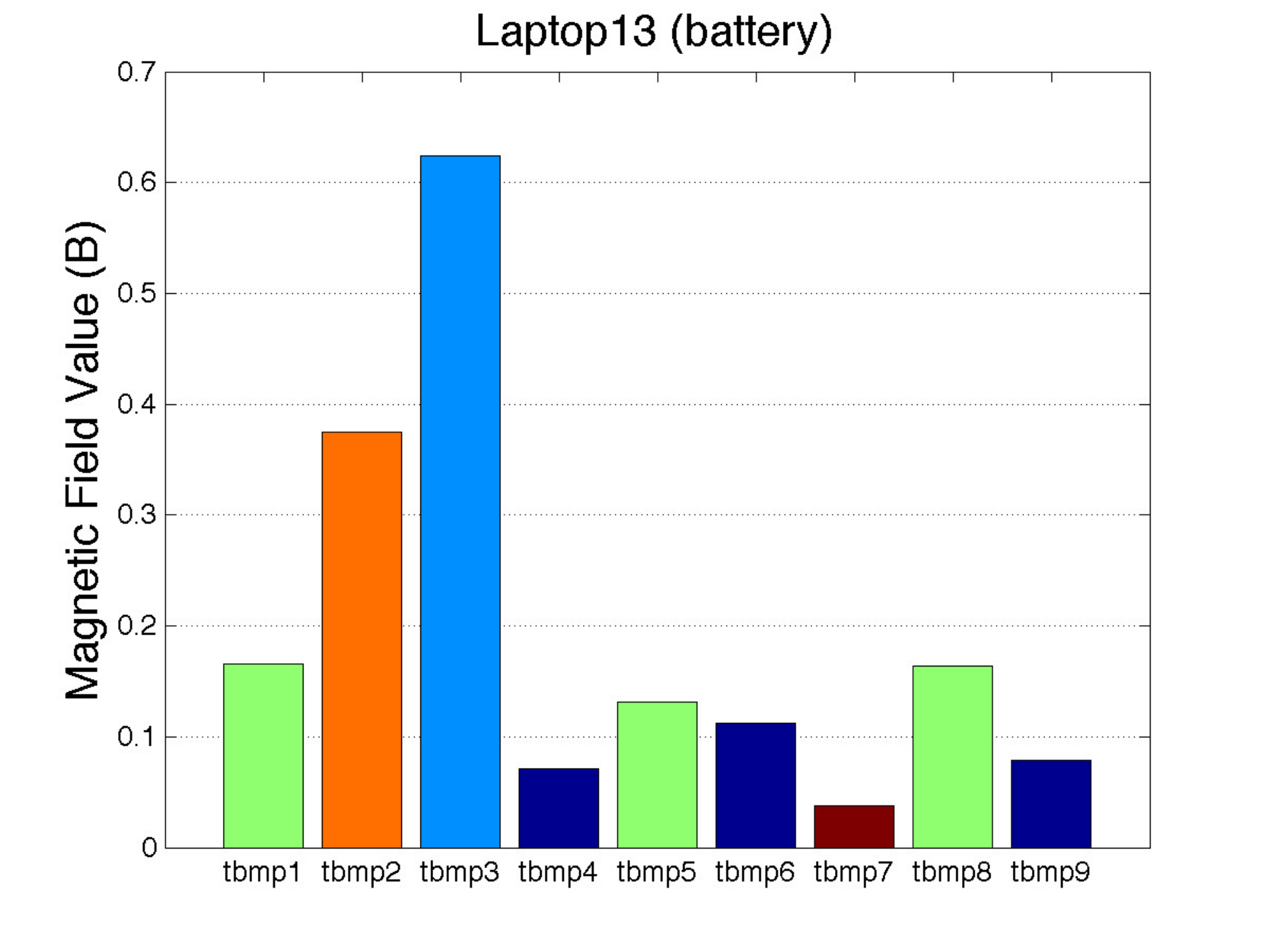}
}
\caption{Results of K-Medians on the dataset of top body points for the 13 laptops supplied by battery (a-m). Each color along the figures represents a cluster of top body points. The names of the top body points included in the cluster are reported in correspondence to the $x$ axis. $Y$ axis gives the magnetic field values measured at the top body points belonging to that cluster.}
\label{Figure7}
\end{center}
\end{figure*}

\begin{figure*}[t]
\begin{center}
\subfigure[]{
\includegraphics[width=3cm, height=3cm, keepaspectratio]{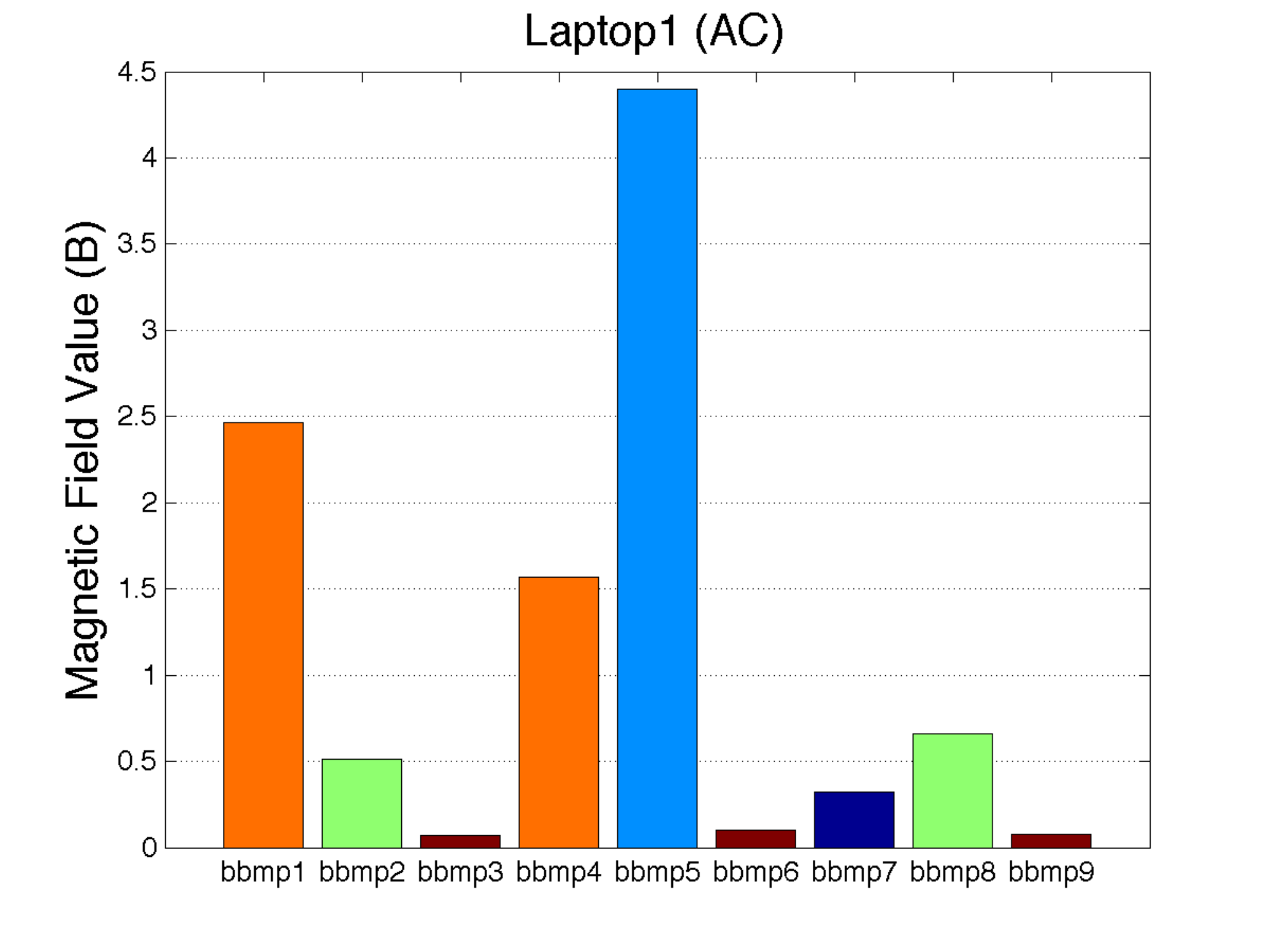}
}
\subfigure[]{
\includegraphics[width=3cm, height=3cm, keepaspectratio]{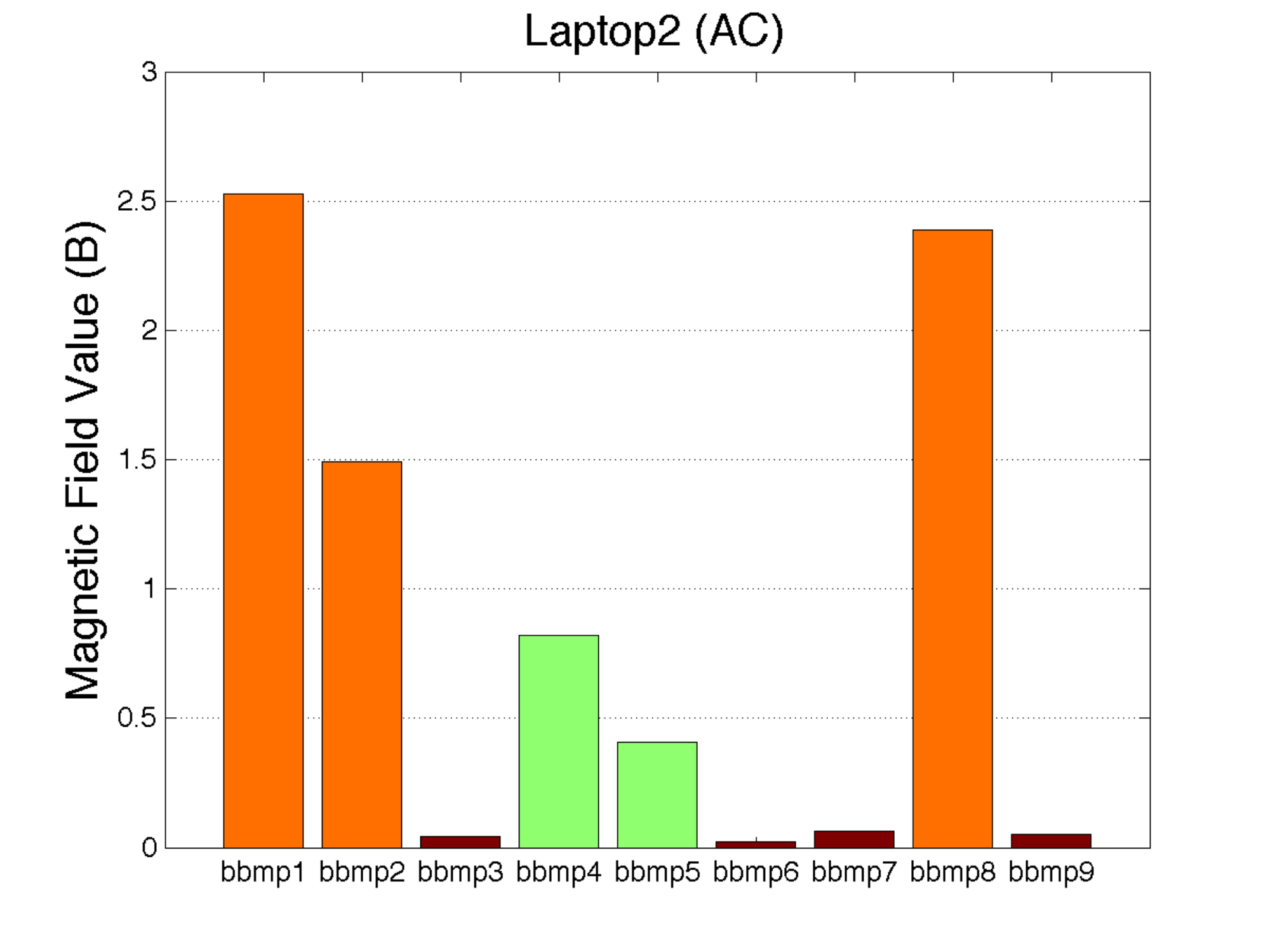}
}
\subfigure[]{
\includegraphics[width=3cm, height=3cm, keepaspectratio]{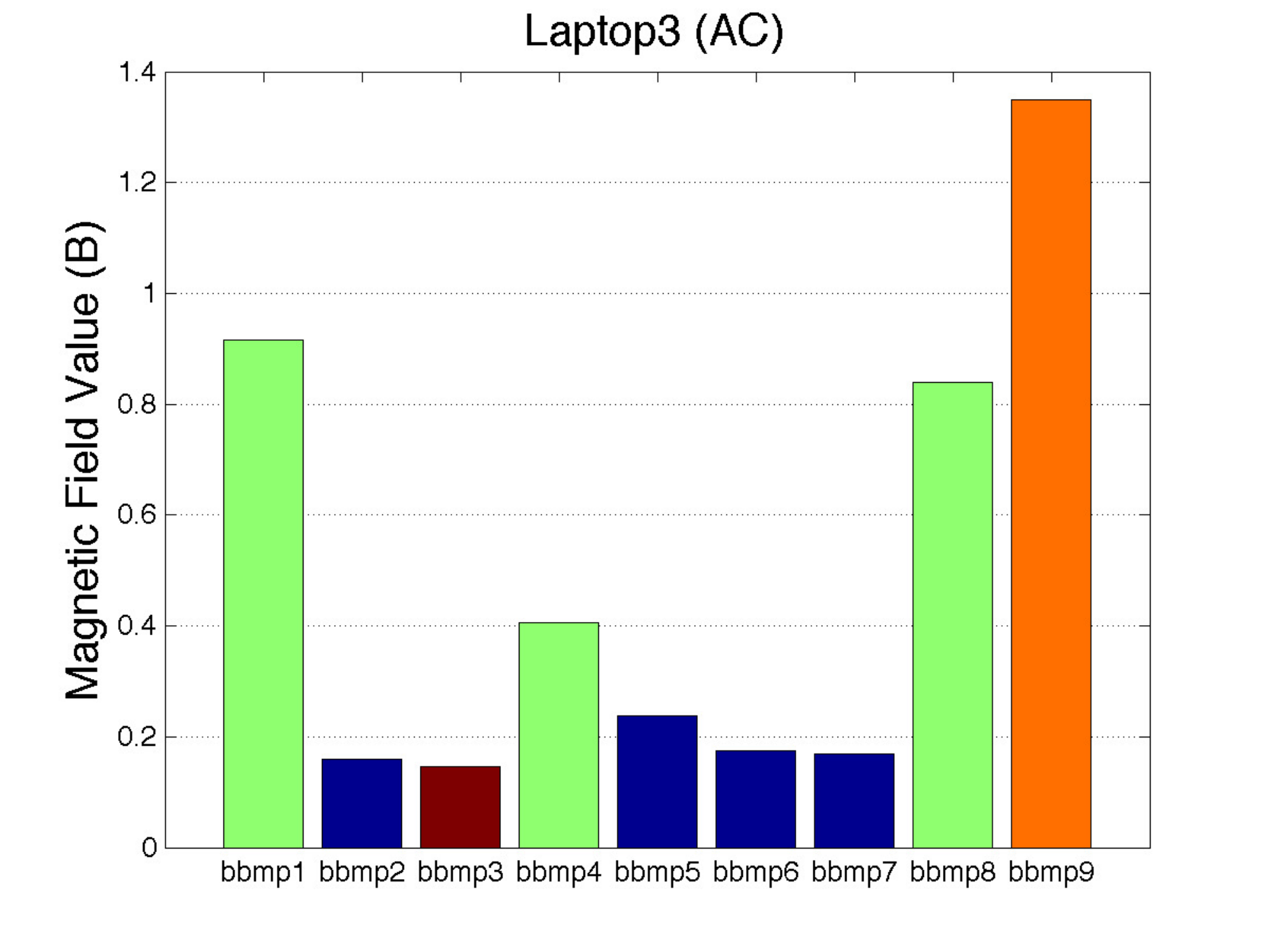}
}
\subfigure[]{
\includegraphics[width=3cm, height=3cm, keepaspectratio]{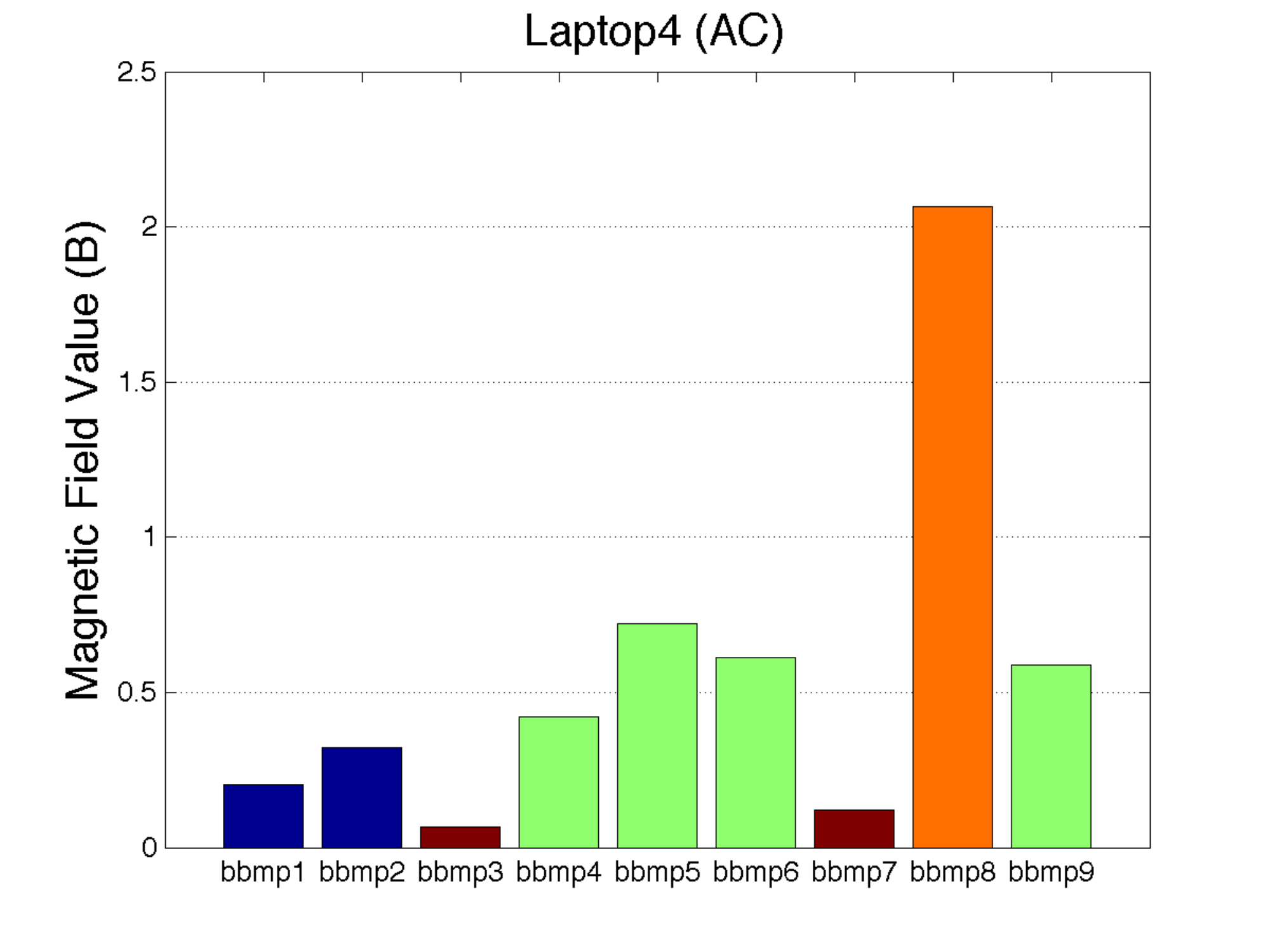}
}
\subfigure[]{
\includegraphics[width=3cm, height=3cm, keepaspectratio]{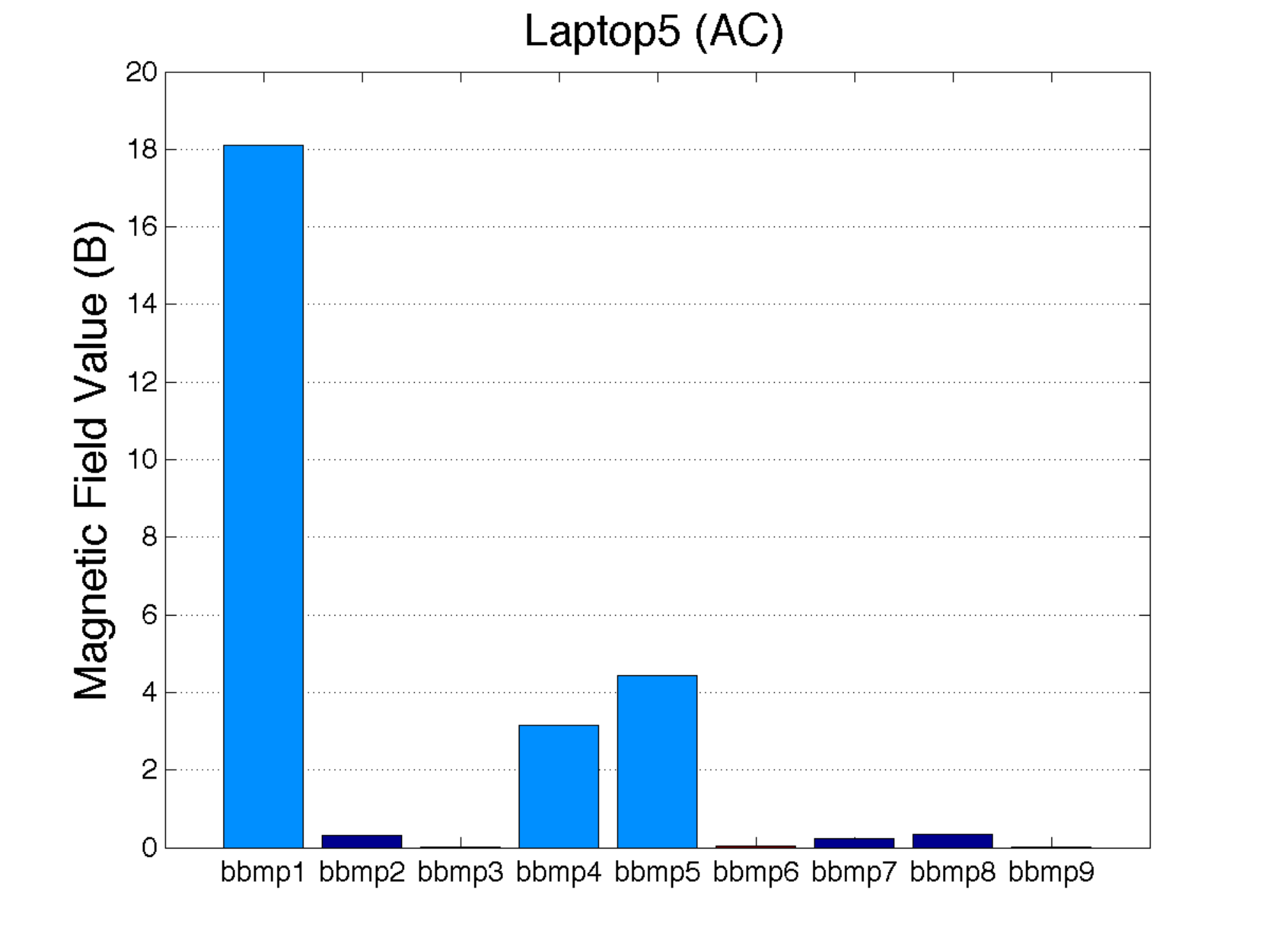}
}
\subfigure[]{
\includegraphics[width=3cm, height=3cm, keepaspectratio]{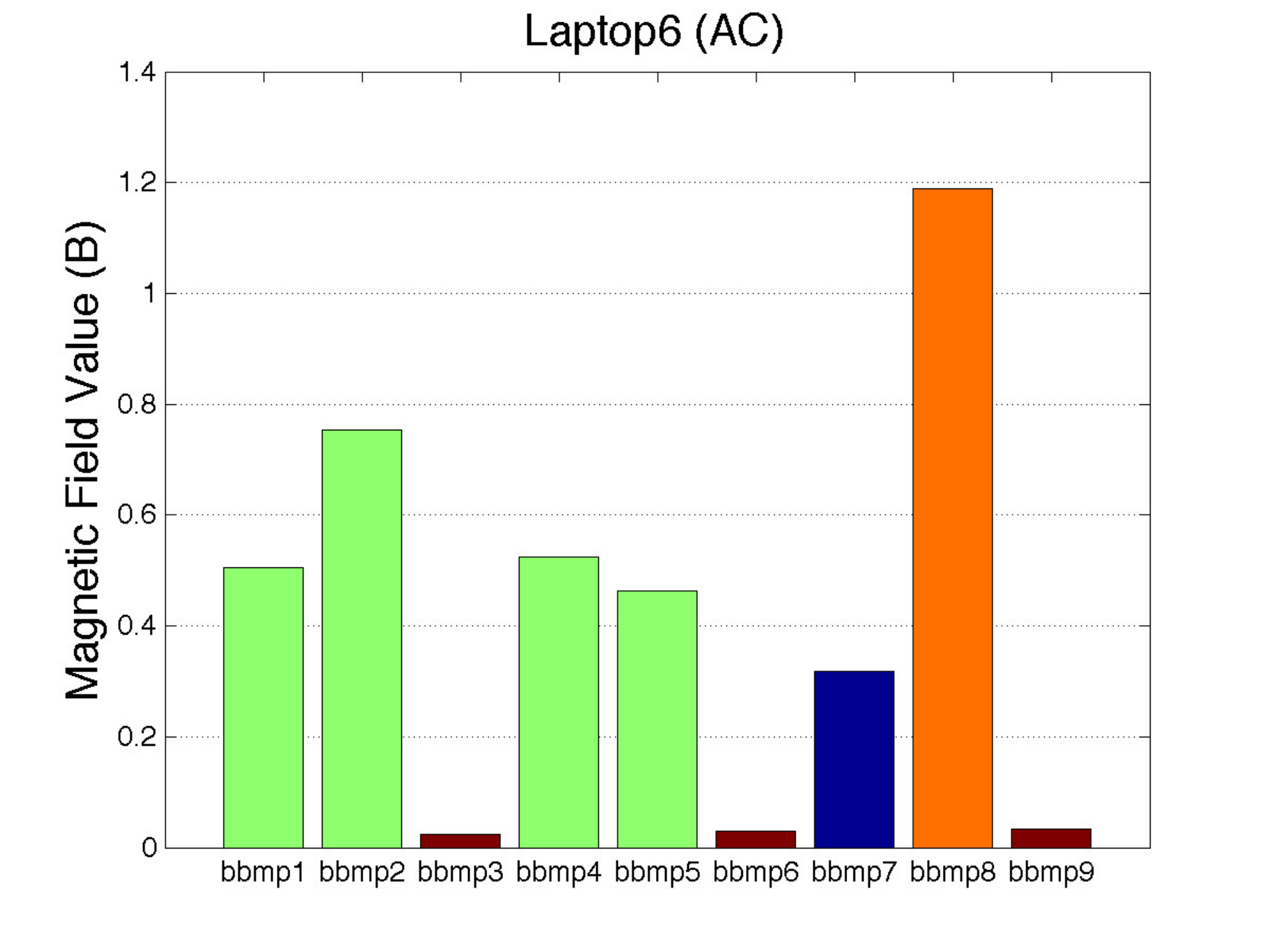}
}
\subfigure[]{
\includegraphics[width=3cm, height=3cm, keepaspectratio]{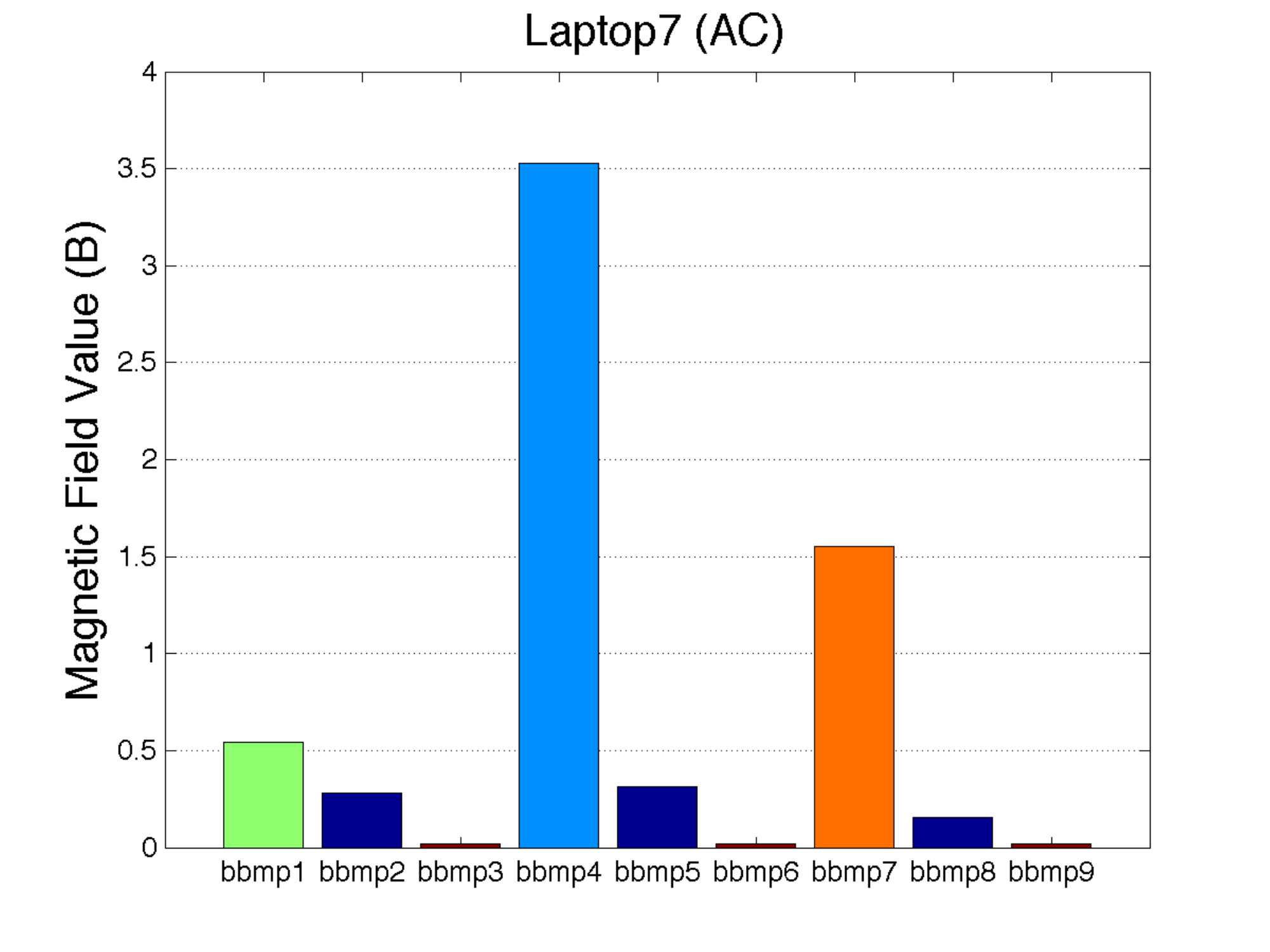}
}
\subfigure[]{
\includegraphics[width=3cm, height=3cm, keepaspectratio]{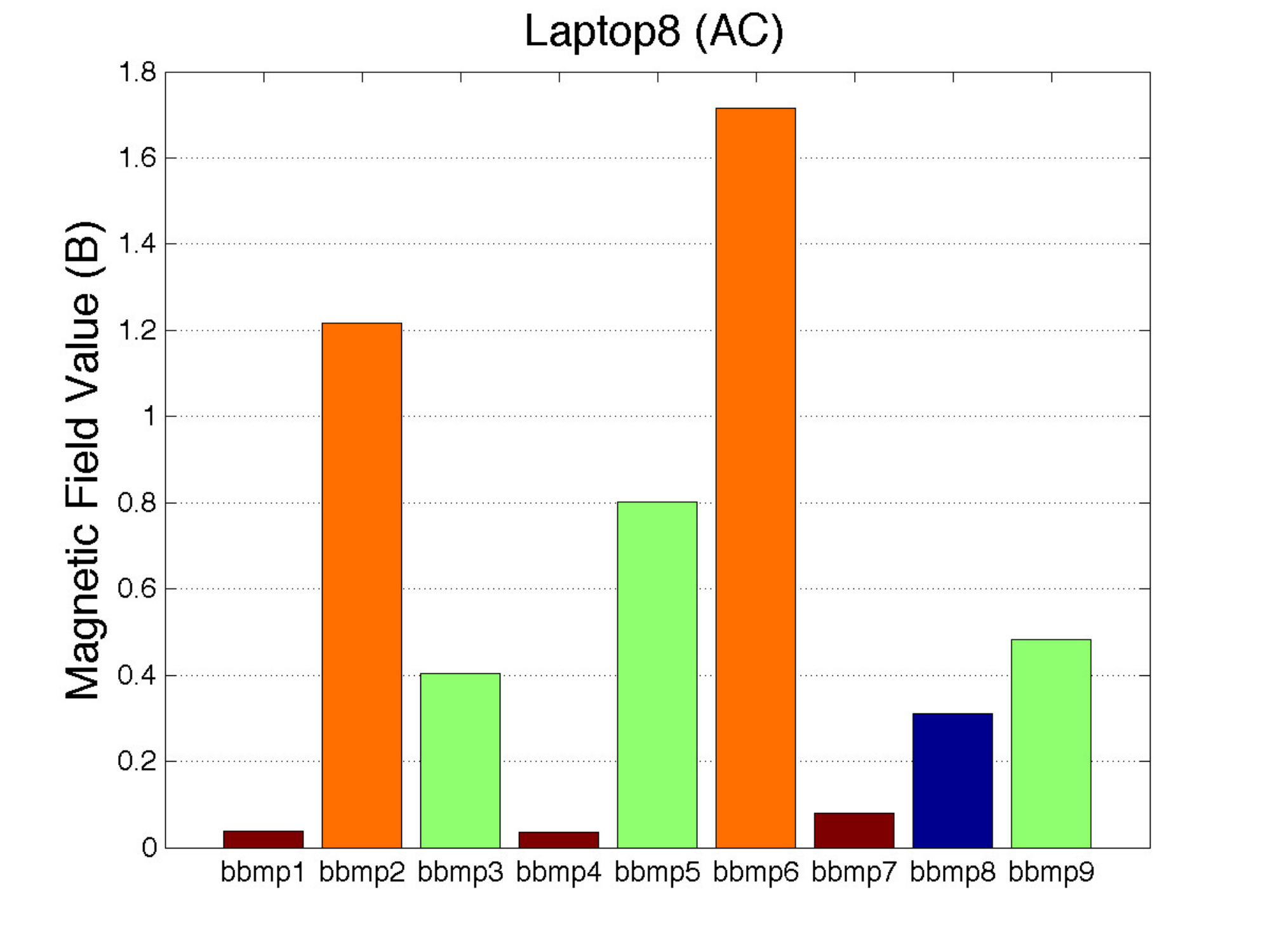}
}
\subfigure[]{
\includegraphics[width=3cm, height=3cm, keepaspectratio]{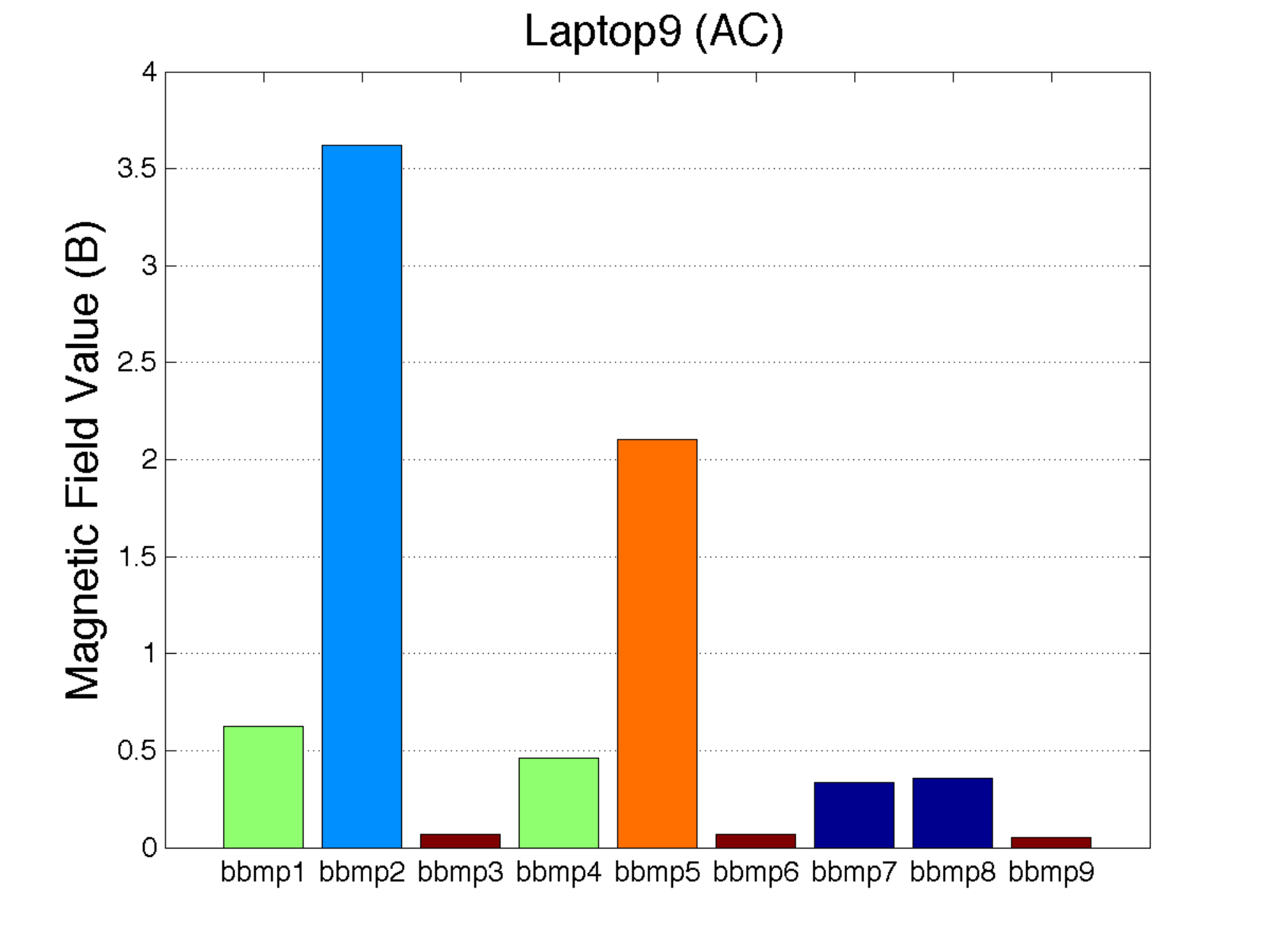}
}
\subfigure[]{
\includegraphics[width=3cm, height=3cm, keepaspectratio]{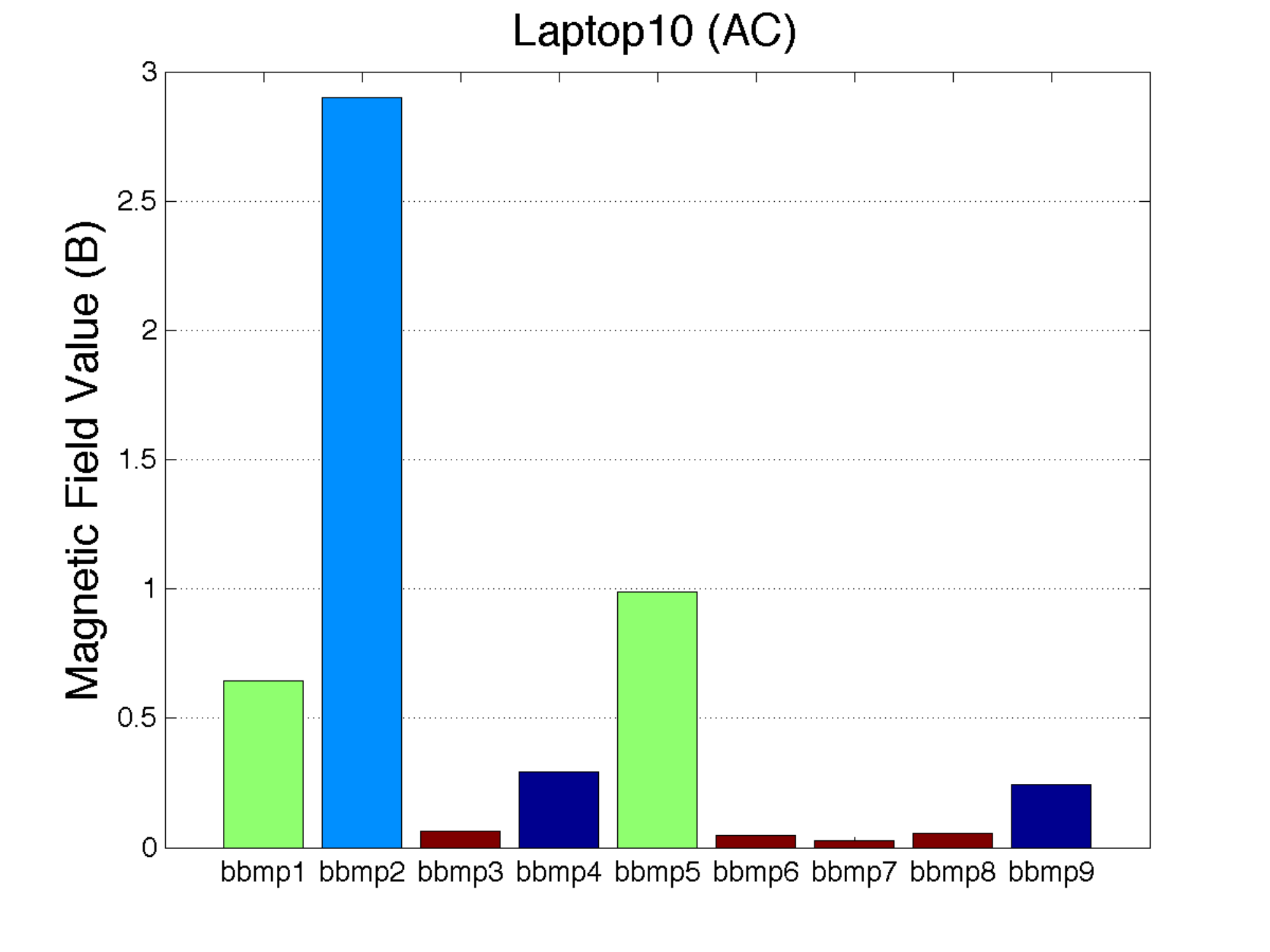}
}
\subfigure[]{
\includegraphics[width=3cm, height=3cm, keepaspectratio]{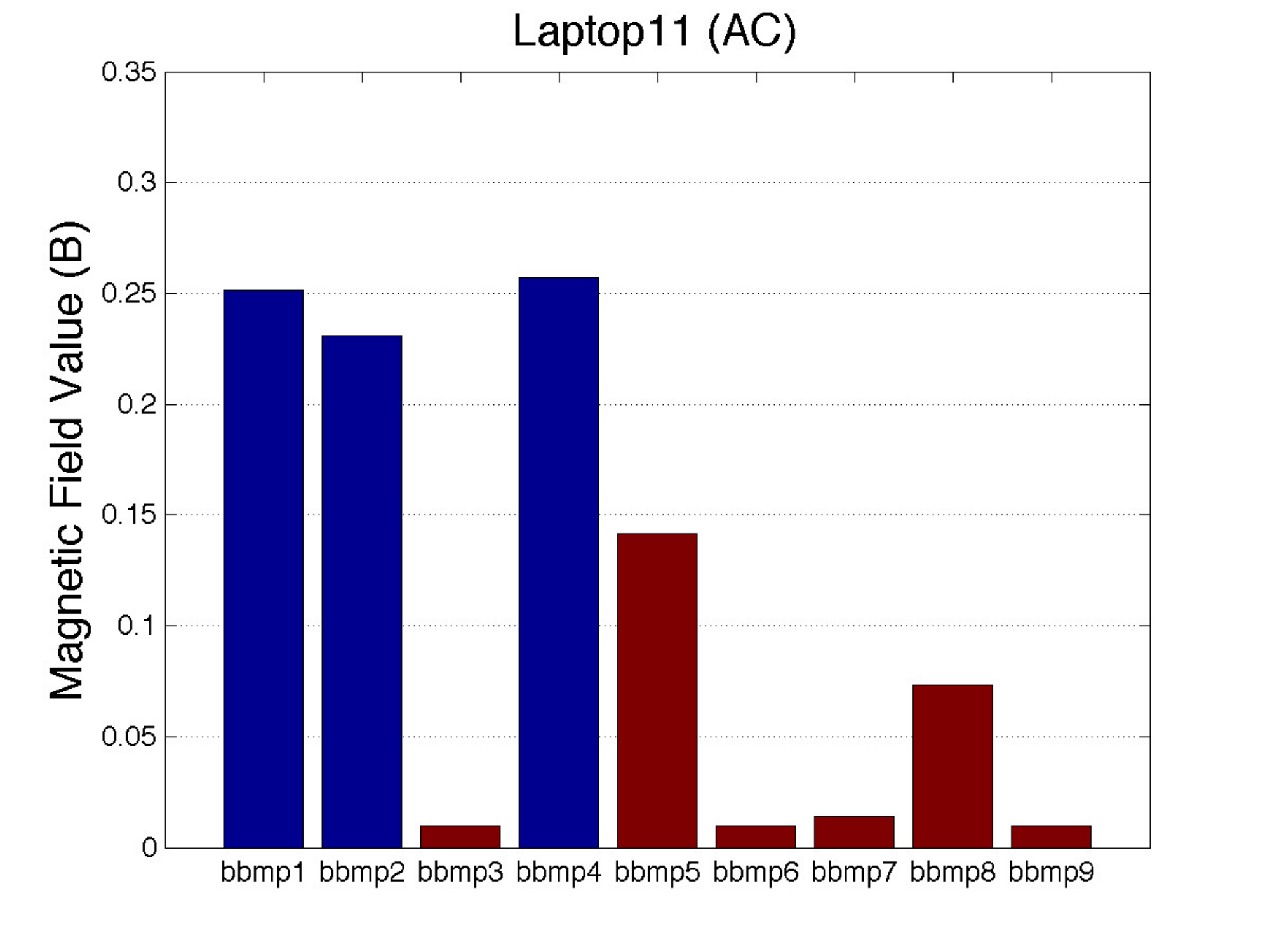}
}
\subfigure[]{
\includegraphics[width=3cm, height=3cm, keepaspectratio]{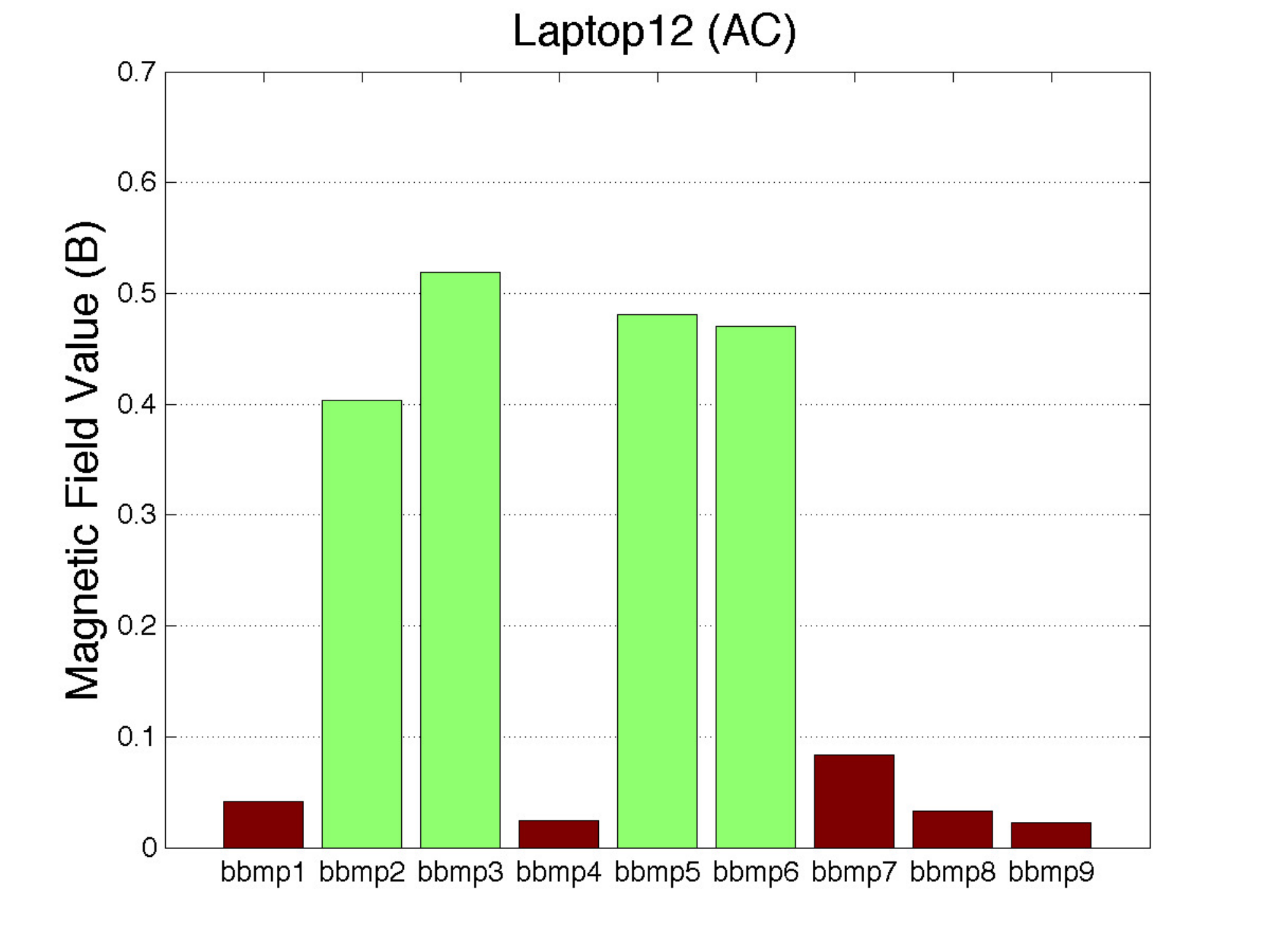}
}
\subfigure[]{
\includegraphics[width=3cm, height=3cm, keepaspectratio]{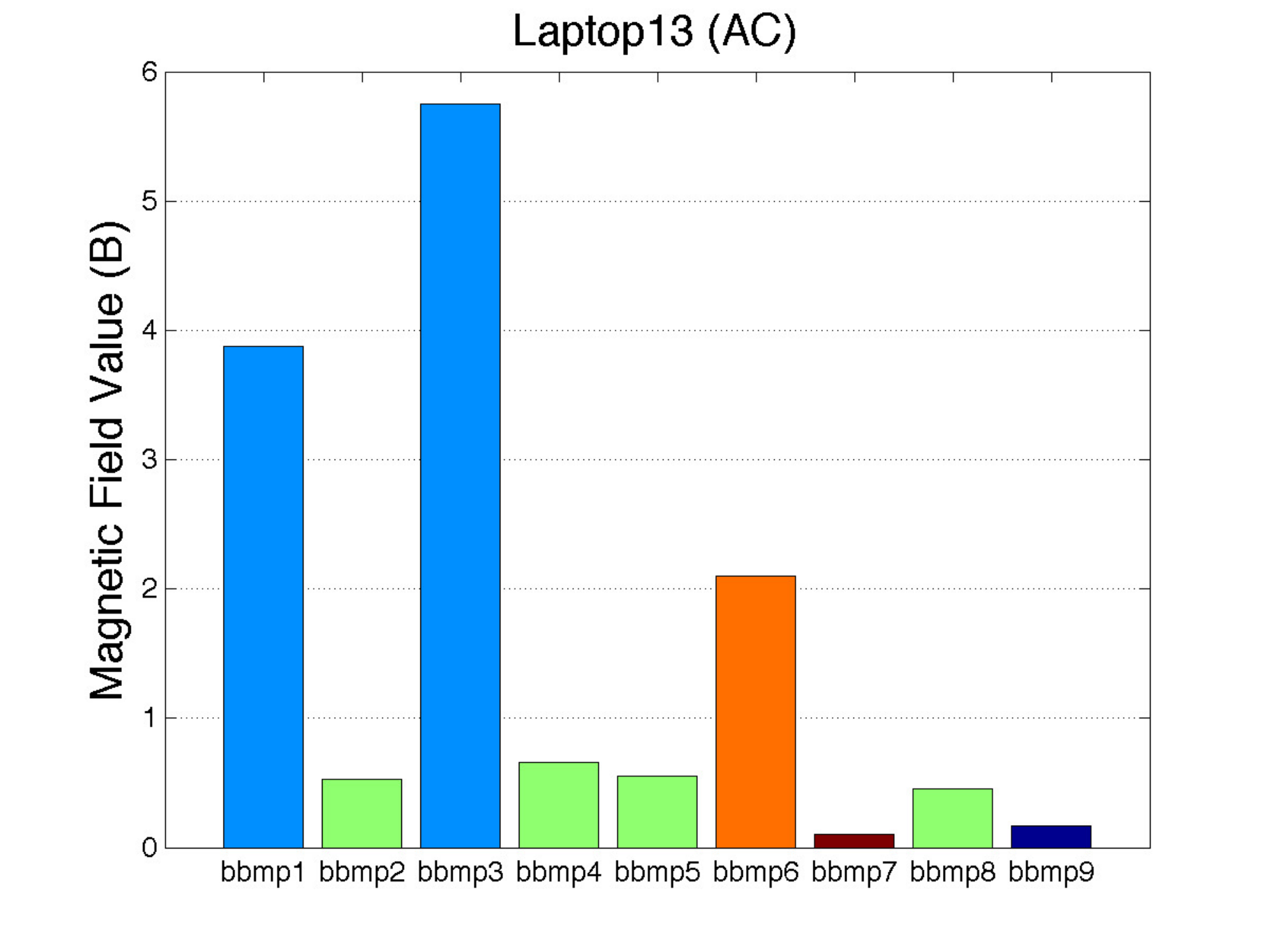}
}
\caption{Results of K-Medians on the dataset of bottom body points for the 13 laptops supplied by AC (a-m). Each color along the figures represents a cluster of bottom body points. The names of the bottom body points included in the cluster are reported in correspondence to the $x$ axis. $Y$ axis gives the magnetic field values measured at the bottom body points belonging to that cluster.}
\label{Figure8}
\end{center}
\end{figure*}

\begin{figure*}[t]
\begin{center}
\subfigure[]{
\includegraphics[width=3cm, height=3cm, keepaspectratio]{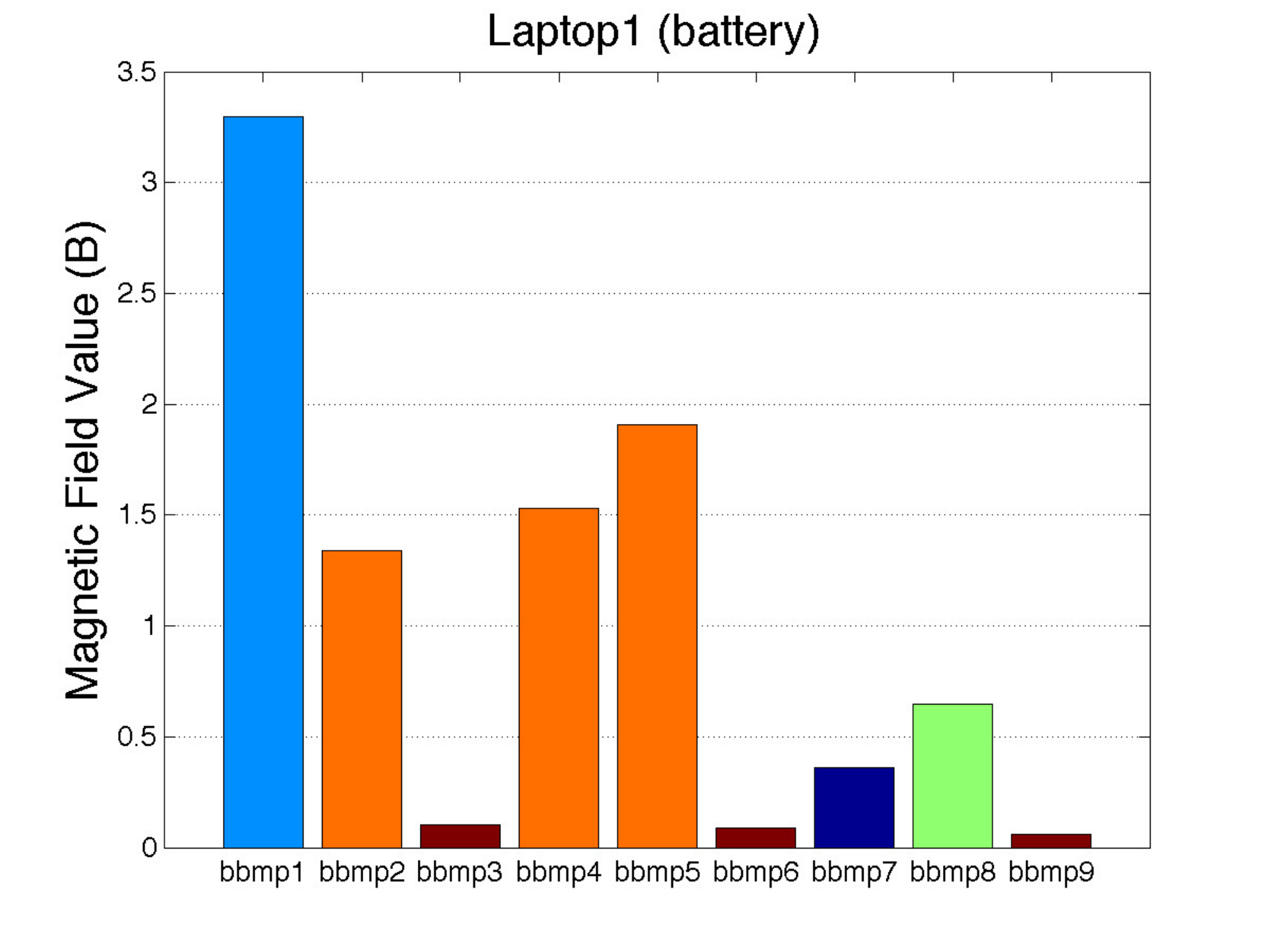}
}
\subfigure[]{
\includegraphics[width=3cm, height=3cm, keepaspectratio]{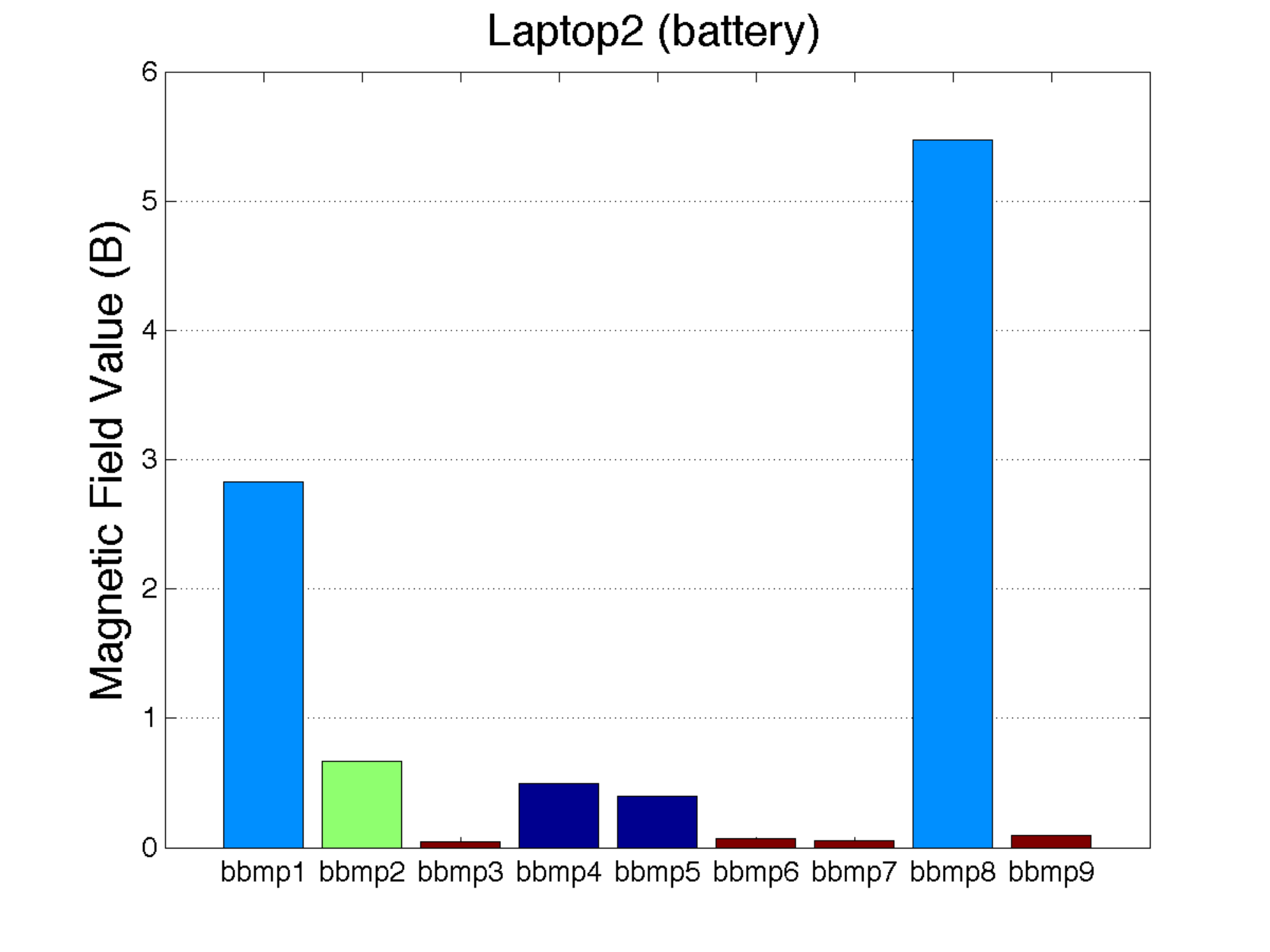}
}
\subfigure[]{
\includegraphics[width=3cm, height=3cm, keepaspectratio]{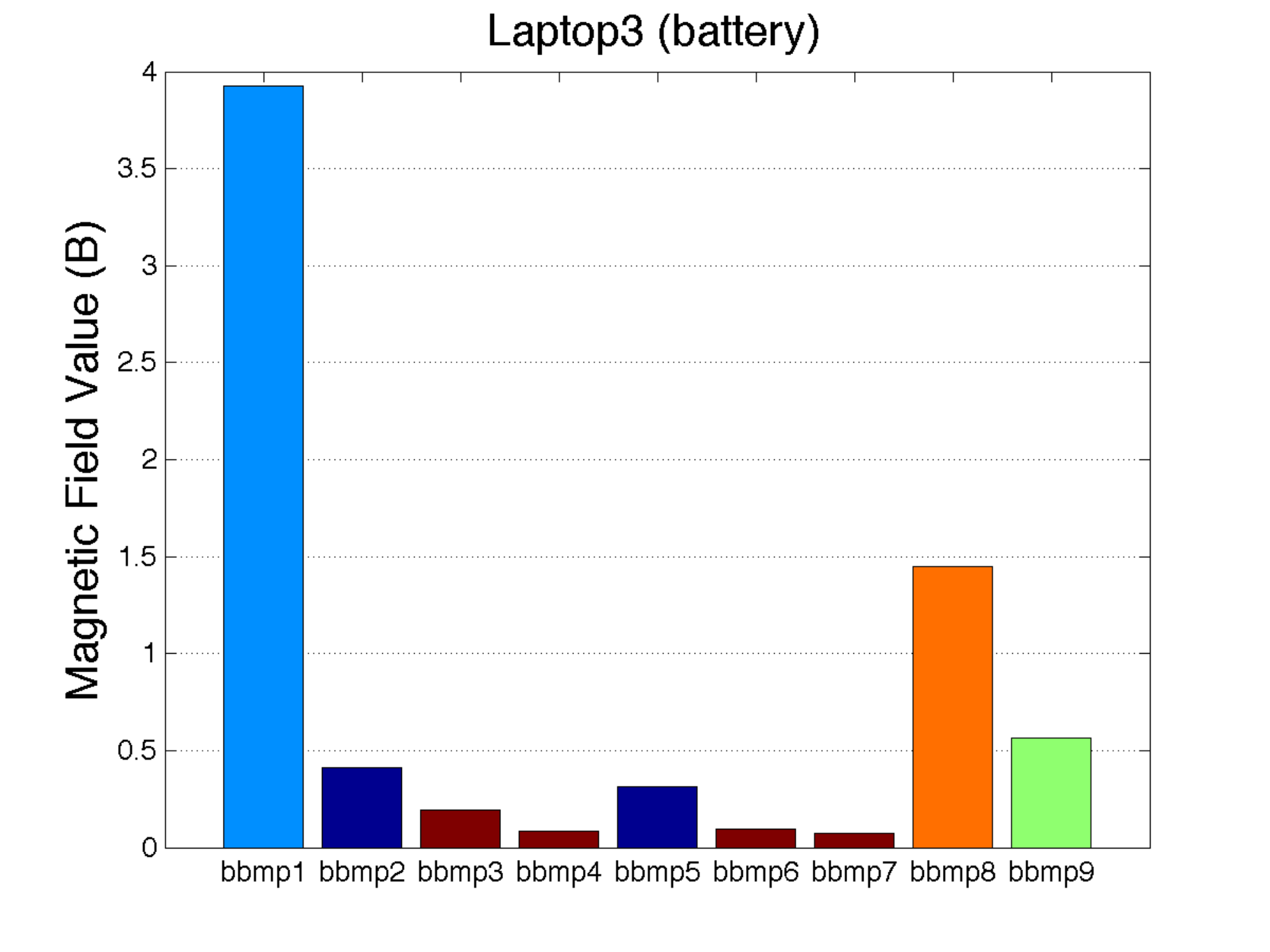}
}
\subfigure[]{
\includegraphics[width=3cm, height=3cm, keepaspectratio]{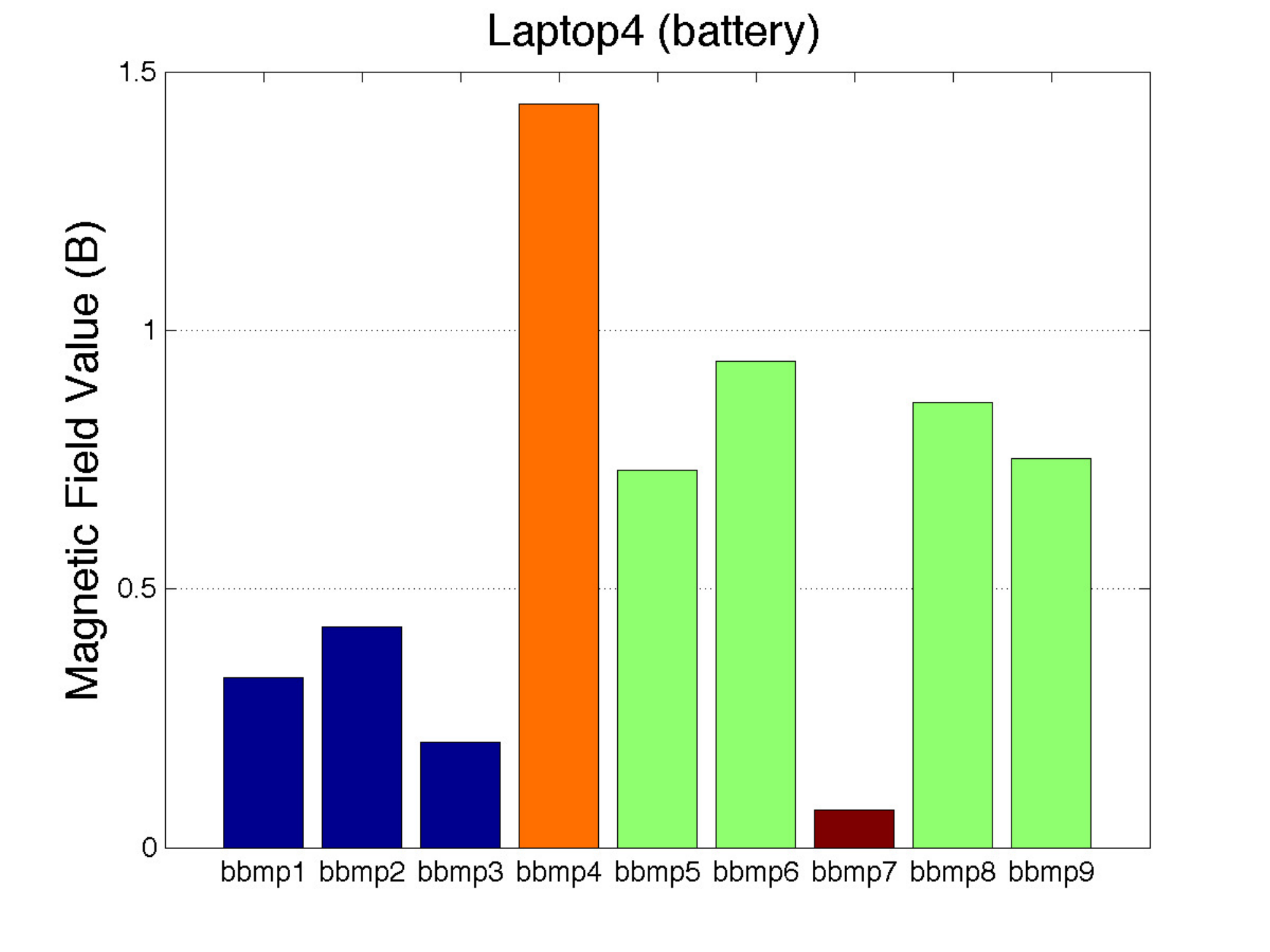}
}
\subfigure[]{
\includegraphics[width=3cm, height=3cm, keepaspectratio]{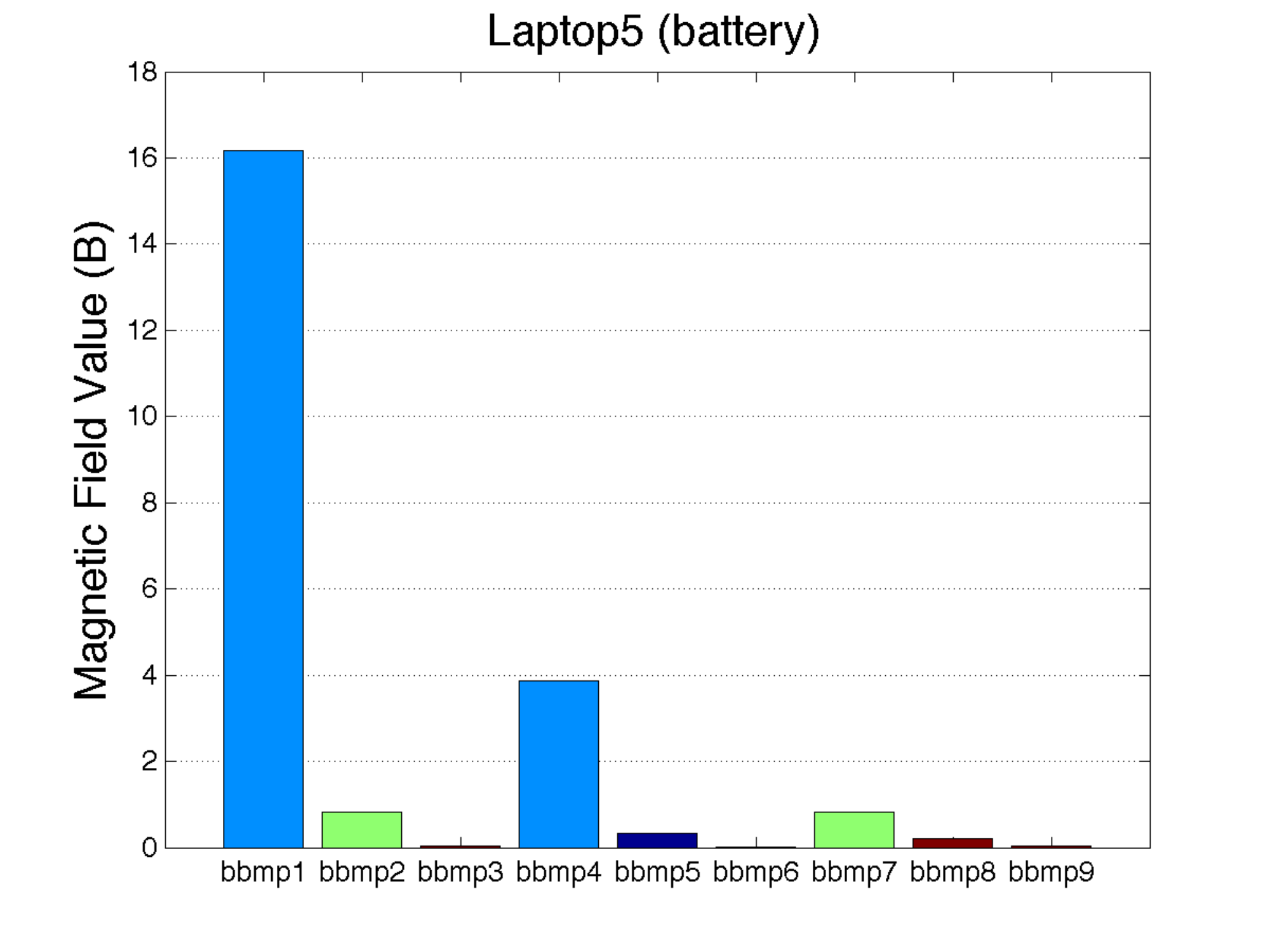}
}
\subfigure[]{
\includegraphics[width=3cm, height=3cm, keepaspectratio]{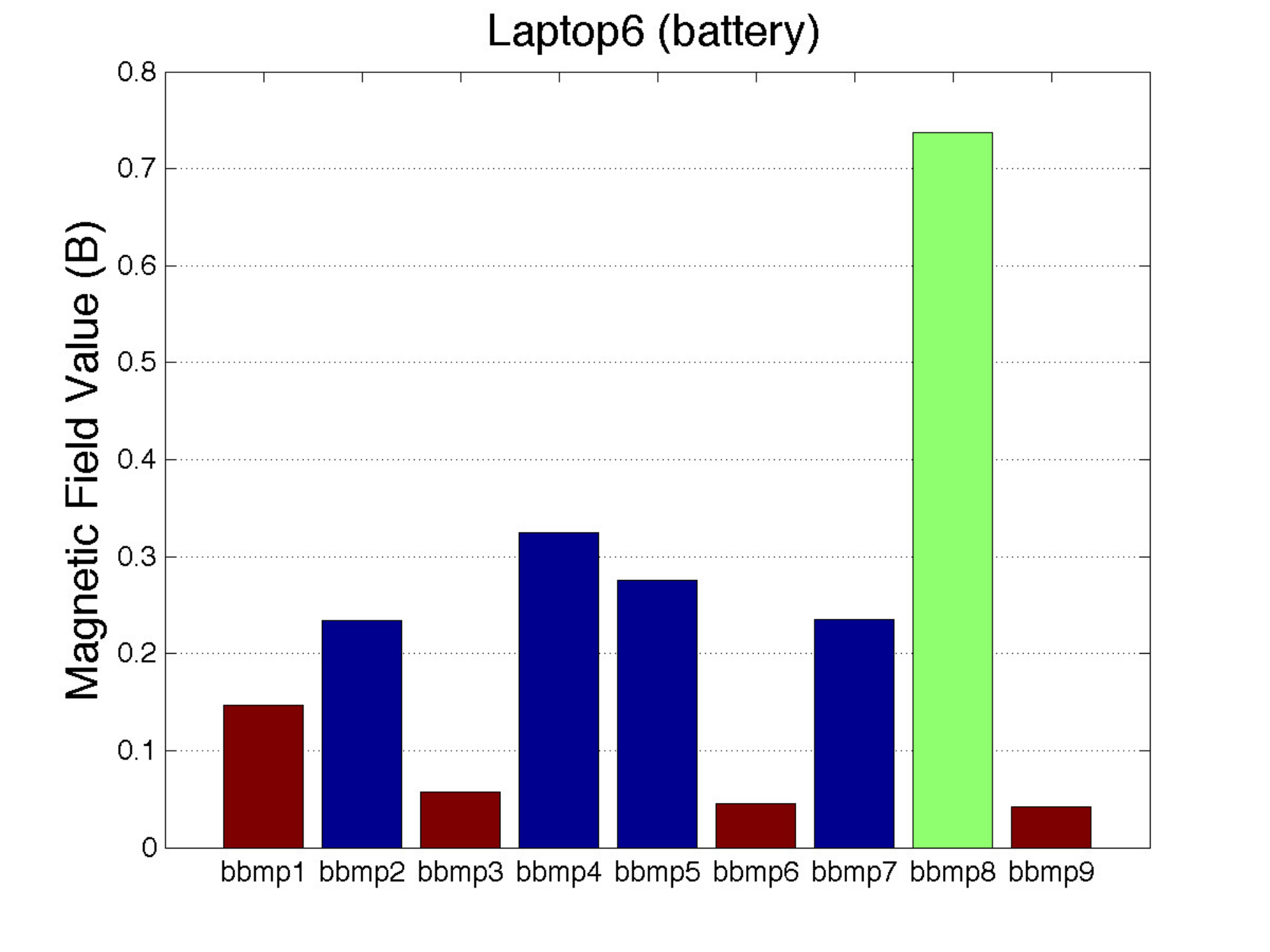}
}
\subfigure[]{
\includegraphics[width=3cm, height=3cm, keepaspectratio]{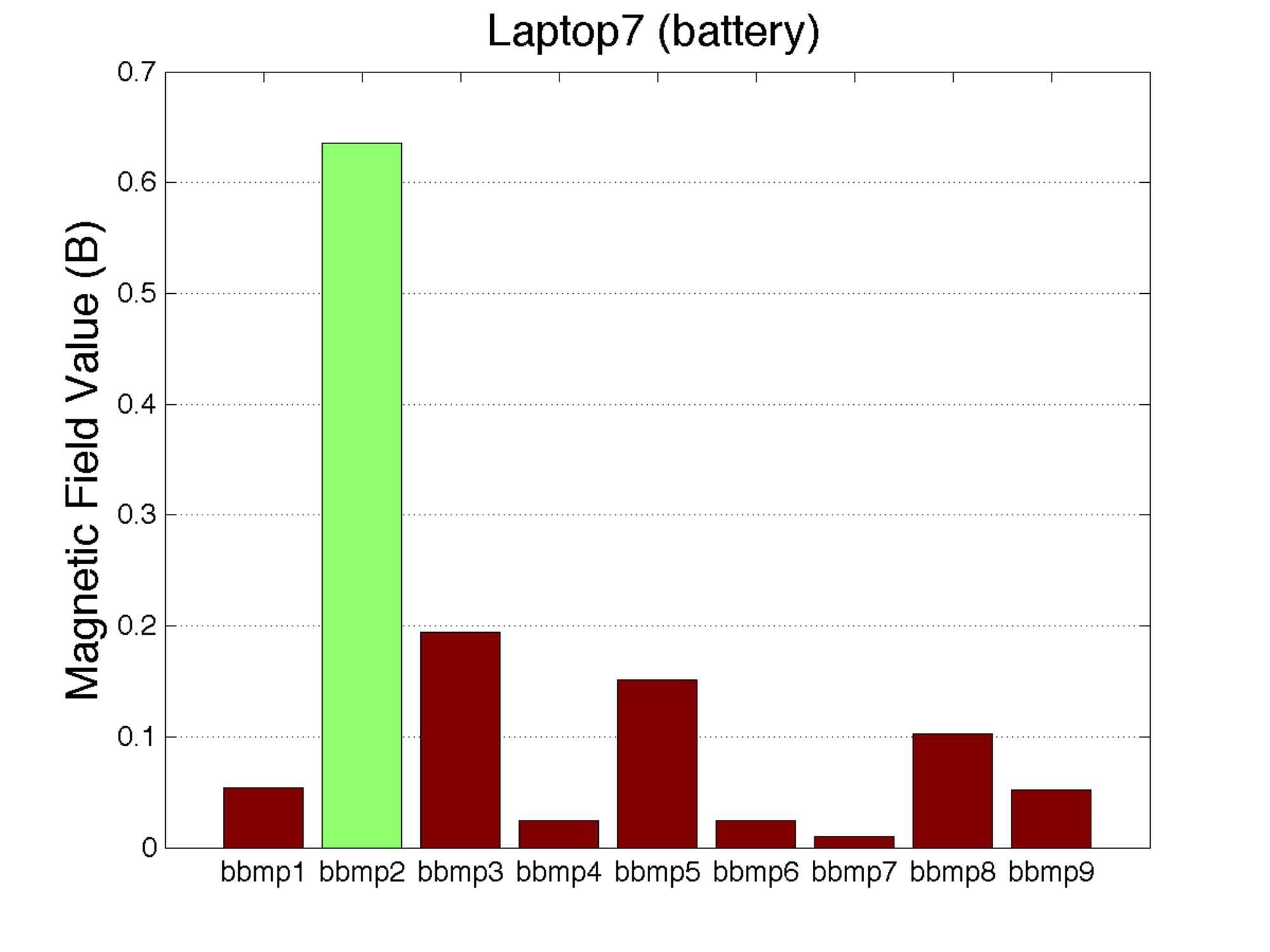}
}
\subfigure[]{
\includegraphics[width=3cm, height=3cm, keepaspectratio]{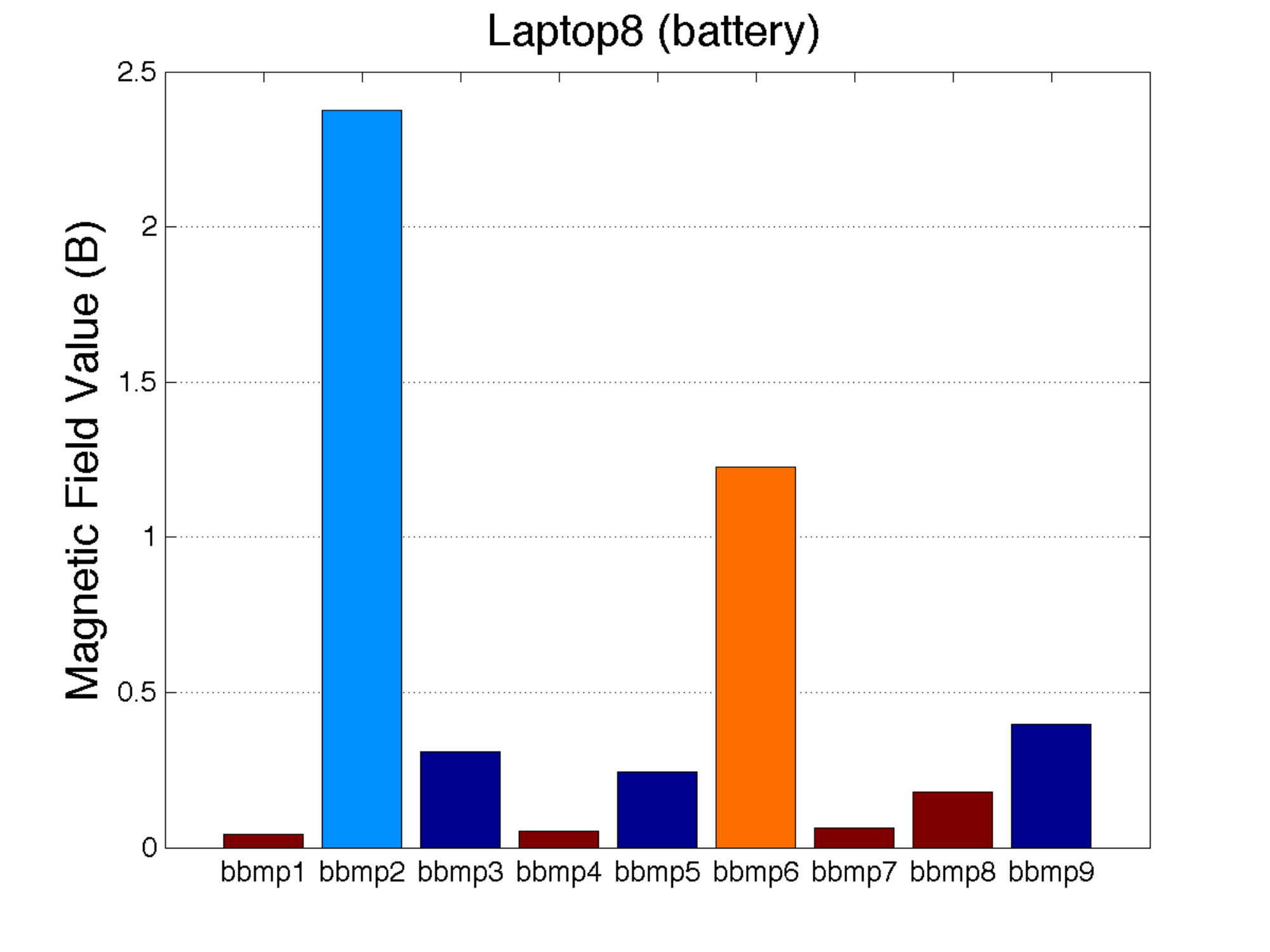}
}
\subfigure[]{
\includegraphics[width=3cm, height=3cm, keepaspectratio]{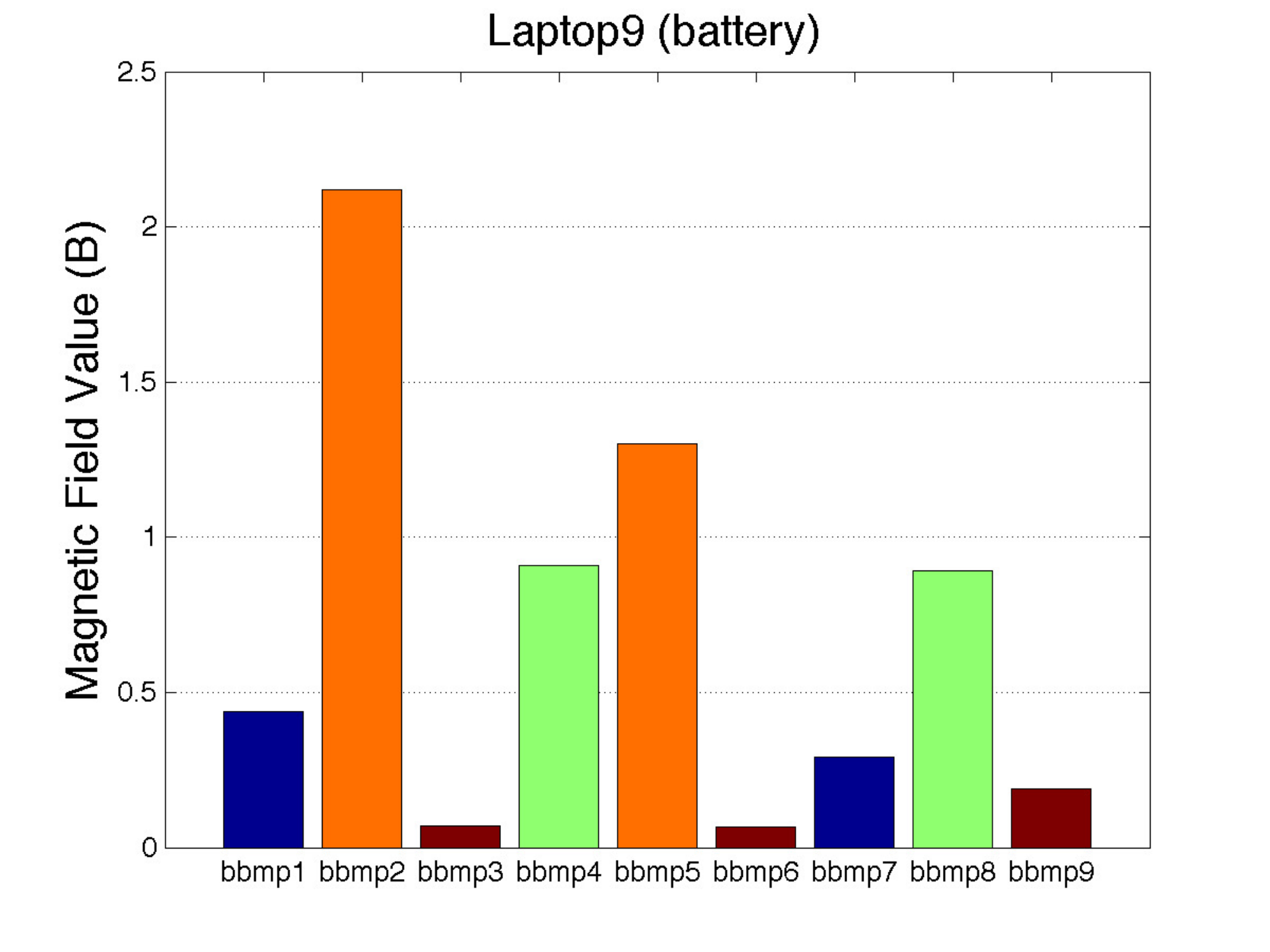}
}
\subfigure[]{
\includegraphics[width=3cm, height=3cm, keepaspectratio]{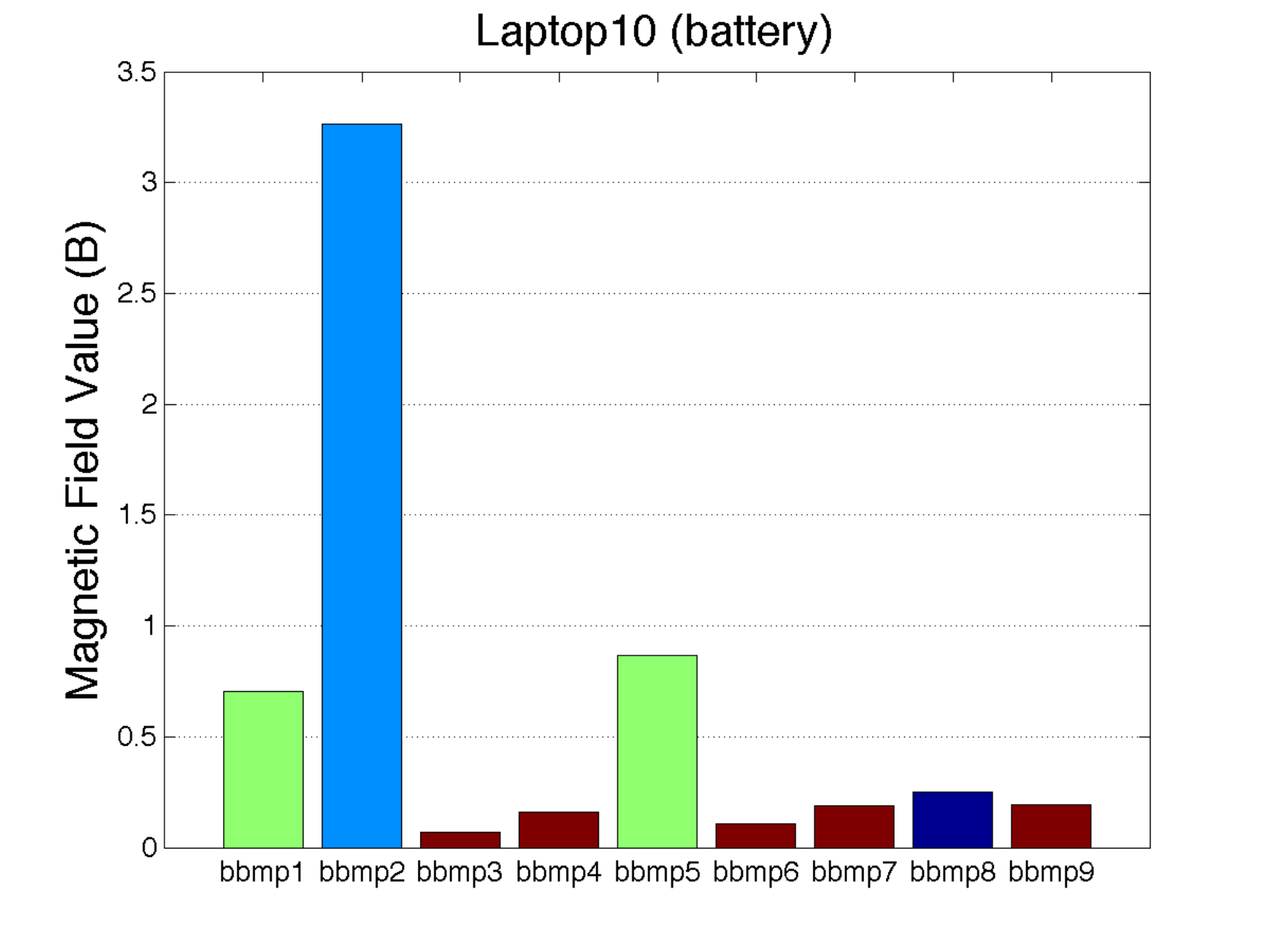}
}
\subfigure[]{
\includegraphics[width=3cm, height=3cm, keepaspectratio]{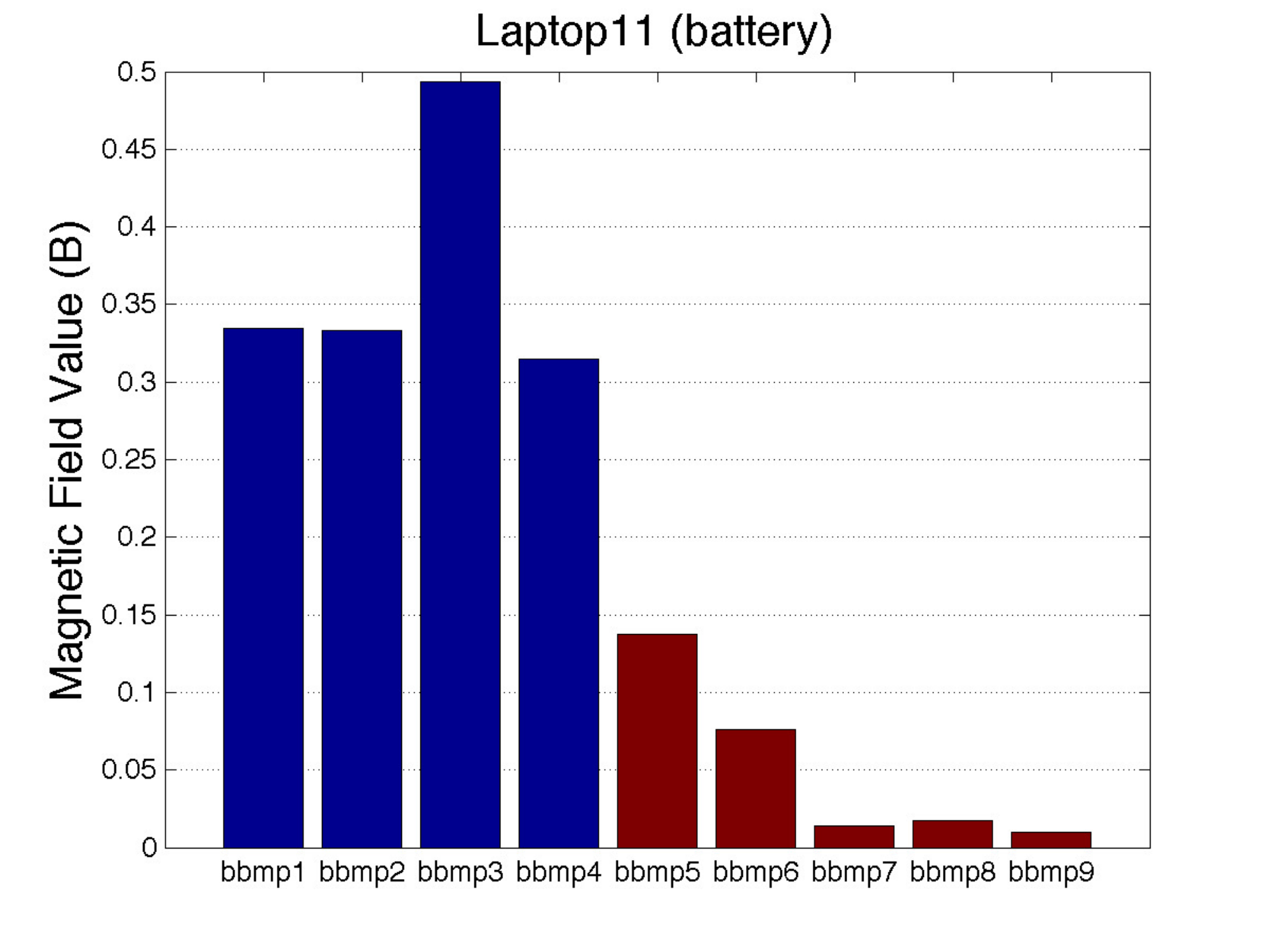}
}
\subfigure[]{
\includegraphics[width=3cm, height=3cm, keepaspectratio]{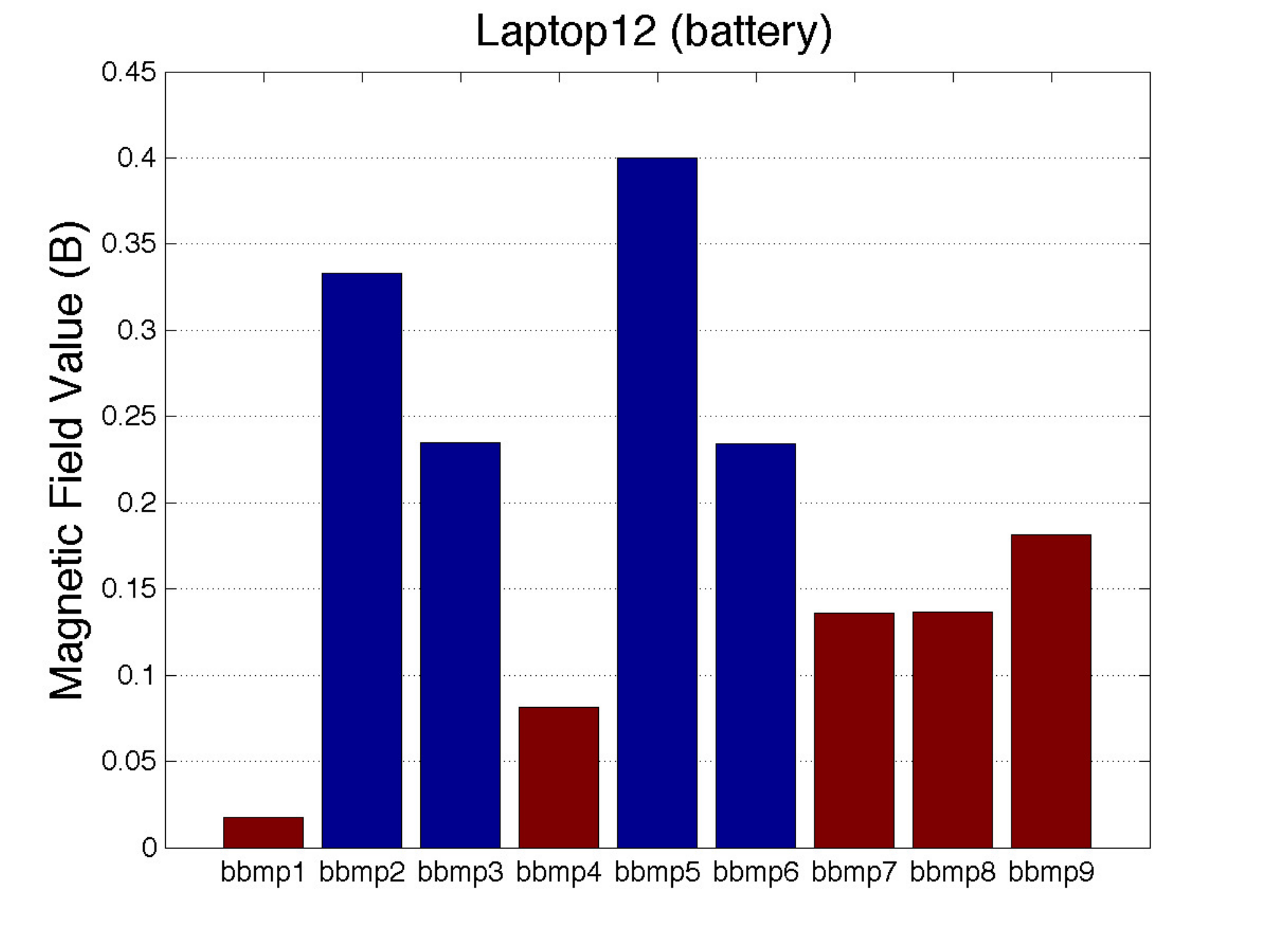}
}
\subfigure[]{
\includegraphics[width=3cm, height=3cm, keepaspectratio]{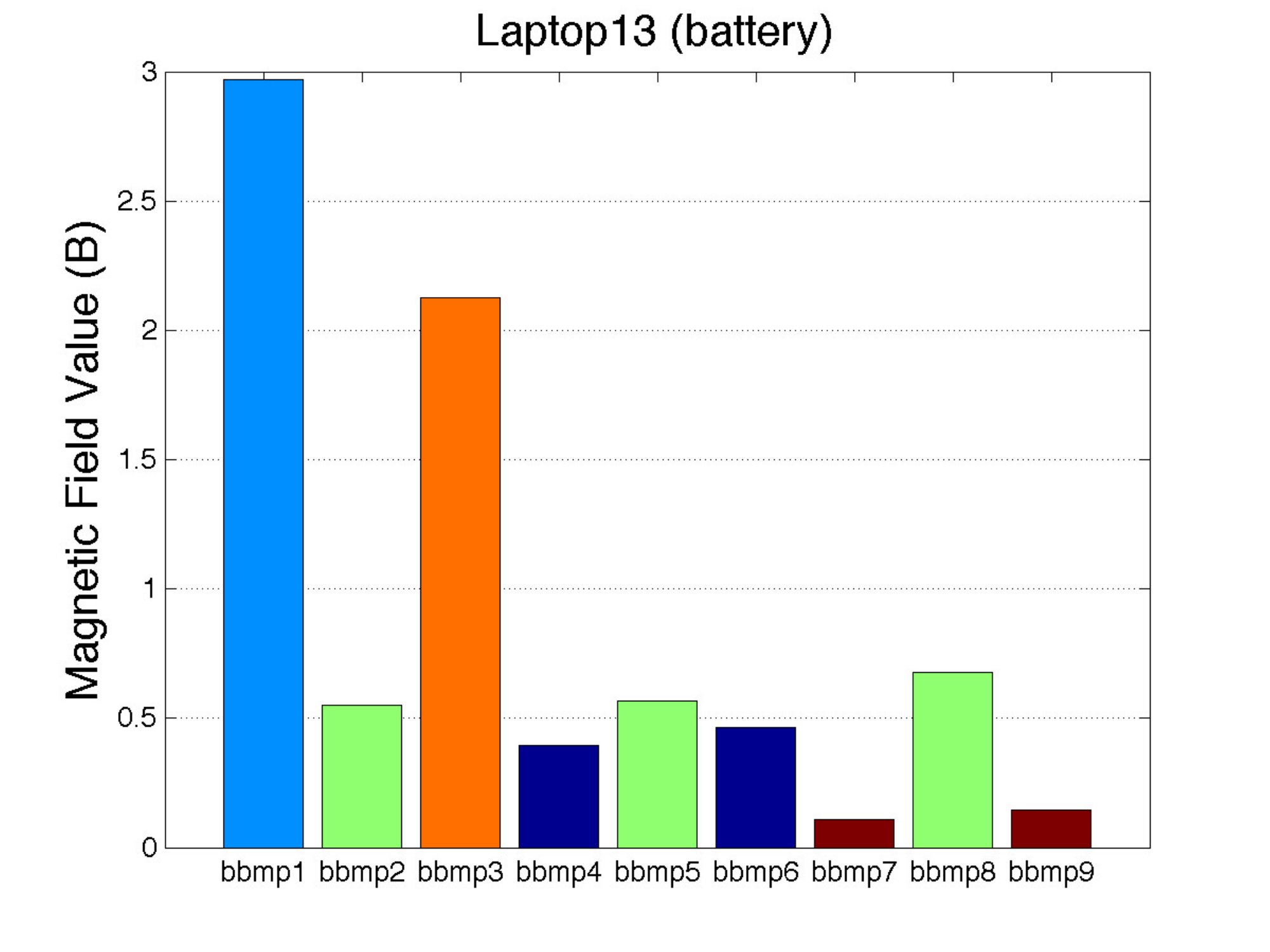}
}
\caption{Results of K-Medians on the dataset of bottom body points for the 13 laptops supplied by battery (a-m). Each color along the figures represents a cluster of bottom body points. The names of the bottom body points included in the cluster are reported in correspondence to the $x$ axis. $Y$ axis gives the magnetic field values measured at the bottom body points belonging to that cluster.}
\label{Figure9}
\end{center}
\end{figure*}

Results of the clustering by K-Medians algorithm are illustrated for each dataset in Figs.\ref{Figure6}-\ref{Figure9}. Given a dataset, colors represent the clusters of points from the 13 different laptops located at a given position ($x$ axis) and with a given magnetic field value ($y$ axis).

Test on the first dataset of top body points of laptops supplied by AC is depicted in Fig.\ref{Figure6}. It provides three clusters of laptop's dangerous points with corresponding magnetic field values higher than limit of 0.3 $\mu$T (cyan, orange and green). It is reasonable because laptops are fed by AC. The first cluster (cyan) contains the peak greater than 1.5 $\mu$T. The second cluster (orange) corresponds to high measures but less than 1.5 $\mu$T. The third cluster (green) groups middle values less than 0.6 $\mu$T. The last two clusters (blue and brown) are composed of points with magnetic field values less than 0.3 $\mu$T. They correspond to no dangerous laptop's positions on top body. The first safe cluster (blue) is associated with values less than 0.25 $\mu$T. The second cluster (brown) has very safe points with magnetic field values less than 0.1 $\mu$T.
 
The second experiment is on the dataset of top body points and laptop supplied by the battery. It is illustrated in Fig.\ref{Figure7}.
Currently, a new class is added to the category of no dangerous (green). It is reasonable because the processor is working at low frequency, less power is used, there is no cooling, hard disk is going to sleep and the intensity of the screen is lower. Consequently, we obtain two clusters of dangerous points (cyan and orange) with respect to the three clusters of the previous experiment. The first cluster (cyan) has the peak greater than 0.60 $\mu$T. The second cluster (orange) has points with high values greater than 0.30 $\mu$T and until 0.60 $\mu$T. The last three clusters (green, blue and brown) have no dangerous middle, safe and very safe measures. The first one (green) contains points corresponding to magnetic field values from 0.13 $\mu$T to 0.30 $\mu$T. The second cluster (blue) has points with safe magnetic field values between 0.06 $\mu$T and 0.12 $\mu$T. The last one (brown) corresponds to very safe values less than 0.06 $\mu$T.

In the last two experiments, the datasets of bottom body points of laptops supplied respectively by AC and battery, are considered. The results are illustrated in Figs.\ref{Figure8}-\ref{Figure9}, respectively. In these circumstances, the measured magnetic field values are higher compared to the points at top body. Consequently, a new class is added to the category of dangerous points (blue), while the very safe class is merged with the safe class of points (brown).
 
In the third experiment performed on the dataset of bottom body points of laptops supplied by AC (See Fig.\ref{Figure8} for reference), we obtain a class of peaks greater than 2.8 $\mu$T (cyan). Another class contains points with magnetic field values between 1 $\mu$T and 2.8 $\mu$T (orange). The third class has points at low-high values between 0.40 $\mu$T and 0.99 $\mu$T (green). The fourth class exhibits values less than 0.40 $\mu$T and greater than 0.15 $\mu$T (blue). The last class contains safe points with magnetic field values less than 0.15 $\mu$T (brown).
 
In the last experiment on the dataset of bottom body points of laptops supplied by the battery (See Fig.\ref{Figure9} for reference), we also obtain four dangerous classes (cyan, orange, green and blue) and only one no dangerous class (brown). The first class (cyan) has points with peaks greater than 2.3 $\mu$T. The second class ranges between 1 $\mu$T and 2.3 $\mu$T (orange). The third one contains points with low-high values between 0.53 $\mu$T and 0.99 $\mu$T (green). The fourth and fifth classes (blue and brown) exhibit respectively middle values less than 0.53 $\mu$T and safe values less than 0.20 $\mu$T.
 
Now an evaluation of the dangerous and no dangerous zones of the laptops based on the aforementioned classification will be performed. Looking at Figs.\ref{Figure6}-\ref{Figure9}, we find that point classes of portable computers supplied by alternating current and battery (top and bottom) with the highest peaks correspond in most cases to top-left and central zones of the laptops (i.e. tbmp1, tbmp2, tbmp4, tbmp5 in Fig.\ref{Figure4}(a) and bbmp1, bbmp4, bbmp7, bbmp8 in Fig.\ref{Figure4}(b)). It is compatible with the inner architecture of a laptop where the processor, GPU and cool system are located. On the other hand, it is also interesting to observe as classes of safe points correspond to no dangerous zones located at "peripheral" positions of the laptop (i.e. tbmp3, tbmp6, tbmp9 in Fig.\ref{Figure4}(a) and bbmp3, bbmp6, bbmp9 in Fig.\ref{Figure4}(b)). They are associated with other inner devices around the core devices, such as the hard disk. This observations track a new important research direction in the design of safe-EMF-emitting control units such as processors and GPU. 
 
To use laptop safely concerning magnetic field radiation, we propose the following: (i) to avoid direct contact with the fingers and hands with a laptop, and (ii) to avoid direct contact of the laptop and put laptops out of her/his lap, stomach or genitals. It can be done by: (i) using the external mouse, and (ii) using the external keyboard. We can realize that many laptop users utilize the external mouse regularly. Also, the integration of the lower quality keyboard into the laptop facilitates the use of an external keyboard, especially in the office. Still, the bad habit of putting the laptop to her/his lap or stomach should be avoided. Hence,  the observation that the magnetic field rapidly decreases after a few cms from the magnetic field emitter represents an encouraging news. Hence, it can be suggested to use the laptop at the office desk anytime, when it is possible.

\section{Conclusion}
The paper presented measurement results of the extremely low-frequency magnetic field below 300 Hz, which is especially dangerous to the laptop users. The testing was performed on 13 different laptop computers in the normal operation condition (typical office operation). The obtained measurement results were classified with unsupervised classifier K-Medians in order to distinguish different levels of the magnetic field radiation, which represents the risk to the laptop users. The classification showed that some points of the laptop computers emitted a very strong magnetic field, especially when laptops are supplied by alternating current. The bottom part of the laptop radiated the highest level of the magnetic field. Hence, the laptop must be used with extreme caution. We proposed to avoid using the laptop by putting it to her/his lap, stomach or genitals. 

Further research will be conducted toward measuring the laptop's magnetic field radiation in the normal and under stress conditions.
\begin{center}
{\bf Acknowledgement}
\end{center}

This work was partially supported by the Grant of the Ministry of Education, Science and Technological Development of the Republic Serbia, as a part of the project TR33037.

\end{document}